\documentclass[12pt]{article}

\usepackage{jheppub, bm, wrapfig,float,array}
\usepackage[utf8]{inputenc}
\numberwithin{equation}{section}
\renewcommand\afterTocRuleSpace{\bigskip}

\usepackage{amsfonts}
\usepackage{amsmath}
\usepackage{color}
\usepackage{graphicx}
\usepackage{float}
\usepackage[T1]{fontenc}
\usepackage[utf8]{inputenc}
\usepackage{alphabeta}
\usepackage{fancyvrb}
\usepackage[dvipsnames]{xcolor}
\usepackage{bold-extra}
\usepackage{lmodern}
\usepackage{multirow}
\usepackage[normalem]{ulem}

\usepackage{tikz-cd}

\newcommand{\Rep}{\mathsf{Rep}}
\newcommand{\Spin}{\text{Spin}}
\newcommand{\Pin}{\text{Pin}}
\newcommand{\TwoRep}{\mathsf{2Rep}}
\newcommand{\TwoVec}{\mathsf{2Vec}}
\newcommand{\ModC}{\mathsf{Mod}}

\usepackage{tikz}
\usetikzlibrary{arrows}
\usetikzlibrary{arrows.meta}
\usetikzlibrary{shapes.geometric,calc,arrows, positioning,shapes.misc,decorations.markings}
\tikzset{
  big arrow/.style={
    decoration={markings,mark=at position 1 with {\arrow[scale=2,#1]{>}}},
    postaction={decorate},
    shorten >=0.4pt},
  big arrow/.default=black}
  
\pgfdeclarelayer{edgelayer}
\pgfdeclarelayer{nodelayer}
\pgfsetlayers{edgelayer,nodelayer,main} 
\tikzstyle{none}=[inner sep=0pt] 
\tikzstyle{Star}=[draw, shape=star, fill=black, star points=8, inner sep=0pt, minimum size=8pt]

\tikzstyle{NodeCross}=[draw, shape=circle, cross out, inner sep=0pt, minimum size=6pt,line width=0.25mm]
\newcommand{\be}{\begin{equation}}
\newcommand{\ee}{\end{equation}}
\newcommand{\bit}{\begin{itemize}}
\newcommand{\eit}{\end{itemize}}
\newcommand{\ben}{\begin{enumerate}}
\newcommand{\een}{\end{enumerate}}

\newcommand{\ba}{\begin{aligned}}
\newcommand{\ea}{\end{aligned}}

\newcommand{\wt}{\widetilde}
\newcommand{\wh}{\widehat}

\newcommand{\Z}{{\mathbb Z}}
\newcommand{\R}{{\mathbb R}}
\newcommand{\C}{{\mathbb C}}

\newcommand{\bG}{{\mathbb G}}

\newcommand{\cA}{\mathcal{A}}
\newcommand{\cB}{\mathcal{B}}
\newcommand{\cC}{\mathcal{C}}

\newcommand{\cG}{\mathcal{G}}
\newcommand{\cH}{\mathcal{H}}
\newcommand{\cI}{\mathcal{I}}

\newcommand{\cS}{\mathcal{S}}

\newcommand{\G}[1]{\Gamma^{(#1)}}

\newcommand{\fT}{\mathfrak{T}}

\newcommand{\id}{\mathsf{id}}
\newcommand{\lid}{\mathsf{1}}
\renewcommand{\Vec}{\mathsf{Vec}}
\newcommand{\SPT}{\mathsf{SPT}}
\newcommand{\TQFT}{\mathsf{T}}

\usepackage{makecell}

\title{Universal Non-Invertible Symmetries}

\author{
Lakshya Bhardwaj, 
Sakura Sch\"afer-Nameki, 
Jingxiang Wu}

\affiliation{Mathematical Institute, University of Oxford,\\Andrew Wiles Building, Woodstock Road, Oxford, OX2 6GG, UK}

\abstract{It is well-known that gauging a finite 0-form symmetry in a quantum field theory leads to a dual symmetry generated by topological Wilson line defects. These are described by the representations of the 0-form symmetry group which form a 1-category. We argue that for a $d$-dimensional quantum field theory the full set of dual symmetries one obtains is in fact much larger and is described by a $(d-1)$-category, which is formed out of lower-dimensional topological quantum field theories with the same 0-form symmetry. We study in detail a 2-categorical piece of this $(d-1)$-category described by 2d topological quantum field theories with 0-form symmetry. We further show that the objects of this 2-category are the recently discussed 2d condensation defects constructed from higher-gauging of Wilson lines. 
Similarly, dual symmetries obtained by gauging any higher-form or higher-group symmetry also form a $(d-1)$-category formed out of lower-dimensional topological quantum field theories with that higher-form or higher-group symmetry.
A particularly interesting case is that of the 2-category of dual symmetries associated to gauging of finite 2-group symmetries, as it describes non-invertible symmetries arising from gauging 0-form symmetries that act on $(d-3)$-form symmetries. Such non-invertible symmetries were studied recently in the literature via other methods, and our results not only agree with previous results, but our approach also provides a much simpler way of computing various properties of these non-invertible symmetries. We describe how our results can be applied to compute non-invertible symmetries of various classes of gauge theories with continuous disconnected gauge groups in various spacetime dimensions. We also discuss the 2-category formed by 2d condensation defects in any arbitrary quantum field theory.
}

\begin{document}

\maketitle

\section{Introduction}

Recent months have seen discovery and new constructions of 
non-invertible symmetries in higher-dimensional Quantum Field Theories (QFTs). There are by now various field theoretically motivated constructions -- which however all in one way or another involve the gauging of symmetries. 
The constructions applicable in general dimension fall into three broad categories: 
duality defects \cite{Kaidi:2021xfk, Choi:2021kmx}, 
outer-automorphism gauging \cite{Bhardwaj:2022yxj, Antinucci:2022eat} and higher-gauging leading to condensation defects  \cite{Gaiotto:2019xmp, Roumpedakis:2022aik, Choi:2022zal}.
The full physical impact of these symmetries is yet to be uncovered, but already numerous phenomenological and other applications have been studied \cite{Heidenreich:2021xpr, Choi:2022jqy, Cordova:2022ieu, Choi:2022rfe, Bashmakov:2022jtl, Kaidi:2022uux}.

All the above constructions rely crucially on gauging of global symmetries in theories with invertible, possibly generalized, symmetries, i.e. 0-form or higher-form \cite{Gaiotto:2014kfa} or higher-group symmetries \cite{Sharpe:2015mja, Tachikawa:2017gyf, Benini:2018reh}. A comprehensive understanding of the process of gauging a generalized symmetry is therefore a pre-requisite to uncovering the full symmetry structure of a QFT.

The main catalyzer that lead to this vast generalization of symmetries is the insight that topological operators in a QFT should be regarded as symmetry generators \cite{Gaiotto:2014kfa}. In a given QFT, topological operators are of varied dimensionality and can have non-trivial inter-relations.
The fusion of these topological operators need not satisfy a group law -- thus the term ``non-invertibility''. The mathematical structure that captures such a collection of topological operators and their fusion, is a higher-fusion category (see e.g. \cite{Cui:2016bmd,douglas2018fusion,Johnson-Freyd:2020usu,decoppet2021weak,johnson-freyd_yu_2021,decoppet2022multifusion,decoppetRelativeDeligneTensor2022}). The recent progress in non-invertible symmetries has pointed sharply towards the fact that higher fusion-categories are the natural  framework for describing symmetries of QFTs. 
A slight bump on the road to fully developing this proposal is that the jury is still out for the precise mathematical definition of such higher categories. 
Nevertheless much can be uncovered by motivating the constructions from a physics point of view, see for example \cite{Kapustin:2010ta,Cui:2016bmd,Gaiotto:2019xmp,Johnson-Freyd:2020usu,Bhardwaj:2022yxj,kong2020algebraic,Kong:2014qka,Kong:2020wmn} for constructions in higher dimensional theories.  
This is the approach which we will take in this paper. 



The standard lore states that gauging a finite abelian 0-form symmetry $\Gamma^{(0)}$ in a $d$-dimensional theory implies a dual $(d-2)$-form symmetry, which is generated by topological Wilson lines. 
As it turns out, this is only the tip of the iceberg, and we will see that the most general dual symmetries arising from such a 0-form gauging are described by a $(d-1)$-category, which has topological defects of all dimensions up to codimension 1. 
This $(d-1)$-category is universal in the sense that it describes symmetries in any $d$-dimensional QFT that can be constructed by gauging the 0-form symmetry of another $d$-dimensional QFT.


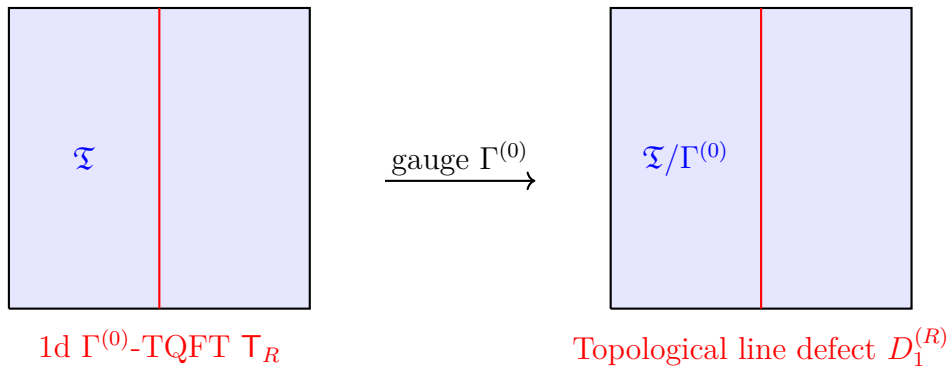
\begin{figure}
\centering
\begin{tikzpicture}
\begin{scope}[shift={(0,0)}]
\draw [fill=blue,opacity=0.1] 
(0,0) -- (4,0) -- (4,4) -- (0,4) -- (0,0);
\draw [thick] (0,0) -- (4,0) -- (4,4) -- (0,4) -- (0,0);
\draw [thick,red] (2,0) -- (2,4) ;
\node[red] at (2, -0.5) {1d $\Gamma^{(0)}$-TQFT $\TQFT_R$}; 
\draw [thick, ->] (5,1.7) -- (7,1.7);
 \node[blue] at (1,2) {$\mathfrak{T}$};
\node[black] at (6,2) {$\text{gauge }\Gamma^{(0)}$};
\end{scope}
\begin{scope}[shift={(8,0)}]
\draw [fill=blue,opacity=0.1] 
(0,0) -- (4,0) -- (4,4) -- (0,4) -- (0,0);
\draw [thick] (0,0) -- (4,0) -- (4,4) -- (0,4) -- (0,0);
\draw [thick,red] (2,0) -- (2,4) ;
\node[red] at (2, -0.5) {Topological line defect $D_1^{(R)}$}; 
 \node[blue] at (1,2) {$\mathfrak{T}/\Gamma^{(0)}$};
\end{scope}
\end{tikzpicture}
\caption{On the left hand side $\mathfrak{T}$ is a $d$-dimensional theory with $\Gamma^{(0)}$ 0-form global symmetry stacked with a 1d TQFT $\TQFT_R$ with $\Gamma^{(0)}$ 0-form symmetry, which is specified by a representation $R$ of $\G0$. After gauging the diagonal $\Gamma^{(0)}$ these two formerly decoupled theories become coupled, and the 1d TQFT becomes a topological line defect $D_1^{(R)}$ (topological Wilson line) of the $d$-dimensional gauged theory $\fT/\G0$. The topological line defect $D_1^{(R)}$ describes a (dual) symmetry of $\mathfrak{T}/\Gamma^{(0)}$. \label{fig:1dTQFT}}
\end{figure}


The idea for this construction is very intuitive and physically motivated. The simplest setup is a theory $\mathfrak{T}$ with a global 0-form symmetry $\Gamma^{(0)}$. Let us stack this theory with a 1d topological TQFT $\TQFT$, which also has a global $\Gamma^{(0)}$ 0-form symmetry, and gauge the diagonal $\Gamma^{(0)}$. This diagonal gauging has the effect of coupling the $d$-dimensional and the 1-dimensional systems together, such that $\TQFT$ becomes a topological line defect $D_1$ i.e. a symmetry generator in the gauged theory $\fT/\G0$. In particular,  the $d$-dimensional system $\fT/\G0$ and the 1-dimensional system $D_1$ are no longer factorized. This gauging is depicted in figure \ref{fig:1dTQFT}. 

This idea has a multitude of implications:
\ben
\item We can take $\TQFT$ to be any TQFT of dimension $p<d$ having $\G0$ 0-form symmetry. After the diagonal gauging,  $\TQFT$ becomes a $p$-dimensional topological defect of the theory $\fT/\G0$, which generates a symmetry of $\fT/\G0$.
\item We can consider a QFT $\fT$ which has 
some other symmetry $\cS$, which is either a higher-form or a higher-group symmetry. Now a $p$-dimensional TQFT $\TQFT$ with $\cS$ symmetry gives rise to a $p$-dimensional topological defect of the theory $\fT/\cS$, which generates a symmetry of $\fT/\cS$.
\een

 \begin{table}\centering
\begin{tabular}{|c|c|c|}\hline
Global Symmetry $\mathcal{S}$ & 2-Category $\mathcal{C}_2^\mathcal{S}$ or 2d TQFTs & Section \cr \hline\hline
 0-form symmetry $\Gamma^{(0)}$ & $\TwoRep(\Gamma^{(0)})$ &  \ref{T0F}  \cr \hline 
 1-form symmetry $\Gamma^{(1)}$ &  $\mathsf{2Vec}_{\wh{\G1}}$ &  \ref{sec:2catGamma1}  \cr \hline 
 2-group split $(\Gamma^{(0)}, \Gamma^{(1)}, \rho)$ & $\TwoRep\big(\cC_{\G0}(\G1,\rho)\big)$ &  \ref{T2GS}  \cr \hline 
 2-group  with $\omega \in H^3_\rho (\Gamma^{(0)}, \Gamma^{(1)})$ & $\TwoRep\big(\cC_{\G0}^\omega(\G1,\rho)\big)$
 & \ref{sec:2gpNonSplit} \cr \hline
\end{tabular}
\caption{Global symmetries $\mathcal{S}$, the dual symmetry 2-categories that are obtained after gauging $\mathcal{S}$ (or equivalently 2d TQFTs which have symmetry $\mathcal{S}$ and interfaces between these) and where we discuss TQFTs with these symmetries. We provide a physical interpretation in terms of TQFTs  in the sections linked.  \label{tab:SCats}}
\end{table}

In section \ref{sec:DualCat} we make the case that there is a universal symmetry sector for a $d$-dimensional QFT with gauged higher-form or higher-group symmetry $\mathcal{S}$. This is a  
 $(d-1)$-category of symmetries  arising as duals to gauging of $\mathcal{S}$, and captures TQFTs of dimension $< d$.  

Central to this endeavour of constructing such universal sectors in the symmetry category is a comprehensive characterization of the TQFTs that carry symmetries $\mathcal{S}$. We will carry out this program for $\mathcal{S}$  either a 0-form, 1-form or 2-group symmetry and for TQFTs of dimension $\le 2$. 
This will construct a 2-category that is a sub-category of the full $(d-1)$-category of dual symmetries.
We devote section \ref{TQFT} to the classification of 2d TQFTs which carry such symmetries, and discuss the fusion that the defects in these categories have. Let us briefly summarize the type of 2-categories that arise for each type of symmetry: we denote the 2-category associated to a symmetry $\mathcal{S}$ by $\mathcal{C}_2^\mathcal{S}$. The cases we study in this paper are summarized in table \ref{tab:SCats}.
We will now provide some physical intuition for the structures appearing in table \ref{tab:SCats}:


\begin{table}\centering
\begin{tabular}{|c|c|c|}\hline
 2-Category &  $\TwoRep(\Gamma^{(0)})$ & TQFT Language   \cr \hline\hline
\makecell{Simple \\ 
Objects} & Module category $\mathcal{M}_{\Vec_{\G0}}({\G0}',\alpha)$  &  
\makecell{$\TQFT_{{\G0}',\alpha}$ \\
2d  $\G0$-TQFT, unbroken ${\G0}'$ \\  with ${\G0}'$-SPT phase $\alpha$ } \cr \hline 
1-morphisms &  Module functors  &  $\G0$-invariant 1d interfaces   \cr \hline 
2-morphisms &  Natural transformations & $\G0$-invariant point junction  \cr \hline
\end{tabular}
\caption{Physical interpretation of the 2-category $\TwoRep(\G0)$ for 0-form symmetry given by a finite group $\mathcal{S}=\G0$. A simple object is labelled by  a subgroup ${\G0}'$ of $\G0$, together with an SPT phase $\alpha\in H^2({\G0}',U(1))$. 
\label{tab:Phys0}}
\end{table}


\paragraph{The symmetries $\mathcal{S}$.} these are $p$-form symmetries labeled by finite groups $\Gamma^{(p)}$, and the 2-group symmetries. The data for a 2-group includes a 0-form and 1-form symmetry with an action of the former on the latter
\be
\rho: \quad \Gamma^{(0)} \ \rightarrow \ \mathsf{Aut} (\Gamma^{(1)}) \,.
\ee
The split/non-split distinction is characterized by the triviality/non-triviality of the Postnikov class which is an element in the $\rho$-twisted group cohomology $\omega \in H^3_\rho (\Gamma^{(0)}, \Gamma^{(1)})$ \cite{Tachikawa:2017gyf, Benini:2018reh}.

\paragraph{The 2-categories $\mathcal{C}_2^{\mathcal{S}}$.} 
The 2-categories appearing in table \ref{tab:SCats} are 2-categories of representations and 2-representations of $\mathcal{S}$, which we physically identify with 2d TQFTs with symmetry $\mathcal{S}$. 
The objects in the 2-categories are mapped to  2d $\mathcal{S}$ symmetric TQFTs, $\mathcal{S}$-invariant interfaces between these 2d TQFTS correspond to $\mathcal{S}$-invariant 1-morphisms and point-junctions between 1d interfaces are $\mathcal{S}$-invariant 2-morphisms.  To uncover more structure of this 2-category requires computing the composition and fusion of these structures. 
To this end, we start by constructing \emph{minimal} 2d TQFTs with symmetry $\mathcal{S}$, corresponding to simple objects in the 2-categories. We then map out the entire 2-categorical structure based on the physical construction.

In table \ref{tab:Phys0} this is exemplified for the case when $\mathcal{S}$ is a 0-form symmetry given by a finite group $\Gamma^{(0)}$. Minimal 2d TQFTs with $\Gamma^{(0)}$-symmetry are characterized by an unbroken subgroup $\Gamma^{(0)'}$ and an SPT phase $\alpha \in H^2 (\Gamma^{(0)'}, U(1))$. These are denoted by $\TQFT_{\Gamma^{(0)'}, \alpha}$, which correspond to indecomposable module categories $\mathcal{M}_{\Vec_{\G0}}({\G0}',\alpha)$. 1-morphisms are interfaces between 2d TQFTs, e.g. the interfaces between the $\TQFT_{\Gamma^{(0)'}, \alpha}$ and the trivial $\Gamma^{(0)}$-symmetric TQFT $\TQFT_{\Gamma^{(0)}, \alpha=0}$ is simply the set of projective representations of $\Gamma^{(0)'}$ with cocyle $\alpha$. These in turn form a 1-category $\Rep_\alpha(\Gamma^{(0)'})$.

\paragraph{Condensation Defects.}
In section \ref{RCD} we study the relation of the above dual symmetries $\mathcal{C}_2^\mathcal{S}$ for $\cS=\G0$ being a finite 0-form symmetry to the condensation defects constructed using higher gauging of other symmetries \cite{Roumpedakis:2022aik, Choi:2022zal}. In summary, the surface defects that arise in the 2-category $\mathcal{C}_2^{\G0}$ are condensation defects and their fusion follows in a straight-forward way from our general analysis. 

In the same section, we also describe that the 2-category formed by two-dimensional condensation defects in any arbitrary quantum field theory $\fT$ is the 2-category $\mathsf{Mod}(\cB_\fT)$ formed by module 1-categories of the braided fusion 1-category $\cB_\fT$ formed by topological line defects of $\fT$.

\paragraph{Relation to Gauge Theories and Symmetry Categories.}
Finally, in section \ref{sec:GaugeTh} we also discuss the relation to gauge theories, where the topological operators generate symmetries. Here, we are able to connect to the non-invertible symmetries in gauge theories in any dimension $d$. 
In particular we compare to recent constructions of non-invertible symmetries including gauging of 0-, 1-form and 2-group symmetries and discuss how they fit into the universal 2-category subsector. We find agreement with the results in \cite{Bhardwaj:2022yxj} for $d=3,4$ gauge theories with gauge groups $\Pin^+(2N)$, $PO(2N)$ and $\Spin(8)\rtimes S_3$. The fusion of the non-condensation defects that are computed in \cite{Bhardwaj:2022yxj} are in full agreement, and using the methods in this paper, extend the results to include the fusion of the condensation defects that are present in these theories, and the fusion of the line defects living on both non-condensation and condensation defects.

An appendix \ref{App:Cats} summarizes various categorical background material, in particular for 2-categories, which the reader may find useful. Finally in appendix \ref{App:Z2Z2} the case of $\mathcal{S}= \Gamma^{(0)}=\mathbb{Z}_2\times \mathbb{Z}_2$ is explained, which is the simplest example with non-trivial SPT.

\section{Dual Symmetries from TQFTs}
\label{sec:DualCat}

In this section we argue that if one gauges a higher-form or higher-group symmetry $\cS$ in a quantum field theory $\fT$ of spacetime dimension $d$, then one obtains a $(d-1)$-category of dual symmetries in the gauged quantum field theory $\fT/\cS$. This $(d-1)$-category describes topological quantum field theories (TQFTs) of spacetime dimension less than $d$ and topological junctions between them.

\subsection{Wilson Lines}
Let us begin with the case where $\cS$ is a 0-form symmetry described by a finite group $\G0$. When $\G0$ is abelian, it is well known that the theory $\fT/\G0$, obtained after gauging the $\G0$ 0-form symmetry, carries a $(d-2)$-form symmetry described by the finite abelian group $\wh{\G0}$, which is known as the Pontryagin dual of $\G0$. The elements of $\wh{\G0}$ are irreducible representations of $\G0$, all of which have dimension 1. The group structure of $\wh{\G0}$ corresponds to the tensor product of these representations. The generators of the $(d-2)$-form symmetry are identified with the Wilson line defects for the gauged $\G0$.

When $\G0$ is non-abelian, we have irreducible representations of dimension bigger than 1. Consequently, the tensor product of irreducible representations forms a ring rather than a group. Thus, the dual symmetry generated by Wilson lines is non-invertible and is sometimes referred to as a non-invertible $(d-2)$-form symmetry. The information regarding this dual symmetry can be captured in the fusion 1-category $\Rep(\G0)$, which is the fusion category formed by all representations of $\G0$. For a summary of categorical notions see appendix \ref{App:Cats}.
For the special case of abelian $\G0$, we have the identification
\be
\Rep(\G0)=\Vec_{\wh{\G0}} \,,
\ee
where the latter category $\Vec_{\wh{\G0}}$ is the fusion category formed by $\wh{\G0}$-graded vector spaces. The structure of the category $\Vec_{\wh{\G0}}$ is completely captured by the group $\wh{\G0}$ and it makes sense to regard the dual symmetry as being described by a group, namely the $(d-2)$-form symmetry group.

The main point utilized in this paper is to note that the category $\Rep(\G0)$ also describes 1d and 0d topological systems with $\G0$ symmetry. More precisely, we have the following identification:
\ben
\item Objects of $\Rep(\G0)$ = 1d TQFTs with $\G0$ 0-form symmetry
\item Morphisms of $\Rep(\G0)$ = $\G0$-invariant topological junction local operators between 1d TQFTs with $\G0$ 0-form symmetry.
\een
Indeed, a 1d TQFT is described by a vector space $V$ which is identified as the state space of the theory. Then, a 1d TQFT with $\G0$ 0-form symmetry is obtained by converting $V$ into a representation $R$ for the group $\G0$. These are precisely the objects of the category $\Rep(\G0)$. On the other hand, topological local operators between two 1d TQFTs associated to vector spaces $V_1$ and $V_2$ are given by linear maps between the two vector spaces. Now, $\G0$-invariant operators are those linear maps which intertwine the $\G0$ actions between the two corresponding representations $R_1$ and $R_2$. These are precisely the morphisms of the category $\Rep(\G0)$.

The above correspondence between dual symmetries obtained after $\G0$ 0-form gauging and 1d TQFTs with $\G0$ 0-form symmetry is quite intriguing. There is a neat physical explanation of this correspondence. Consider the original ungauged $d$-dimensional theory $\fT$ having $\G0$ 0-form symmetry. One can stack on top of it a 1d TQFT $\TQFT_R$ with $\G0$ 0-form symmetry associated to a representation $R$ of $\G0$. So far the two systems are decoupled. Now, one can gauge the diagonal $\G0$ 0-form symmetry of the combined system. This gauging procedure couples the two systems, and $\TQFT_R$ becomes a topological line defect of the gauged theory $\fT/\G0$, which is then identified with the Wilson line defect in representation $R$. See figure \ref{fig:1dTQFT}. Similarly, a $\G0$ invariant local operator between two 1d TQFTs with $\G0$ symmetry becomes a topological local operator between the corresponding Wilson line defects in the theory $\fT/\G0$ after gauging.

The above physical procedure suggests that $\Rep(\G0)$ is not the full list of dual symmetries that one obtains in $\fT/\G0$. One should also obtain, for every $p<d$, dual symmetries arising from $p$-dimensional TQFTs having $\G0$ 0-form symmetry that become $p$-dimensional topological defects of the theory $\fT/\G0$ after the gauging. In total, as discussed in section \ref{DS} below, the dual symmetries form a $(d-1)$-category of which the 1-category $\Rep(\G0)$ is only a small piece.

\subsection{Wilson Surfaces}
Now consider the case where $\cS$ is a $p$-form symmetry for $0<p<d$. Such a symmetry is described by a finite group $\G p$ which must be abelian. It is well known that the theory $\fT/\G p$, obtained after gauging the $\G p$ $p$-form symmetry, carries a $(d-p-2)$-form symmetry described by the finite abelian group $\wh{\G p}$. 

The generators of the $(d-p-2)$-form symmetry $\wh{\G p}$ are $(p+1)$-dimensional topological defects, which can be recognized as $(p+1)$-dimensional Wilson surface defects that can be expressed as
\be
\text{exp}\left(2\pi i\int \chi(b_{p+1})\right) \,,
\ee
where $b_{p+1}$ is the gauge field in $\fT/\G p$ obtained by promoting the background field associated to $p$-form symmetry of $\fT$ to a dynamical field under the gauging process. It is a $(p+1)$-cocycle on spacetime valued in the group $\G p$. On the other hand,
\be
\chi:~\G p\to\R/\Z
\ee
is a homomorphism, which can be identified as an element of $\wh{\G p}$. Thus, such Wilson surfaces are labeled by elements of $\wh{\G p}$.

In the previous subsection we observed that the Wilson lines obtained by gauging $\G0$ 0-form symmetry can be identified with 1d TQFTs with $\G0$ 0-form symmetry. Similarly, the $(p+1)$-dimensional Wilson surfaces obtained by gauging $\G p$ $p$-form symmetry can be identified with $(p+1)$-dimensional TQFTs with $\G p$ $p$-form symmetry. In particular, the Wilson surfaces can be identified with $(p+1)$-dimensional SPT phases protected by $\G p$ $p$-form symmetry, which are $(p+1)$-dimensional TQFTs with $\G p$ $p$-form symmetry whose underlying $(p+1)$-dimensional TQFT, obtained after forgetting the $\G p$ $p$-form symmetry, is the trivial TQFT. 

Such SPT phases are labeled by elements $\chi\in\wh{\G p}$, which describes the partition function $Z_\chi(\Sigma_{p+1},B_{p+1})$ on a $(p+1)$-dimensional manifold $\Sigma_{p+1}$ in the presence of a background $B_{p+1}$ of the $\G p$ $p$-form symmetry as
\be
Z_\chi(\Sigma_{p+1},B_{p+1})=\text{exp}\left(2\pi i\int_{\Sigma_{p+1}} \chi(B_{p+1})\right) \,.
\ee
The identification between the SPT phases protected by $\G p$ $p$-form symmetry and Wilson surfaces obtained by gauging $\G p$ $p$-form symmetry works simply by identifying the character $\chi$ of $\G p$.


\begin{figure}
\centering

\begin{tikzpicture}
\begin{scope}[shift={(0,0)}]
 \node[blue] at (0.7,0.5) {$\mathfrak{T}$};
 \node[red] at (0.5,-1.2) {$\SPT_\chi$};
\draw [blue, fill=blue,opacity=0.1] (0,1.5) -- (0,-3) -- (3,-3) -- (3,1.5) -- (0,1.5);
\draw [blue, fill=blue,opacity=0.1] (-2,0) -- (0,1.5) -- (0,-3) -- (-2,-4.5) -- (-2,0);
  \draw [blue, fill=blue,opacity=0.1] (-2,0) -- (1,0) -- (3,1.5) -- (0,1.5) -- (-2,0);
      \draw [blue, fill=blue,opacity=0.1] (-2,-4.5) -- (1,-4.5) -- (3,-3) -- (0,-3) -- (-2,-4.5);
      \draw [blue, fill=blue,opacity=0.1] (1,0) -- (3,1.5) -- (3,-3) -- (1,-4.5) -- (1,0);
    \draw [red, fill=red,opacity=0.2] (-2,-2) -- (1,-2) -- (3,-0.5) -- (0,-0.5) -- (-2,-2); 
    \draw [black, thick] (-2,-2) -- (1,-2) -- (3,-0.5) -- (0,-0.5) -- (-2,-2);
 \end{scope}
\begin{scope}[shift={(-1,-3)}]
\draw [thick, ->] (5,1.7) -- (7,1.7);
\node[black] at (6,2) {$\text{gauge }\Gamma^{(p)}$};
\end{scope}
\begin{scope}[shift={(9,0)}]
 \node[blue] at (0.7,0.5) {$\mathfrak{T}/\Gamma^{(p)}$};
 \node[red] at (0.5,-1.2) {$D_{p+1}^{(\chi)}$};
\draw [blue, fill=blue,opacity=0.1] (0,1.5) -- (0,-3) -- (3,-3) -- (3,1.5) -- (0,1.5);
\draw [blue, fill=blue,opacity=0.1] (-2,0) -- (0,1.5) -- (0,-3) -- (-2,-4.5) -- (-2,0);
  \draw [blue, fill=blue,opacity=0.1] (-2,0) -- (1,0) -- (3,1.5) -- (0,1.5) -- (-2,0);
      \draw [blue, fill=blue,opacity=0.1] (-2,-4.5) -- (1,-4.5) -- (3,-3) -- (0,-3) -- (-2,-4.5);
      \draw [blue, fill=blue,opacity=0.1] (1,0) -- (3,1.5) -- (3,-3) -- (1,-4.5) -- (1,0);
    \draw [red, fill=red,opacity=0.2] (-2,-2) -- (1,-2) -- (3,-0.5) -- (0,-0.5) -- (-2,-2); 
    \draw [black, thick] (-2,-2) -- (1,-2) -- (3,-0.5) -- (0,-0.5) -- (-2,-2);
 \end{scope}
\end{tikzpicture}
\caption{On the left hand side $\mathfrak{T}$ is a $d$-dimensional theory with $\Gamma^{(p)}$  $p$-form global symmetry stacked with a $(p+1)$-dimensional TQFT $\SPT_\chi$ which is an SPT phase protected by $\G p$ $p$-form symmetry associated to a character $\chi$ of $\G p$. After gauging $\Gamma^{(p)}$, these two formerly decoupled theories become coupled, and the $p+1$ dimensional SPT becomes a $p+1$ dimensional topological defect $D_{p+1}^{(\chi)}$ (Wilson surface) in the $d$-dimensional gauged theory $\mathfrak{T}/\Gamma^{(p)}$, which generates a dual $(d-p-2)$-form symmetry of $\mathfrak{T}/\Gamma^{(p)}$. \label{fig:2dTQFT}}
\end{figure}
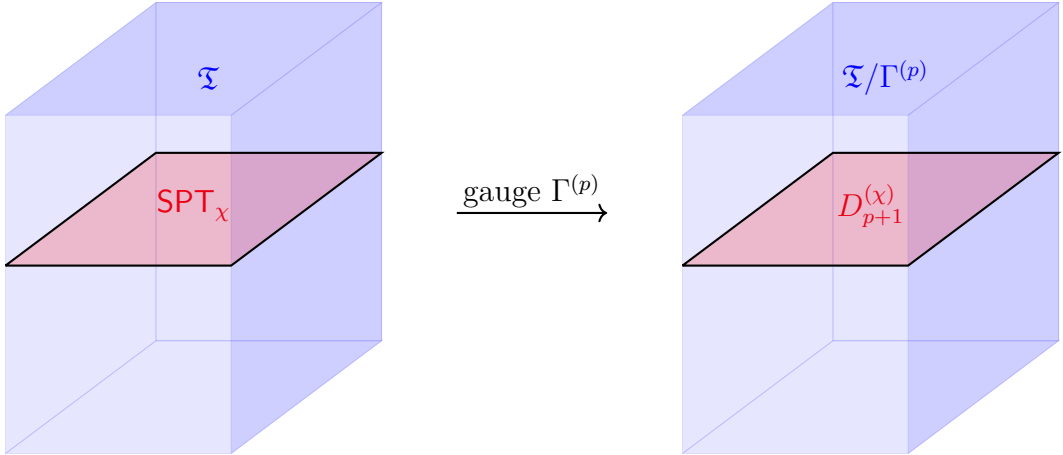


The physical explanation for this identification is similar to the one described in the previous subsection. Consider the original ungauged $d$-dimensional theory $\fT$ having $\G p$ $p$-form symmetry. One can stack on top of it a $(p+1)$-dimensional SPT phase $\SPT_\chi$ protected by $\G p$ $p$-form symmetry associated to a character $\chi$ of $\G p$. So far the two systems are decoupled. Now, one can gauge the diagonal $\G p$ $p$-form symmetry of the combined system. This gauging procedure couples the two systems, and $\SPT_\chi$ becomes a topological $(p+1)$-dimensional defect of the gauged theory $\fT/\G p$, which is then identified with the $(p+1)$-dimensional Wilson surface defect associated to character $\chi$. See figure \ref{fig:2dTQFT}. 

Again, as in the previous subsection, the above physical procedure suggests that Wilson surfaces valued in $\wh{\G p}$ do not comprise the full list of dual symmetries that one obtains in $\fT/\G p$. One should also obtain, for every $q<d$, dual symmetries arising from $q$-dimensional TQFTs having $\G p$ $p$-form symmetry that become $q$-dimensional topological defects of the theory $\fT/\G p$ after the gauging. In total, as discussed in section \ref{DS} below, the dual symmetries form a $(d-1)$-category containing a subcategory describing the Wilson surfaces valued in $\wh{\G p}$.

\subsection{Higher-Category of General Dual Symmetries}\label{DS}
In the previous two subsections, we saw that the known dual symmetries arising from the gauging of a $p$-form symmetry can all be identified with special types of TQFTs carrying the $p$-form symmetry. This perspective suggests that there are many more dual symmetries, corresponding to allowing arbitrary TQFTs carrying the $p$-form symmetry. In this subsection, we discuss the most general dual symmetries obtained in this way.

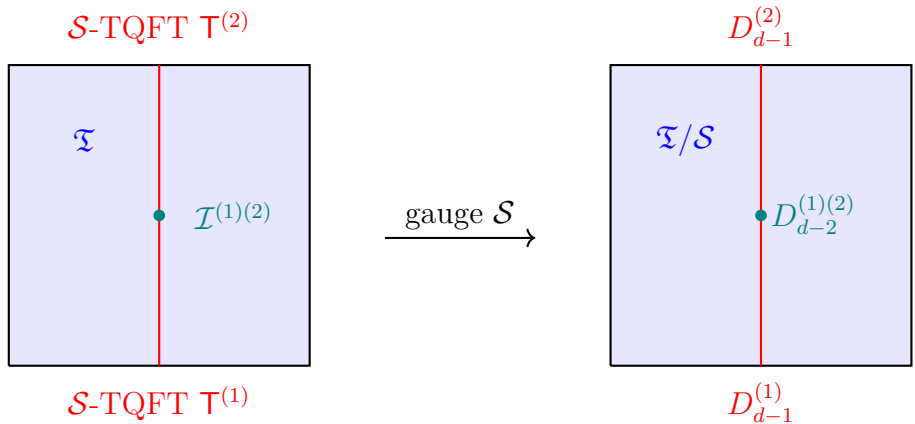
\begin{figure}
\centering
\begin{tikzpicture}
\begin{scope}[shift={(0,0)}]
\draw [fill=blue,opacity=0.1] 
(0,0) -- (4,0) -- (4,4) -- (0,4) -- (0,0);
\draw [thick] (0,0) -- (4,0) -- (4,4) -- (0,4) -- (0,0);
\draw [thick,red] (2,0) -- (2,4) ;
\node[red] at (2, -0.5) {$\mathcal{S}$-TQFT $\TQFT^{(1)}$};
\node[red] at (2, 4.5) {$\mathcal{S}$-TQFT $\TQFT^{(2)}$};
\draw [thick, ->] (5,1.7) -- (7,1.7);
 \node[blue] at (1,3) {$\mathfrak{T}$};
 \draw [teal,fill=teal] (2,2) ellipse (0.07 and 0.07);
 \node[teal] at (3,2) {$\mathcal{I}^{(1)(2)}$};
\node[black] at (6,2) {$\text{gauge }\mathcal{S}$};
\end{scope}
\begin{scope}[shift={(8,0)}]
\draw [fill=blue,opacity=0.1] 
(0,0) -- (4,0) -- (4,4) -- (0,4) -- (0,0);
\draw [thick] (0,0) -- (4,0) -- (4,4) -- (0,4) -- (0,0);
\draw [thick,red] (2,0) -- (2,4) ;
\node[red] at (2, -0.5) {$D_{d-1}^{(1)}$}; 
\node[red] at (2, 4.5) {$D_{d-1}^{(2)}$}; 
 \node[blue] at (1,3) {$\mathfrak{T}/\mathcal{S}$};
  \draw [teal,fill=teal] (2,2) ellipse (0.07 and 0.07);
 \node[teal] at (2.7,2) {$D_{d-2}^{(1)(2)}$};
\end{scope}
\end{tikzpicture}
\caption{On the left hand side, we have an $\cS$-symmetric $d$-dimensional QFT $\fT$ stacked with decoupled $\cS$-symmetric $(d-1)$-dimensional TQFTs $\TQFT^{(i)}$, along with an $\cS$-symmetric $(d-2)$-dimensional topological interface $\cI^{(1)(2)}$ between $\TQFT^{(i)}$, where $\cS$ is either a higher-form or a higher-group symmetry.
Upon gauging $\cS$ diagonally, $\TQFT^{(i)}$ become $(d-1)$-dimensional topological defects $D_{d-1}^{(i)}$ of the gauged $d$-dimensional theory $\fT/\cS$ and $\cI^{(1)(2)}$ becomes a topological $(d-2)$-dimensional defect $D_{d-2}^{(1)(2)}$ of $\fT/\cS$ living at the interface between defects $D_{d-1}^{(i)}$.
\label{fig:dTQFT} }
\end{figure}

Let us consider the general case where $\fT$ is a $d$-dimensional theory with a higher-form or higher-group symmetry $\cS$. Then, the dual symmetries obtained in the theory $\fT/\cS$ after gauging $\cS$ form a $(d-1)$-category $\cC^\cS_{d-1}$ which is comprised of the following information:
\ben
\item The objects of $\cC^\cS_{d-1}$ are $(d-1)$-dimensional TQFTs with $\cS$ symmetry. Such a TQFT leads to $(d-1)$-dimensional topological defect of $\fT/\cS$, which is constructed by first stacking the TQFT with $\fT$ and then gauging the diagonal $\cS$ symmetry. {This situation is depicted in figure \ref{fig:dTQFT}.}

\item The 1-morphisms of $\cC^\cS_{d-1}$ are $(d-2)$-dimensional topological interfaces with $\cS$ symmetry interpolating between $(d-1)$-dimensional TQFTs with $\cS$ symmetry. Such a topological interface leads to a $(d-2)$-dimensional topological defect in $\fT/\cS$ interpolating between $(d-1)$-dimensional topological defects of $\fT/\cS$ arising from $(d-1)$-dimensional TQFTs with $\cS$ symmetry as above. The construction involves first stacking the topological interface between TQFTs with $\fT$, and then gauging the diagonal $\cS$ symmetry. {These 1-morphisms are depicted in figure \ref{fig:dTQFT} as well.}

\item $2$-morphisms of $\cC^\cS_{d-1}$ are $(d-3)$-dimensional topological interfaces with $\cS$ symmetry interpolating between $(d-2)$-dimensional topological interfaces with $\cS$ symmetry interpolating between $(d-1)$-dimensional TQFTs with $\cS$ symmetry. Such a topological interface leads to a $(d-3)$-dimensional topological defect in $\fT/\cS$ interpolating between $(d-2)$-dimensional topological defects in $\fT/\cS$ interpolating between $(d-1)$-dimensional topological defects of $\fT/\cS$ arising from $(d-1)$-dimensional TQFTs with $\cS$ symmetry. The construction involves first stacking the topological interface between topological interfaces between TQFTs with the theory $\fT$, and then gauging the diagonal $\cS$ symmetry. 
\item The story is similar for higher-morphisms which are identified with $\cS$ symmetric higher-codimensional topological interfaces between TQFTs with $\cS$ symmetry, and describe higher-codimensional topological defects (in general interpolating between other topological defects) in $\fT/\cS$.
\een
Note that $\cC^\cS_p$ is a sub-$p$-category of the $(d-1)$-category $\cC^\cS_{d-1}$ for $p<d-1$. In section \ref{TQFT} we study the 2-category $\cC^\cS_2$ in detail for various possible $\cS$.

\section{Dual 2-Categorical Symmetries}
\label{TQFT}

In this section, we study the 2-categories $\cC^\cS_2$ of dual symmetries defined in section \ref{DS}. The objects of such a 2-category are 2d TQFTs with $\cS$ symmetry. As far as higher-form symmetries are concerned, 2d theories can only carry 0-form or 1-form symmetries. And as far as higher-group symmetries are concerned, 2d theories can only carry 2-group symmetries. Thus, $\cC^\cS_2$ is non-trivial only if $\cS$ is one of the following:
\ben
\item 0-form symmetry.
\item 1-form symmetry.
\item 2-group symmetry.
\een

\subsection{2d TQFTs}
\paragraph{Classification.}
2d TQFTs have a very simple classification, being classified\footnote{More precisely any unitary 2d TQFT is related to the 2d TQFTs we study by adding local `Euler number' counterterms.} by a positive integer $n$. This integer characterizes the number of vacua of the 2d TQFT.

For $n=1$, we have the trivial TQFT which has a single vacuum. For general $n$, the theory is such that in each vacuum, it behaves like the trivial theory. One can also represent this fact as an equation of the form
\be
\TQFT_n=\bigoplus_{i=1}^n \mathsf{Trivial}_i \,,
\ee
where $\TQFT_n$ denotes a 2d TQFT with $n$ vacua, and $\mathsf{Trivial}_i$ denotes a copy of the trivial 2d TQFT, describing the behavior of $\TQFT_n$ in its $i$-th vacuum.

\paragraph{Local operators.}
The vector space of local operators of $\TQFT_n$ is $n$-dimensional, with a distinguished basis given by operators $\id_i$ for $1\le i\le n$. The fusion of these operators is given by
\be\label{fus0}
\id_i\id_j=\delta_{ij}\id_i \,,
\ee
which converts the space of local operators into an algebra. The distinguished operator $\id_i$ projects a generic operator to the $i$-th vacuum, which justifies the fusion rule (\ref{fus0}). Note that
\be
\id:=\sum_i \id_i
\ee
is the identity local operator.

\paragraph{Line operators.}
The line operators of $\TQFT_n$ are described as follows. First let us discuss line operators with a single vacuum. There is a line operator $\lid_{ii}$ for each $i$, which describes the identity line in the $i$-th vacuum. There are also line operators $\lid_{ij}$ for $i\neq j$, which describes an interface from the vacuum $i$ to vacuum $j$, and is interpreted as the domain wall between the two vacua. The fusions of these lines are
\be
\lid_{ij}\circ \lid_{kl}=\delta_{jk}\lid_{il}
\ee
and
\be
\lid:=\bigoplus_i\lid_{ii}
\ee
is the identity line for the full theory $\TQFT_n$. More general line operators are constructed by taking direct sums of the above line operators.

All line operators and local operators of $\TQFT_n$ combine to form a multi-fusion category known as $\mathsf{Mat}_n(\Vec)$. For $n=1$, the multi-fusion category $\mathsf{Mat}_1(\Vec)$ is identified with the fusion category of complex vector spaces, which will be denoted as $\Vec$.

\paragraph{2-category $\TwoVec$ formed  by 2d TQFTs.}
2d TQFTs sit in a fusion 2-category known as $\mathsf{2Vec}$, whose objects are `2-vector spaces' i.e. finite semi-simple 1-categories. This 1-category is the category of boundary conditions of the associated 2d TQFT. For $\TQFT_n$, this 1-category has $n$ simple objects corresponding to the boundary conditions $\lid_i$ for each vacuum $1\le i\le n$.

The 1-morphisms from $\TQFT_n$ to $\TQFT_m$ are topological interfaces between the two 2d TQFTs. Such 1-morphisms form a 1-category which is the category of functors from a semi-simple 1-category with $n$ simples to a semi-simple 1-category with $m$ simples. This 1-category of 1-morphisms can be recognized as $\mathsf{Mat}_{m\times n}(\Vec)$, whose objects are $m\times n$ matrices with elements being vector spaces. Concretely, the simple objects of $\mathsf{Mat}_{m\times n}(\Vec)$ are interfaces $\lid_{i,a}$ (having single vacuum) with $1\le i\le n$ and $1\le a\le m$ from vacuum $i$ of $\TQFT_n$ to vacuum $a$ of $\TQFT_m$. The fusion category $\mathsf{Mat}_{n}(\Vec)$ thus describes the category of 1-endomorphisms of $\TQFT_n$.

The composition of 1-morphisms works via fusion rules
\be
\lid_{i,a}\circ \lid_{b,x}=\delta_{ab}\lid_{i,x} \,,
\ee
where $\lid_{i,a}$ goes from $\TQFT_n$ to $\TQFT_m$, $\lid_{b,x}$ goes from $\TQFT_m$ to $\TQFT_p$, and $\lid_{i,x}$ goes from $\TQFT_n$ to $\TQFT_p$. This composition provides the monoidal structure for the fusion category $\mathsf{Mat}_{n}(\Vec)$ of 1-endomorphisms.

The monoidal structure on the 2-category is obtained by stacking TQFTs. We have
\be
\TQFT_n\otimes\TQFT_m=\TQFT_{nm} \,.
\ee
The vacua of $\TQFT_{nm}$ are given by pairs $(i,a)$ where $i$ is a vacuum for $\TQFT_n$ and $a$ is a vacuum for $\TQFT_m$. For 1-morphisms, the fusion is
\be
\lid_{i,a}\otimes\lid_{x,u}=\lid_{(i,x),(a,u)} \,,
\ee
where $\lid_{i,a}$ goes from $\TQFT_n$ to $\TQFT_m$, $\lid_{x,u}$ goes from $\TQFT_p$ to $\TQFT_q$, and $\lid_{(i,x),(a,u)}$ goes from $\TQFT_{np}$ to $\TQFT_{mq}$.

\subsection{2d TQFTs with 0-Form Symmetry}\label{T0F}

In this section we construct 2d TQFTs, which have a 0-form symmetry $\Gamma^{(0)}$, and provide the 2-categorical structure, which is given by $\TwoRep(\Gamma^{(0)})$. See table \ref{tab:Phys0} for a summary of the basic ingredients of this 2-category.

\subsubsection{2-Categorical Structure and $\TwoRep(\Gamma^{(0)})$}
\label{sec:2catGamma0}

\begin{figure}
\centering

\begin{tikzpicture}
\draw [blue,thick](-2.5,-1.5) -- (-0.5,0.5) -- (-0.5,2) (-1.5,-0.5) -- (-0.5,-1.5) (-0.5,0.5) -- (1.5,-1.5);
\draw [teal,fill=teal] (-0.5,0.5) ellipse (0.05 and 0.05);
\draw [teal,fill=teal] (-1.5,-0.5) ellipse (0.05 and 0.05);
\draw [blue,thick,-stealth](-2.125,-1.125) -- (-2,-1);
\draw [blue,thick,-stealth](-0.875,-1.125) -- (-1,-1);
\draw [blue,thick,-stealth](-1.125,-0.125) -- (-1,0);
\draw [blue,thick,-stealth](0.5,-0.5) -- (0.375,-0.375);
\draw [blue,thick,-stealth](-0.5,1.125) -- (-0.5,1.25);
\node[blue] at (-2.5,-2) {$L_{\gamma_1}$};
\node[blue] at (-0.5,-2) {$L_{\gamma_2}$};
\node[blue] at (1.5,-2) {$L_{\gamma_3}$};
\node[blue] at (-0.5,2.5) {$L_{\gamma_1\gamma_2\gamma_3}$};
\node[blue] at (-1.55,0.2247) {$L_{\gamma_1 \gamma_2}$};
\node[teal] at (-2.25,-0.5) {{$O_{\gamma_1,\gamma_2}$}};
\node[teal] at (0.3,0.5) {{$O_{\gamma_1\gamma_2, \gamma_3}$}};
\node at (2.5,0.5) {=};
\begin{scope}[shift={(6,0)}]
\draw [blue,thick](-2.5,-1.5) -- (-0.5,0.5) -- (-0.5,2) (0.5,-0.5) -- (-0.5,-1.5) (-0.5,0.5) -- (1.5,-1.5);
\draw [teal, fill=teal] (-0.5,0.5) ellipse (0.05 and 0.05);
\draw [teal,fill=teal] (0.5,-0.5) ellipse (0.05 and 0.05);
\draw [blue,thick,-stealth](-1.5,-0.5) -- (-1.375,-0.375);
\draw [blue,thick,-stealth](-0.125,-1.125) -- (0,-1);
\draw [blue,thick,-stealth](0.125,-0.125) -- (0,0);
\draw [blue,thick,-stealth](1.125,-1.125) -- (1,-1);
\draw [blue,thick,-stealth](-0.5,1.125) -- (-0.5,1.25);
\node[blue] at (-2.5,-2) {$L_{\gamma_1}$};
\node[blue] at (-0.5,-2) {$L_{\gamma_2}$};
\node[blue] at (1.5,-2) {$L_{\gamma_3}$};
\node[blue] at (-0.5,2.5) {$L_{\gamma_1\gamma_2\gamma_3}$};
\node[blue] at (0.5,0.25) {$L_{\gamma_2\gamma_3}$};
\node[teal] at (1.125,-0.375) {$O_{\gamma_2,\gamma_3}$};
\node[teal] at (-1.5,0.5) {$O_{\gamma_1, \gamma_2\gamma_3}$};
 \end{scope}
\end{tikzpicture}
\caption{Associativity condition satisfied by line operators $L_\gamma$ and local operators $O_{\gamma,\gamma'}$.
\label{fig:Assoc}}

\end{figure}
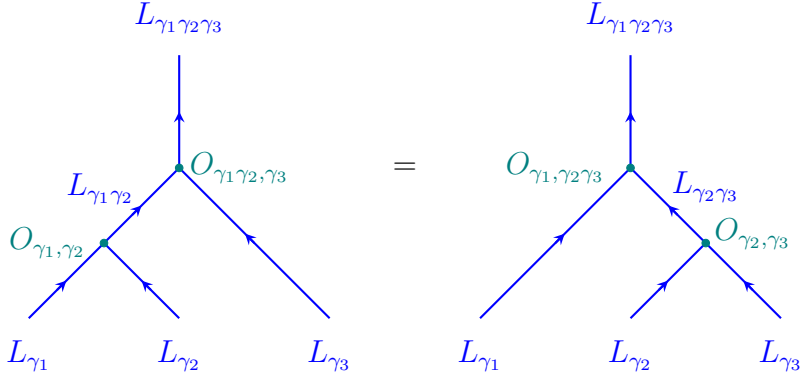

\paragraph{Making a 2d TQFT 0-form symmetric.}
Let $\G0$ be a finite group. A 2d TQFT $\TQFT_n$ characterized by an integer $n$ acquires a $\G0$ 0-form symmetry if one specifies a monoidal functor
\be
\cC_{\G0}\to\mathsf{Mat}_n(\Vec) \,,
\ee
where $\cC_{\G0}$ is the monoidal category defined in appendix \ref{App:Cats}, see also \cite{etingof2016tensor} sections 2.3 and 2.11. This means that we specify, in $\TQFT_n$, a line operator $L_\gamma$ for every $\gamma\in\G0$ such that
\be
L_\gamma\circ L_{\gamma'}=L_{\gamma\gamma'}
\ee
and local operators $O_{\gamma,\gamma'}$ living at the junction of $L_\gamma$, $L_{\gamma'}$ and $L_{\gamma\gamma'}$, such that these operators satisfy an associativity condition as shown in figure \ref{fig:Assoc}.


\paragraph{2d SPT phases protected by $\Gamma^{(0)}$.}
Let us study 2d TQFTs with a single vacuum and having $\G0$ 0-form symmetry. In other words, we are studying a phase in which all of $\G0$ is spontaneously preserved. These are known as \textit{2d SPT phases protected by a $\G0$ 0-form symmetry}. In this case, we are studying monoidal functors
\be
\cC_{\G0}\to\Vec \,,
\ee
which are known to be classified by elements of
\be
H^2\big(\G0,U(1)\big) \,.
\ee
Let us see quickly how this classification roughly comes about. We must have
\be
L_\gamma=\lid \,,
\ee
which is the identity line, for all $\gamma\in\G0$. Each local operator $O_{\gamma,\gamma'}$ is thus a local operator in the trivial theory, and hence can be identified with a non-zero complex number. The associativity condition imposes that these numbers describe a group 2-cocycle for $\G0$.

A slight extension of the above is provided by 2d TQFTs with $n$ vacua, but $\G0$ is spontaneously preserved in each vacuum. Such a theory can be described as a direct sum
\be
\bigoplus_{i=1}^n \SPT_i \,,
\ee
where $\SPT_i$ describes the SPT phase appearing in the $i$-th vacuum, which is labeled by an element $\alpha_i\in H^2\big(\G0,U(1)\big)$.

\paragraph{Spontaneous symmetry breaking phases.}
We can study more general phases which involve \textit{spontaneous breaking of $\G0$ 0-form symmetry}. Let us assume that we have a vacuum $v$ in which a subgroup ${\G0}'<\G0$ is spontaneously preserved, while the rest of $\G0/{\G0}'$ symmetry is spontaneously broken. Then, the TQFT must contain at least $\left|\G0/{\G0}'\right|$ number of vacua related to $v$ by (action of) elements of $\G0/{\G0}'$. The spontaneously preserved subgroup of $\G0$ in a vacuum $[\gamma]\cdot v$ obtained by acting on $v$ by an element $[\gamma]\in \G0/{\G0}'$ can be described as
\be
\gamma{\G0}'\gamma^{-1} \,,
\ee
where $\gamma\in\G0$ is a representative of the class $[\gamma]\in \G0/{\G0}'$. Note that $\gamma{\G0}'\gamma^{-1}$ is isomorphic to ${\G0}'$ even though it is a different subgroup of $\G0$.

\paragraph{`Minimal' 2d TQFTs with 0-form symmetry.}
Consider a minimal TQFT which has precisely $\left|\G0/{\G0}'\right|$ number of vacua including the vacuum $v$. Since, ${\G0}'$ is spontaneously preserved in $v$, it can carry an SPT phase for ${\G0}'$ associated to an element $\alpha\in H^2\big({\G0}',U(1)\big)$. But since $\G0/{\G0}'$ relates every other vacuum to $v$, every other vacuum must also carry the same SPT phase described by the element $\alpha$. Thus, such a minimal TQFT having $\G0$ 0-form symmetry is described by the following two pieces of data:
\ben
\item A subgroup ${\G0}'$ of $\G0$ up to conjugation by elements of $\G0$.
\item An element $\alpha\in H^2\big({\G0}',U(1)\big)$.
\een
The above discussed SPT phases for $\G0$ provide examples of such minimal TQFTs for which ${\G0}'=\G0$.

The most general 2d TQFT with $\G0$ 0-form symmetry can be expressed as a finite direct sum
\be
\bigoplus_i \TQFT_{{\G0_i}',\alpha_i} \,,\label{eq:generalTFTasdirectsum}
\ee
where $\TQFT_{{\G0_i}',\alpha_i}$ denotes minimal TQFT specified by ${\G0_i}'\subseteq\G0$ and $\alpha_i\in H^2\big({\G0_i}',U(1)\big)$. Such a TQFT has a total of
\be
n=\sum_i \left|\G0/{\G0_i}'\right|
\ee
number of vacua.

\paragraph{2-category of 2d TQFTs with 0-form symmetry.}
2d TQFTs with $\G0$ symmetry are naturally  objects of a 2-category. The 1-morphisms in this 2-category correspond to $\G0$ symmetric topological interfaces between 2d TQFTs with $\G0$ symmetry, and 2-morphisms correspond to $\G0$ invariant local operators between $\G0$ symmetric topological interfaces.

\begin{figure}
\centering
\begin{tikzpicture}[rotate=-90]
\draw [thick](1.5,-1.5) -- (-1.5,-1.5);
\draw [thick,blue](-0.5,-1.5) -- (1.5,-3.5) (0.5,-1.5) -- (1.5,-2.5);
\draw [thick,-stealth](1.125,-1.5) -- (1,-1.5);
\draw [thick,-stealth](0.125,-1.5) -- (0,-1.5);
\draw [thick,-stealth](-0.875,-1.5) -- (-1,-1.5);
\draw [thick,-stealth,blue](0.5,-2.5) -- (0.375,-2.375);
\draw [thick,-stealth,blue](1.125,-2.125) -- (1,-2);
\draw [black,fill=black] (-0.5,-1.5) ellipse (0.05 and 0.05);
\draw [black,fill=black] (0.5,-1.5) ellipse (0.05 and 0.05);
\node[blue] at (2,-2.5) {{$L_{\gamma_2}$}};
\node[blue] at (2,-3.5) {{$L_{\gamma_1}$}};
\node at (0.5,-1) {{$O_{\gamma_2}$}};
\node at (-0.5,-1) {{$O_{\gamma_1}$}};
\node at (0,0.5) {=};
\begin{scope}[shift={(0,5)}]
\draw [thick](1.5,-1.5) -- (-1.5,-1.5);
\draw [thick,blue](-0.5,-1.5) -- (1.5,-3.5) (0.5,-2.5) -- (1.5,-2);
\draw [thick,-stealth](0.625,-1.5) -- (0.5,-1.5);
\draw [blue,thick,-stealth](0.1255,-2.1335) -- (0,-2.008);
\draw [thick,-stealth](-0.875,-1.5) -- (-1,-1.5);
\draw [thick,-stealth,blue](1,-3) -- (0.875,-2.875);
\draw [thick,-stealth,blue](1.25,-2.125) -- (1,-2.25);
\draw [black,fill=black] (-0.5,-1.5) ellipse (0.05 and 0.05);
\draw [blue,fill=blue] (0.5,-2.5) ellipse (0.05 and 0.05);
\node[blue] at (2.008,-2.008) {$L_{\gamma_2}$};
\node[blue] at (2,-3.5) {$L_{\gamma_1}$};
\node[blue] at (0.3523,-3.2324) {\scriptsize{$\alpha(\gamma_1,\gamma_2)$}};
\node at (-0.5,-0.75) {{$O_{\gamma_1\gamma_2}$}};
\node[blue] at (-0.5,-2.5) {$L_{\gamma_1\gamma_2}$};
\end{scope}
\end{tikzpicture}
\caption{The operators $O_{\gamma_i}$ for $\gamma_i\in{\G0}'$ acting on the vector space carried by the boundary condition need to describe a projective representation of ${\G0}'$, because the junction local operator where lines $L_{\gamma_i}$ meet contributes the phase $\alpha_{\gamma_1,\gamma_2}\in U(1)$. \label{fig:Proj}}
\end{figure}
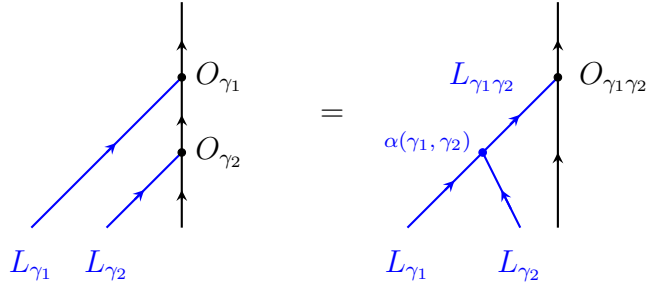

We can recognize this 2-category as $\TwoRep(\G0)$, which is the 2-category formed by module categories of the category $\Vec_{\G0}$ of $\G0$-graded complex vector spaces. The identification between a $\G0$ symmetric 2d TQFT and a module category of $\Vec_{\G0}$ goes via the boundary conditions of the 2d TQFT as follows:\\
Let us consider the theory $\TQFT_{{\G0}',\alpha}$. Recall that if we do not care about symmetries, then its boundary conditions are described by a finite semi-simple category with $n=\left|\G0/{\G0}'\right|$ number of simple objects $\lid_i$ labeled by the vacua of $\TQFT_{{\G0}',\alpha}$. The $\G0$ action permutes all these simple objects into each other, and hence $\G0$ invariant boundary conditions must be of the form
\be
m\bigoplus_{i=1}^{n}\lid_i
\ee
for some positive integer $m$. Additionally, the $m$ dimensional vector space carried by this boundary condition must form a projective representation of ${\G0}'$ lying in the class $\alpha$. See figure \ref{fig:Proj}. 


Such a boundary condition carrying an irreducible projective representation describes an indecomposable module category of $\Vec_{\G0}$. Moreover, the indecomposable module categories corresponding to two different irreducible representations are equivalent as module categories. Thus, an equivalence class of module categories captures a $\G0$ symmetric 2d TQFT by recognizing it as the equivalence class of its minimal $\G0$ symmetric boundary conditions. This is a rather neat example of what is known as \textit{bulk-boundary correspondence}.

It is possible to describe the full structure of this fusion 2-category $\TwoRep(\G0)$ by performing calculations in the underlying 2d TQFTs without $\G0$ symmetry. We describe these computations and the resulting structure of $\TwoRep(\G0)$ below.

\paragraph{Fusion of objects.}
Let us describe the monoidal structure on the objects of this 2-category, which is physically the operation of obtaining a $\G0$ symmetric 2d TQFT by stacking two $\G0$ symmetric 2d TQFTs. Let us begin with the case of abelian $\G0$ for simplicity. We want to determine the fusion of $\TQFT_{{\G0_1}',\alpha_1}$ and $\TQFT_{{\G0_2}',\alpha_2}$. Let $i$ parametrize vacua of $\TQFT_{{\G0_1}',\alpha_1}$ and $a$ parametrize vacua of $\TQFT_{{\G0_2}',\alpha_2}$. The resulting 2d TQFT has vacua parametrized by pairs $(i,a)$. In each such vacuum, only 
\be\label{G012}
{\G0_{12}}':={\G0_1}'\cap {\G0_2}'
\ee
is spontaneously preserved, while the rest of $\G0/{\G0_{12}}'$ is spontaneously broken. Thus each vacuum sits in an orbit of $\G0/{\G0_{12}}'$ comprising of a total of $\left|\G0/{\G0_{12}}'\right|$ number of vacua. There are a total of
\be\label{n12}
n_{12}:=\frac{\left|\G0/{\G0_{1}}'\right|\times\left|\G0/{\G0_{2}}'\right|}{\left|\G0/{\G0_{12}}'\right|}=\frac{\left|\G0\right|\times\left|{\G0_{12}}'\right|}{\left|{\G0_{1}}'\right|\times\left|{\G0_{2}}'\right|} 
\ee
number of such orbits. The SPT phase for ${\G0_{12}}'$ is the same in each vacuum and is described by the element
\be\label{a12}
\alpha_{12}=\alpha_1|_{12}+\alpha_2|_{12}\in H^2\big({\G0_{12}}',U(1)\big) \,,
\ee
where $\alpha_i|_{12}\in H^2\big({\G0_{12}}',U(1)\big)$ is obtained as the image of $\alpha_i$ under the pull-back map
\be
H^2\big({\G0_i}',U(1)\big)\to H^2\big({\G0_{12}}',U(1)\big)
\ee
associated to the inclusion map
\be
{\G0_{12}}'\to {\G0_i}' \,.
\ee
Thus, the fusion can be summarized as
\be\label{G0fus}
\TQFT_{{\G0_1}',\alpha_1}\otimes \TQFT_{{\G0_2}',\alpha_2}=n_{12}\TQFT_{{\G0_{12}}',\alpha_{12}} \,,
\ee
where the right hand side describes $n_{12}$ copies of $\TQFT_{{\G0_{12}}',\alpha_{12}}$.

The fusion rules for the case of an arbitrary, possibly non-abelian, $\G0$ can be deduced similarly. These fusion rules convert the set of objects of $\TwoRep(\G0)$ into a ring, which can be identified with what is known as the \textit{twisted Burnside ring}, which we discuss in section \ref{sec:Burn}. 

Some general remarks can be made here. $\TQFT_{\Gamma^{(0)},0}$, corresponding to the trivial SPT phase of $\Gamma^{(0)}$ is always the monoidal unit. On the other hand, $\TQFT_{\id,0}$ corresponds to the untwisted gauge theory with gauge group $\widehat{\Gamma^{(0)}}$ and has fusion
\begin{equation}
    \TQFT_{\id,0} \otimes \TQFT_{{\G0}',0} = |\Gamma^{(0)}:{{\G0}'}| \, \TQFT_{\id,0} \,,
\end{equation}
where ${\G0}'$ is a subgroup of $\G0$ and $|\G0:{{\G0}'}| $ is the index of ${{\G0}'}$ in $\G0$.

\paragraph{1-morphisms.}
Let us discuss 1-morphisms in the category $\TwoRep(\G0)$. 
First of all, consider  1-morphisms from a minimal $\G0$ symmetric TQFT $\TQFT_{{\G0}',\alpha}$ to the trivial $\G0$ symmetric TQFT $\TQFT_{{\G0},0}$. These are $\G0$ symmetric topological boundary conditions of $\TQFT_{{\G0}',\alpha}$. We can describe such a boundary condition by a collection of boundary conditions for each vacuum $i$ of $\TQFT_{{\G0}',\alpha}$. Since each vacuum $i$ carries the SPT phase corresponding to $\alpha$ for ${\G0}'$, the boundary condition for every vacuum must carry a topological quantum mechanical system having ${\G0}'$ 0-form symmetry with 't Hooft anomaly given by $\alpha$. 
Such a quantum mechanical system is described by its Hilbert space, which is a vector space forming a projective representation of ${\G0}'$ of type $\alpha$. Moreover, the projective representation for all vacua $i$ must be the same as all vacua are related by $\G0$ action. The 2-morphisms from a 1-morphism carrying a projective representation $R_1^\alpha$ to a 1-morphism carrying a projective representation $R_2^\alpha$ correspond to the ${\G0}'$-intertwiners from $R_1^\alpha$ to $R_2^\alpha$. Thus, 1-morphisms from $\TQFT_{{\G0}',\alpha}$ to $\TQFT_{{\G0},0}$ in $\TwoRep(\G0)$ are described by the finite semi-simple category
\be
\Rep_\alpha\big({\G0}'\big)
\ee
of projective representations of ${\G0}'$ of type $\alpha$, with the simple 1-morphisms being the irreducible projective representations of this type. See again figure \ref{fig:Proj}.

It is easy to generalize the above analysis to determine 1-morphisms from $\TQFT_{{\G0_1}',\alpha_1}$ to $\TQFT_{{\G0_2}',\alpha_2}$ in $\TwoRep(\G0)$. These are $\G0$ symmetric topological interfaces from $\TQFT_{{\G0_1}',\alpha_1}$ to $\TQFT_{{\G0_2}',\alpha_2}$ and are comprised as collections of topological interfaces between vacua of the two TQFTs. Consider an interface from a vacuum $i$ of $\TQFT_{{\G0_1}',\alpha_1}$ to a vacuum $a$ of $\TQFT_{{\G0_2}',\alpha_2}$. Let ${\G0_{1,i}}'\cong {\G0_1}'$ be the subgroup of $\G0$ preserved in vacuum $i$, and ${\G0_{2,a}}'\cong {\G0_2}'$ be the subgroup of $\G0$ preserved in vacuum $a$. Then, the interface from $i$ to $a$ must carry a topological quantum mechanical system carrying 0-form symmetry
\be
{\G0_{12,(i,a)}}':={\G0_{1,i}}'\cap {\G0_{2,a}}' \label{eq:G012ia}
\ee
with 't Hooft anomaly given by
\be
\alpha_{12,(i,a)}=\alpha_1|_{12,i}-\alpha_2|_{12,a}\,,
\ee
where $\alpha_1|_{12,i}\in H^2\big({\G0_{12,(i,a)}}',U(1)\big)$ is obtained as the image of $\alpha_1$ under the pull-back map
\be
H^2\big({\G0_{1,i}}',U(1)\big)\to H^2\big({\G0_{12,(i,a)}}',U(1)\big)
\ee
and $\alpha_2|_{12,a}\in H^2\big({\G0_{12,(i,a)}}',U(1)\big)$ is obtained as the image of $\alpha_2$ under the pull-back map
\be
H^2\big({\G0_{2,a}}',U(1)\big)\to H^2\big({\G0_{12,(i,a)}}',U(1)\big)\,.
\ee
The interface for any other vacuum pair $(i',a')$ related to the vacuum pair $(i,a)$ by $\gamma$ action for some $\gamma\in\G0$ needs to carry the same quantum mechanical system as the interface for the vacuum pair $(i,a)$. The orbits of vacuum pairs are parametrized by the double coset ${\G0_1}' \backslash \G0 \slash {\G0_2}'$. This completely determines the category of 1-morphisms from $\TQFT_{{\G0_1}',\alpha_1}$ to $\TQFT_{{\G0_2}',\alpha_2}$ in $\TwoRep(\G0)$. For the special case of abelian $\G0$, the above category of 1-morphisms from $\TQFT_{{\G0_1}',\alpha_1}$ to $\TQFT_{{\G0_2}',\alpha_2}$ in $\TwoRep(\G0)$ can be described simply as
\be
n_{12}\times\Rep_{\alpha'_{12}}\big({\G0_{12}}'\big) \,,\label{eq:1morphisms}
\ee
that is as $n_{12}$ copies of $\Rep_{\alpha'_{12}}\big({\G0_{12}}'\big)$, where ${\G0_{12}}'$ is given in (\ref{G012}), $n_{12}$ is given in (\ref{n12}), and
\be\label{ap12}
\alpha'_{12}=\alpha_1|_{12}-\alpha_2|_{12}\,,
\ee
where $\alpha_i|_{12}$ for $i=1,2$ have been defined above.

\paragraph{1d TQFTs with $\Gamma^{(0)}$ 0-form symmetry and Wilson line defects.}
Consider now 1-morphisms for the particular case of ${\G0_1}'={\G0_2}'=\G0$ and $\alpha_1=\alpha_2=0$. This corresponds to studying $\G0$ symmetric topological interfaces from the trivial 2d TQFT to itself, or in other words 1d TQFTs with $\G0$ 0-form symmetry. Indeed, we find that such 1-morphisms are described by $\Rep(\G0)$.
These 1-morphisms give rise to Wilson line defects in a $d$-dimensional theory $\fT/\G0$ obtained after gauging $\G0$ 0-form symmetry.

\paragraph{Composition of 1-morphisms.}
Let us now describe composition of morphisms for abelian $\G0$. For a general non-abelian $\G0$, the computation works similarly but is more complicated, and we leave it to the interested reader to figure it out. We will discuss a non-abelian example $\G0=S_3$ later on. Consider a 1-morphism $\{R_I^{\alpha'_{12}}\}$ from $\TQFT_{{\G0_1}',\alpha_1}$ to $\TQFT_{{\G0_2}',\alpha_2}$, and a 1-morphism $\{R_A^{\alpha'_{23}}\}$ from $\TQFT_{{\G0_2}',\alpha_2}$ to $\TQFT_{{\G0_3}',\alpha_3}$, where $R_I^{\alpha'_{12}}$ for $1\le I\le n_{12}$ are projective representations of ${\G0_{12}}'$ of type $\alpha'_{12}$ and $R_A^{\alpha'_{23}}$ for $1\le A\le n_{23}$ are projective representations of ${\G0_{23}}'$ of type $\alpha'_{23}$. The composition of 1-morphisms takes the form
\be
\{R_I^{\alpha'_{12}}\}\circ\{R_A^{\alpha'_{23}}\}=\{R_X^{\alpha'_{13}}\}\,,
\ee
where $\{R_X^{\alpha'_{13}}\}$ denotes a 1-morphism from $\TQFT_{{\G0_1}',\alpha_1}$ to $\TQFT_{{\G0_3}',\alpha_3}$, where $R_X^{\alpha'_{13}}$ for $1\le X\le n_{13}$ are projective representations of ${\G0_{13}}'$ of type $\alpha'_{13}$. Let us describe how $R_X^{\alpha'_{13}}$ is computed for a particular orbit $X$. Let $(i,x)$ be a vacuum pair in the orbit $X$, where $i$ is a vacuum of $\TQFT_{{\G0_1}',\alpha_1}$ and $x$ is a vacuum of $\TQFT_{{\G0_3}',\alpha_3}$. The underlying vector space $V_X^{\alpha'_{13}}$ carrying the representation $R_X^{\alpha'_{13}}$ can be described as
\be
V_X^{\alpha'_{13}}=\bigoplus_a V_{I(a)}^{\alpha'_{12}}\otimes V_{A(a)}^{\alpha'_{23}}\,, \label{eq:compositionof1morphismvs}
\ee
where $a$ parametrizes the vacua of $\TQFT_{{\G0_2}',\alpha_2}$, $V_{I(a)}^{\alpha'_{12}}$ is the underlying vector space for the representation $R_{I(a)}^{\alpha'_{12}}$, where $I(a)$ is the orbit in which the vacuum pair 
$(i,a)$ lies, and $V_{A(a)}^{\alpha'_{23}}$ is the underlying vector space for the representation $R_{A(a)}^{\alpha'_{23}}$ where $A(a)$ is the orbit in which the vacuum pair $(a,x)$ lies. An element of ${\G0_{13}}'$ which also lies in ${\G0_{2}}'$ acts on each individual term $V_{I(a)}^{\alpha'_{12}}\otimes V_{A(a)}^{\alpha'_{23}}$ in the direct sum defining $V_X^{\alpha'_{13}}$. On the other hand, an element of ${\G0_{13}}'$ which does not lie in ${\G0_{2}}'$ acts by permuting different terms in the direct sum defining $V_X^{\alpha'_{13}}$. In this way, the elements of ${\G0_{13}}'$ act on $V_X^{\alpha'_{13}}$ and convert it into the required representation $R_X^{\alpha'_{13}}$.

\paragraph{Fusion of 1-morphisms.}
The determination of the monoidal structure on 1-morphisms is quite similar to the above determination of the composition of 1-morphisms. Consider fusing a 1-morphism $\cI_{12}$ from $\TQFT_{{\G0_1}',\alpha_1}$ to $\TQFT_{{\G0_2}',\alpha_2}$ with a 1-morphism $\cI_{34}$ from $\TQFT_{{\G0_3}',\alpha_3}$ to $\TQFT_{{\G0_4}',\alpha_4}$. The resulting 1-morphism from $\TQFT_{{\G0_1}',\alpha_1}\otimes\TQFT_{{\G0_3}',\alpha_3}$ to $\TQFT_{{\G0_2}',\alpha_2}\otimes\TQFT_{{\G0_4}',\alpha_4}$ is determined in terms of appropriate representations between pairs $(ij,ab)$ of vacua between $\TQFT_{{\G0_1}',\alpha_1}\otimes\TQFT_{{\G0_3}',\alpha_3}$ and $\TQFT_{{\G0_2}',\alpha_2}\otimes\TQFT_{{\G0_4}',\alpha_4}$ where $ij$ is a vacuum of $\TQFT_{{\G0_1}',\alpha_1}\otimes\TQFT_{{\G0_3}',\alpha_3}$ and $ab$ is a vacuum of $\TQFT_{{\G0_2}',\alpha_2}\otimes\TQFT_{{\G0_4}',\alpha_4}$. The representation assigned to $(ij,ab)$ is obtained essentially from the information of the representation assigned to the pair of vacua $(i,a)$ by $\cI_{12}$ where $i$ is a vacuum of $\TQFT_{{\G0_1}',\alpha_1}$ and $a$ is a vacuum of $\TQFT_{{\G0_3}',\alpha_3}$ tensored with the representation assigned to the pair of vacua $(j,b)$ by $\cI_{34}$, where $j$ is a vacuum of $\TQFT_{{\G0_2}',\alpha_2}$ and $b$ is a vacuum of $\TQFT_{{\G0_4}',\alpha_4}$.


\subsubsection{Relation to the Twisted Burnside Ring}
\label{sec:Burn}

The fusion structure of the objects in $\TwoRep(\Gamma^{(0)})$ that we encountered has an alternative representation-theoretic description in terms of the twisted Burnside ring \cite{greenoughMonoidal2structureBimodule2010}. 
Recall first the Burnside ring $\Omega(G)$ \cite{burnside2012theory, boucBurnsideRings2000,kerberAppliedFiniteGroup2013} for $G$ a finite (not necessarily abelian) group.

\paragraph{G-sets.} A $G$-set is a set where $G$ acts by permutation. Any action of $G$ on a set can be decomposed into a union of transitive actions. Any $G$-orbit is simply isomorphic\footnote{Recall that given any (conjugacy class of) subgroup $H$ of $G$, the set $G/H$ of left cosets have a natural $G$-action by left multiplication, where $H$ acts trivially.} as a $G$-set to the set of left cosets $G/H$, where $H$ acts trivially on this orbit. So finding (isomorphism classes of) $G$-orbit is the same as (conjugacy classes of) subgroups $H$ of $G$. 

\paragraph{Burnside Ring.} Analogous to the ring $\mathrm{Rep}(G)$ of representations of $G$, it is natural to consider the analogous representation ring in the category of finite sets, which is the Burnside ring $\Omega(G)$: 
\paragraph{Definition (Burnside Ring).} 
The Burnside ring $\Omega(G)$ of a  finite group $G$ is:
\begin{enumerate}
	\item The elements are isomorphism classes of finite $G$-sets. A basis of $\Omega(G)$ is given by {the set of all left cosets}. Namely, as a $\mathbb{Z}$-module, any element of $\Omega(G)$ can be written as
     \begin{equation}
        \sum_i a_i [H_i], \quad a_i \in \mathbb{Z}\,, \label{eq:basisofOmegaG}
    \end{equation}
    where the sum is over all {conjugacy classes $[H]$} of subgroups of $G$.
	\item The additive structure\footnote{This definition of the addition as disjoint union only gives a semi-ring, namely given a set $S$, we do not know what is $-S$. However, we do want $\Omega(G)$ to be a ring with additive inverse, which can be done by including formal symbols like $-S$.} is disjoint union of $G$-sets.
	\item The multiplicative  structure is defined by the Cartesian product of $G$-sets.
	\item The additive identity is defined to be the empty set.
	\item The multiplicative identity is the one element set $[G]$, {which has a trivial $G$ action.} 
\end{enumerate}
Alternatively, we can simply say that for a finite group $G$, the Burnside ring $\Omega(G)$ is the Grothendieck ring of the monoidal category $\mathsf{GSet}$ of finite $G$-sets.

The multiplication of two elements in $\Omega(G)$ can be decomposed into the basis \eqref{eq:basisofOmegaG}, which is given by a Mackey-like formula as 
\be\label{eq:mackey}
[H] \times[K]=\sum_{HaK\in H \backslash G / K}\left[\left(H \cap\ ^a K\right) \right]  \,.
\ee
Here $^xH:= xHx^{-1}$ and $H^x:= x^{-1} Hx$.
This follows by consider the bijection between the set of double cosets $H\backslash G/K$ and $G\backslash \left(G/H\times G/K \right)$
\begin{align}
	G/H \times G/K &\ \rightarrow \ H\backslash G /K\\
	(xH,yK) &\ \mapsto\  H x^{-1}y K \,.
\end{align}
Now consider the double coset $HaK$. One of the pre-images is $(H,aK)$. Under the left $G$ action, $(H,aK)$ gets sent to $(gH,gaK)$ therefore the stabilizer
of $(H,aK)$ is $H \cap\ ^aK$. In other words, the $G$-orbit of $(H,aK)$ is $G/ \left(H \cap\ ^aK\right)$. Therefore, we find \eqref{eq:mackey}.

Computationally, the fusion in the  Burnside ring can be extracted straight forwardly from the knowledge of the marks $m_S(H)$. The marks are the number of $H$-invariant elements in $S$. These satisfy  $m_{[S_1]}(H)m_{[S_2]}(H) = m_{[S_1]\times [S_2]}(H)$ for all $H$. Let 
\be
    M_{ij} = m_{[H_i]} (H_j)= |[H_i]^{H_j}|
\ee
be the set of $H_j$ invariant elements in $[H_i]$, then the fusion coefficients of the Burnside ring are computed from these using a Verlinde-like formula 
\be\label{eq:fusionfromtableofmarks}
N_{i j}^{l}=\sum_{m} M_{i m} \cdot M_{j m} \cdot\left(M^{-1}\right)_{m l} \,.
\ee
We will apply this in the case of $\Gamma^{(0)} = S_3$ later on.

\paragraph{Twisted Burnside Ring and $\TwoRep(G)$.}
Here we are interested in a twisted version of the Burnside ring which includes an $\mathbb{R}/\mathbb{Z}$-{cocycles}, introduced in \cite{greenoughMonoidal2structureBimodule2010}. Elements of the twisted Burnside ring are labeled by pairs $H<G$ and $\mu\in H^2 (H, U(1))$. 

\paragraph{Theorem.\cite{greenoughMonoidal2structureBimodule2010}} Let $G$ be a finite group. The Grothendieck ring of the 2-category $\TwoRep(G)$ is isomorphic to a twisted version of the Burnside ring, defined as follows. As a $\mathbb{Z}$-module, the basis elements are labelled by $([H],\mu)$, i.e. a subgroup $H$ of $G$ and $[H]$ as above the isomorphism class of $G/H$ $G$-sets, as well as  a 2-cocycle $\mu\in H^2(H, U(1))$.
The generalization of the multiplication structure \eqref{eq:mackey} to the twisted Burnside ring is 
\begin{equation}
	( [H], \mu) \times  ([K], \sigma) =\sum_{H a K \in H \backslash G / K} \left( [H \cap\, { }^{a}K], \mu \sigma^{a}\right) \,. \label{eq:fusionofobjectsBurnside}
\end{equation}
Here the {cocycle} $\sigma^a$ is defined as a map
\be\ba
\sigma^a: \qquad \,^aK \times \,^aK &\ \rightarrow \ \mathbb{R}/\mathbb{Z}\cr 
(k_1, k_2) &\ \mapsto \  \sigma (a^{-1}k_1 a, a^{-1}k_2 a) \,.
\ea\ee
This theorem identifies structure of $\TwoRep(G)$ for $G= \Gamma^{(0)}$ can be equally obtained from the structure of the twisted Burnside ring on pairs of subgroups $\Gamma^{(0)'}$ and cocycles. The fusion 
that we discussed in section \ref{sec:2catGamma0}, in particular \eqref{G0fus}, agrees with the multiplicative structure on the twisted Burnside ring where an element $([H],\mu)$ in the twisted Burnside ring is identified with a 2d TQFT $\TQFT_{H,\mu}$ defined around \eqref{eq:generalTFTasdirectsum}.

\subsubsection{Example: $\G0=\Z_2$}
\paragraph{Minimal 2d TQFTs with $\Z_2$ 0-form symmetry.}
Let us describe minimal 2d TQFTs with $\G0=\Z_2$ 0-form symmetry. There are two choices for subgroup ${\G0}'\subseteq\G0$, namely ${\G0}'=\id$ and ${\G0}'=\Z_2$, where the former is the trivial group containing only the identity element $\id$. For both values of ${\G0}'$, the cohomology groups $H^2\big({\G0}',U(1)\big)$ are trivial. Thus, there are two minimal 2d TQFTs with $\Z_2$ 0-form symmetry
\be
\TQFT^{(\id)},~\TQFT^{(\Z_2)}
\ee
that we label by the corresponding values of $\G0/{\G0}'$. The theory $\TQFT^{(\id)}$ is the trivial $\Z_2$ symmetric 2d TQFT, while $\TQFT^{(\Z_2)}$ is non-trivial and has two underlying vacua. In fact, $\TQFT^{(\Z_2)}$ can be recognized as the 2d $\Z_2$ gauge theory, which has a dual $\Z_2$ 0-form symmetry.

\paragraph{Fusion of objects.}
$\TQFT^{(\id)}$ is the monoidal unit and the non-trivial fusion rules are
\be
\TQFT^{(\Z_2)}\otimes\TQFT^{(\Z_2)}=2\TQFT^{(\Z_2)}
\ee
following from (\ref{G0fus}).

\paragraph{1-morphisms.}
Let us now discuss the 1-morphisms. We have two simple 1-endomorphisms of $\TQFT^{(\id)}$ corresponding to two representations of ${\G0}'=\Z_2$, which we label as
\be
\cI^{(\id)},~\cI^{(-)} \,,
\ee
where $\cI^{(\id)}$ corresponds to the trivial representation and $\cI^{(-)}$ corresponds to the non-trivial representation. These 1-endomorphisms give rise to Wilson line defects in a $d$-dimensional theory $\fT/\Z_2^{(0)}$ obtained after gauging a $\Z_2$ 0-form symmetry.

We also have two simple 1-endomorphisms of $\TQFT^{(\Z_2)}$ corresponding to two orbits of vacuum pairs, which we label as
\be
\cI^{(\Z_2)},~\cI^{(\Z_2,-)} \,,
\ee
where $\cI^{(\Z_2)}$ corresponds to orbit $\{(v_1,v_1),(v_2,v_2)\}$ and $\cI^{(\Z_2,-)}$ corresponds to orbit $\{(v_1,v_2),(v_2,v_1)\}$. Finally, we have a single simple 1-morphism
\be
\cI^{(\id),(\Z_2)}
\ee
from $\TQFT^{(\id)}$ to $\TQFT^{(\Z_2)}$, and a single simple 1-morphism
\be
\cI^{(\Z_2),(\id)}
\ee
from $\TQFT^{(\Z_2)}$ to $\TQFT^{(\id)}$, as in both cases we have a single orbit of vacuum pairs.

\paragraph{Composition of 1-morphisms.}
Let us work out the compositions of these 1-morphisms. The composition of $\cI^{(\id)}$ and $\cI^{(-)}$ just follows the tensor product of representations of $\Z_2$, leading to
\begin{align}
\cI^{(\id)}\circ\cI^{(\id)}&=\cI^{(\id)}\\
\cI^{(-)}\circ\cI^{(\id)}=\cI^{(\id)}\circ\cI^{(-)}&=\cI^{(-)}\\
\cI^{(-)}\circ\cI^{(-)}&=\cI^{(\id)} \,.
\end{align}
Therefore the 1-endomorphisms of the object $\TQFT^{(\id)}$ has the structure of a monoidal 1-category. Similar fusion rules are also obeyed by the 1-endomorphisms of $\TQFT^{(\Z_2)}$, but now for reasons having to do with compatibility of orbits
\begin{align}
\cI^{(\Z_2)}\circ\cI^{(\Z_2)}&=\cI^{(\Z_2)}\\
\cI^{(\Z_2,-)}\circ\cI^{(\Z_2)}=\cI^{(\Z_2)}\circ\cI^{(\Z_2,-)}&=\cI^{(\Z_2,-)}\\
\cI^{(\Z_2,-)}\circ\cI^{(\Z_2,-)}&=\cI^{(\Z_2)} \,.
\end{align}
 These 1-endomorphisms are line operators and act on the boundaries of $\TQFT^{(\Z_2)}$ as
\begin{align}
\cI^{(\id)}\circ\cI^{(\id),(\Z_2)}=\cI^{(-)}\circ\cI^{(\id),(\Z_2)}&=\cI^{(\id),(\Z_2)}\\
\cI^{(\Z_2),\id}\circ\cI^{(\id)}=\cI^{(\Z_2),\id}\circ\cI^{(-)}&=\cI^{(\Z_2),\id}\\
\cI^{(\Z_2)}\circ\cI^{(\Z_2),(\id)}=\cI^{(\Z_2,-)}\circ\cI^{(\Z_2),(\id)}&=\cI^{(\Z_2),(\id)}\label{eq:compositionof1morphismZ2Z2id}\\
\cI^{\id,(\Z_2)}\circ\cI^{(\Z_2)}=\cI^{\id,(\Z_2)}\circ\cI^{(\Z_2,-)}&=\cI^{\id,(\Z_2)} \,.\label{eq:compositionof1morphismidZ2Z2}
\end{align}
Finally the compositions of the two boundaries are
\begin{align}
\cI^{(\Z_2),(\id)}\circ\cI^{(\id),(\Z_2)}&=\cI^{(\Z_2)}\oplus \cI^{(\Z_2,-)}\\
\cI^{(\id),(\Z_2)}\circ\cI^{(\Z_2),(\id)}&=\cI^{(\id)}\oplus \cI^{(-)} \,. \label{eq:fusionZ2xy}
\end{align}

\paragraph{Fusion of 1-morphisms.}
The monoidal product of 1-endomorphisms of the identity object are the same as their compositions:
\begin{align}
\cI^{(\id)}\otimes\cI^{(\id)}&=\cI^{(\id)}\\
\cI^{(-)}\otimes\cI^{(\id)}=\cI^{(\id)}\otimes\cI^{(-)}&=\cI^{(-)}\\
\cI^{(-)}\otimes\cI^{(-)}&=\cI^{(\id)} \,.
\end{align}
Their product with 1-endomorphisms of $\TQFT^{(\Z_2)}$ are
\begin{align}
\cI^{(i)}\otimes\cI^{(\Z_2)}=\cI^{(\Z_2)}\otimes\cI^{(i)}&=\cI^{(\Z_2)}\\
\cI^{(i)}\otimes\cI^{(\Z_2,-)}=\cI^{(\Z_2,-)}\otimes\cI^{(i)}&=\cI^{(\Z_2,-)}
\end{align}
for $i\in\{\id,-\}$.
Their product with the boundary conditions of $\TQFT^{(\Z_2)}$ are also the same as compositions
\begin{align}
\cI^{(i)}\otimes\cI^{(\Z_2),(\id)}=\cI^{(\Z_2),(\id)}\otimes\cI^{(i)}&=\cI^{(\Z_2),(\id)}\\
\cI^{(i)}\otimes\cI^{(\id),(\Z_2)}=\cI^{(\id),(\Z_2)}\otimes\cI^{(i)}&=\cI^{(\id),(\Z_2)}
\end{align}
for $i\in\{\id,-\}$.
The product of 1-endomorphisms of $\TQFT^{(\Z_2)}$ are
\begin{align}
\cI^{(\Z_2)}\otimes\cI^{(\Z_2)}&=\cI^{(\Z_2)}_{1,1}\oplus\cI^{(\Z_2)}_{2,2}\\
\cI^{(\Z_2)}\otimes\cI^{(\Z_2,-)}&=\cI^{(\Z_2)}_{1,2}\oplus\cI^{(\Z_2)}_{2,1}\\
\cI^{(\Z_2,-)}\otimes\cI^{(\Z_2)}&=\cI^{(\Z_2,-)}_{1,2}\oplus\cI^{(\Z_2,-)}_{2,1}\\
\cI^{(\Z_2,-)}\otimes\cI^{(\Z_2,-)}&=\cI^{(\Z_2,-)}_{1,1}\oplus\cI^{(\Z_2,-)}_{2,2} \,,
\end{align}
where $\cI^{(\Z_2)}_{i,i}$ and $\cI^{(\Z_2,-)}_{i,i}$ denote 1-endomorphisms of the $i$-th copy of $\TQFT^{(\Z_2)}$ arising in the fusion $\TQFT^{(\Z_2)}\otimes\TQFT^{(\Z_2)}$. In particular, $\cI^{(\Z_2)}_{i,i}$ is the identity 1-endomorphism while $\cI^{(\Z_2,-)}_{i,i}$ is the non-identity simple 1-endomorphism. On the other hand, $\cI^{(\Z_2)}_{1,2}$ and $\cI^{(\Z_2,-)}_{1,2}$ are simple 1-morphisms from the first copy of $\TQFT^{(\Z_2)}$ to the second copy of $\TQFT^{(\Z_2)}$. Note that in this case both $\cI^{(\Z_2)}_{1,2}$ and $\cI^{(\Z_2,-)}_{1,2}$ are on an equal footing, and the naming we have adopted is simply a choice. The two 1-morphisms $\cI^{(\Z_2)}_{1,2}$ and $\cI^{(\Z_2,-)}_{1,2}$ form a torsor over the $\Z_2$ group generated by $\cI^{(\Z_2)}_{2,2}$ and $\cI^{(\Z_2,-)}_{2,2}$ under composition from the right and a torsor over the $\Z_2$ group generated by $\cI^{(\Z_2)}_{1,1}$ and $\cI^{(\Z_2,-)}_{1,1}$ under composition from the left. Similar remarks hold for $\cI^{(\Z_2)}_{2,1}$ and $\cI^{(\Z_2,-)}_{2,1}$ which are simple 1-morphisms from the second copy of $\TQFT^{(\Z_2)}$ to the first copy of $\TQFT^{(\Z_2)}$.

Some of the products of boundary conditions of $\TQFT^{(\Z_2)}$ are the same as compositions
\begin{align}
\cI^{(\Z_2),(\id)}\otimes\cI^{(\id),(\Z_2)}&=\cI^{(\Z_2)}\oplus\cI^{(\Z_2,-)}\\
\cI^{(\id),(\Z_2)}\otimes\cI^{(\Z_2),(\id)}&=\cI^{(\Z_2)}\oplus\cI^{(\Z_2,-)} \,.
\end{align}
The other products of boundary conditions of $\TQFT^{(\Z_2)}$ are
\begin{align}
\cI^{(\id),(\Z_2)}\otimes\cI^{(\id),(\Z_2)}&=\cI^{(\id),(\Z_2)}_1\oplus\cI^{(\id),(\Z_2)}_2\\
\cI^{(\Z_2),(\id)}\otimes\cI^{(\Z_2),(\id)}&=\cI^{(\Z_2),(\id)}_1\oplus\cI^{(\Z_2),(\id)}_2 \,,
\end{align}
where $\cI^{(\id),(\Z_2)}_1$ is a simple 1-morphism from $\TQFT^{(\id)}$ to the $i$-th copy of $\TQFT^{(\Z_2)}$ and $\cI^{(\Z_2),(\id)}_1$ is a simple 1-morphism from $i$-th copy of $\TQFT^{(\Z_2)}$ to $\TQFT^{(\id)}$.

The products of boundary conditions of $\TQFT^{(\Z_2)}$ with line operators on $\TQFT^{(\Z_2)}$ are
\begin{align}
\cI^{(\id),(\Z_2)}\otimes\cI^{(\Z_2)}&=\cI^{(\Z_2)}_{0,1}\oplus\cI^{(\Z_2,-)}_{0,2}\\
\cI^{(\id),(\Z_2)}\otimes\cI^{(\Z_2,-)}&=\cI^{(\Z_2,-)}_{0,1}\oplus\cI^{(\Z_2)}_{0,2}\\
\cI^{(\Z_2)}\otimes\cI^{(\id),(\Z_2)}&=\cI^{(\Z_2)}_{0,1}\oplus\cI^{(\Z_2)}_{0,2}\\
\cI^{(\Z_2,-)}\otimes\cI^{(\id),(\Z_2)}&=\cI^{(\Z_2,-)}_{0,1}\oplus\cI^{(\Z_2,-)}_{0,2}\\
\cI^{(\Z_2),(\id)}\otimes\cI^{(\Z_2)}&=\cI^{(\Z_2)}_{1,0}\oplus\cI^{(\Z_2,-)}_{2,0}\\
\cI^{(\Z_2),(\id)}\otimes\cI^{(\Z_2,-)}&=\cI^{(\Z_2,-)}_{1,0}\oplus\cI^{(\Z_2)}_{2,0}\\
\cI^{(\Z_2)}\otimes\cI^{(\Z_2),(\id)}&=\cI^{(\Z_2)}_{1,0}\oplus\cI^{(\Z_2)}_{2,0}\\
\cI^{(\Z_2,-)}\otimes\cI^{(\Z_2),(\id)}&=\cI^{(\Z_2,-)}_{1,0}\oplus\cI^{(\Z_2,-)}_{2,0} \,,
\end{align}
where $\cI^{(\Z_2)}_{0,i}$ and $\cI^{(\Z_2,-)}_{0,i}$ are simple 1-morphisms from $\TQFT^{(\Z_2)}$ to $i$-th copy of $\TQFT^{(\Z_2)}$, and $\cI^{(\Z_2)}_{i,0}$ and $\cI^{(\Z_2,-)}_{i,0}$ are simple 1-morphisms from $i$-th copy of $\TQFT^{(\Z_2)}$ to $\TQFT^{(\Z_2)}$.

The underlying 2-category structure $\TwoRep(\mathbb{Z}_2)$ can be conveniently summarized into the following diagram
\begin{equation}
\begin{tikzcd}
{\begin{array}{c}\bullet \\ \TQFT^{(\id)}\end{array}} \arrow["\Rep(\Z_2) = {\{\cI^{(\id)},~\cI^{(-)}\}}"', loop, distance=2em, in=215, out=145] \arrow[r, "\cI^{(\id),(\Z_2)}\in\Vec",bend left] & {\begin{array}{c}\bullet\\ \TQFT^{(\Z_2)}\end{array}} \arrow["\Vec(\mathbb{Z}_2) = {\{\cI^{(\Z_2)},~\cI^{(\Z_2,-)} \}}"', loop, distance=2em, in=35, out=325] \arrow[l, "\cI^{(\Z_2),(\id)}\in\Vec",bend left]
\end{tikzcd}\label{tikz:celldiagram2RepZ2}
\end{equation}
where $0$-cells denote the objects, which are 2d TQFTs. $1$-cells denote 1-morphisms, which are $\Gamma^{(0)}$ symmetric topological interfaces. Note that from \eqref{eq:1morphisms}, the category of 1-morphisms from $\TQFT_{{\G0_1}',\alpha_1}$ to $\TQFT_{{\G0_2}',\alpha_2}$ differs from the category of 1-morphisms from $\TQFT_{{\G0_2}',\alpha_2}$ to $\TQFT_{{\G0_1}',\alpha_1}$ by flipping the sign of the Schur multiplier. Therefore, to reduce clutter, we can simply employ a left-right arrow notation and the diagram above can be simplified to be
\begin{equation}
\begin{tikzcd}
{\begin{array}{c}\bullet \\ \TQFT^{(\id)}\end{array}} \arrow["\Rep(\mathbb{Z}_2) "', loop, distance=2em, in=215, out=145] \arrow[r, "\Vec",leftrightarrow] & {\begin{array}{c}\bullet\\ \TQFT^{(\Z_2)}\end{array}} \arrow["\Vec(\mathbb{Z}_2) "', loop, distance=2em, in=35, out=325] 
\end{tikzcd}
\end{equation}
In the physics literature, such a 2-category has appeared in the studies of toric code, see e.g. \cite{Kong:2020wmn,kong2020algebraic}.

\subsubsection{Example: $\G0=S_3$}\label{eg3p}
Let us now consider 2d TQFTs with $S_3$ 0-form symmetry. 

\paragraph{Presentation of $S_3$.}
We represent $S_3$ as
\be
S_3=\{1,a,a^2,b,ab,a^2b\}
\ee
where $a$ is an order 3 element, $b$ is an order 2 element and we have the commutation relation
\be
ba=a^2b
\ee
The subgroups of $S_3$ are
\be
\mathbb{Z}_3 = \{1, a, a^2\} \,,\qquad 
\mathbb{Z}_2^{b} = \{1, b\} \,,\qquad 
\mathbb{Z}_2^{ab}=\{1, ab\} \,,\qquad 
\mathbb{Z}_2^{a^2b}= \{1, a^2b\} \,,
\ee
along with the trivial subgroup and the whole group. 

\paragraph{Minimal 2d TQFTs with $S_3$ 0-form symmetry.}
The minimal 2d TQFTs with $S_3$ symmetry are as follows. First of all, we have a theory 
\be
\TQFT^{(\id)} \,,
\ee
which has a single vacuum in which all of $S_3$ is preserved. Then, we have a theory
\be
\TQFT^{(\Z_2)}\,,
\ee
which has two vacua, both preserving the $\Z_3$ subgroup; a theory
\be
\TQFT^{(\Z_3)} \,,
\ee
which has three vacua, and in each vacuum $v_i$ a different $\Z_2$ subgroup $\mathbb{Z}_2^{a^{i-1} b}$ for $i=1, 2, 3$ is preserved; and a theory
\be
\TQFT^{(S_3)}
\ee
having six vacua and no subgroup is preserved in any of the vacua.


\paragraph{Fusion of objects.}
To compute the fusion, we follow the same logic that led to the result (\ref{G0fus}) for abelian $\G0$, but now for non-abelian $\G0=S_3$. First of all, the fusion of $\TQFT^{(\mathbb{Z}_2)}$ with itself is computed in the same way as in the $\G0=\mathbb{Z}_2$ example discussed previously
\be
\TQFT^{(\mathbb{Z}_2)} \otimes \TQFT^{(\mathbb{Z}_2)} = 2 \TQFT^{(\mathbb{Z}_2)} \,.
\ee
Now consider $\TQFT^{(\Z_2)}\otimes\TQFT^{(\Z_3)}$. Since the intersection of the $\Z_3$ subgroup of $S_3$ with any of the $\Z_2$ subgroups of $S_3$ is trivial, no non-trivial element of $S_3$ is preserved in any of the six vacua of the resulting theory. Moreover, these six vacua lie in a single orbit of $S_3$, leading to the fusion rule
\be
\TQFT^{(\mathbb{Z}_2)} \otimes \TQFT^{(\mathbb{Z}_3)} = \TQFT^{(\mathbb{Z}_3)} \otimes \TQFT^{(\mathbb{Z}_2)}= \TQFT^{(S_3)}\,.
\ee
For $\TQFT^{(\Z_2)}\otimes\TQFT^{(S_3)}$, no non-trivial element of $S_3$ is preserved in any of the 12 vacua, and these vacua form two orbits of six vacua each, leading to
\be
\TQFT^{(\mathbb{Z}_2)} \otimes \TQFT^{(S_3)} =  \TQFT^{(S_3)} \otimes \TQFT^{(\mathbb{Z}_2)}= 2\TQFT^{(S_3)}\,.
\ee
Now, consider $\TQFT^{(\Z_3)}\otimes\TQFT^{(\Z_3)}$. We have 9 vacua, which we describe as pairs $(v_i,v_j)$, where such a pair describes the vacuum obtained by fusing the vacuum $v_i$ of the first $\TQFT^{(\Z_3)}$ the with the vacuum $v_j$ of the first $\TQFT^{(\Z_3)}$. In any of the three vacua $(v_i,v_i)$, the $\Z_2^{a^{i-1}b}$ subgroup of $S_3$ is preserved, and these vacua lie in a single orbit. The remaining six vacua $(v_i,v_j)$ for $i\neq j$ have no preserved symmetry and form a single orbit. Thus, we obtain the fusion rule
\be
\TQFT^{(\mathbb{Z}_3)} \otimes \TQFT^{(\mathbb{Z}_3)} = \TQFT^{(\mathbb{Z}_3)} \oplus \TQFT^{(S_3)} \,.
\ee
Following the same logic as before, the remaining fusion rules are
\begin{align}
\TQFT^{(\mathbb{Z}_3)} \otimes \TQFT^{(S_3)} = \TQFT^{(S_3)}\otimes\TQFT^{(\mathbb{Z}_3)}   & = 3 \TQFT^{(S_3)} \\
\TQFT^{(S_3)} \otimes \TQFT^{(S_3)} & = 6 \TQFT^{(S_3)} \,.
\end{align}


\paragraph{Burnside-ring perspective.}
Let us briefly comment on the related computation using the Burnside ring (there is no twisting in this case). The isomorphism classes of $S_3$-sets is given by translating $\TQFT^{(G/H)}$ into $[H]$. Then e.g. we compute the fusion $[\mathbb{Z}_2]\times [\mathbb{Z}_2]$ by applying the definitions. The $G$-orbit for $[\mathbb{Z}_2]$ is simply a $3$-element set acted upon by $S_3$ by permutations. Writing out $[\mathbb{Z}_2]\times [\mathbb{Z}_2]$ explicitly, we can easily see this 9-element set consists of two $G$-orbits, namely
\begin{equation}
	[\mathbb{Z}_2]\times [\mathbb{Z}_2] = [\mathbb{Z}_2]+ [\langle e\rangle]\,.
\end{equation}
From the table of marks {${m_{S_3/H_i}(H_j)}$}  computed using \cite{GAP4}, we can immediately derive the fusion from (\ref{eq:fusionfromtableofmarks}) to be: 
\be\label{S3Fusion}
		\begin{array}{|c||c|c|c|c|}\hline
			{m_{S_3/H_i}(H_j)} & 1 & \mathbb{Z}_2 & \mathbb{Z}_3 & {S}_3\\\hline
			\hline
			S_3/1 & 6 & & &\\
			\hline
			S_3/\mathbb{Z}_2 & 3 & 1 & &\\
			\hline
			S_3/\mathbb{Z}_3 & 2 & 0 & 2 &\\
			\hline
			S_3/S_3 & 1 & 1& 1&1\\ \hline
		\end{array}
\qquad \quad 
    \begin{array}{|c||c|c|c|c|} \hline
        [H]\times [K] & [\langle e \rangle] & [\mathbb{Z}_2] & [\mathbb{Z}_3] & [S_3]\\
        \hline\hline
        [\langle e \rangle] & 6 [\langle e \rangle]  & 3 [\langle e \rangle ] & 2[\langle e \rangle ] & [\langle e \rangle] \\\hline
        [\mathbb{Z}_2] &  3 [\langle e \rangle ]& [\mathbb{Z}_2]+ [\langle e \rangle] & [\langle e \rangle] & [\mathbb{Z}_2] \\\hline
        [\mathbb{Z}_3]  & 2 [\langle e \rangle ] & [\langle e \rangle]& 2[\mathbb{Z}_3] & [\mathbb{Z}_3] \cr \hline
        [S_3]   &   [\langle e \rangle ] &  [\mathbb{Z}_2]&  [\mathbb{Z}_3]& [S_3] \\\hline
    \end{array}
\ee
These are in agreement with the fusions discussed above.

\paragraph{1-morphisms.}
Let us first discuss the 1-endomorphisms. 
\ben
\item The 1-endomorphisms of $\TQFT^{(\id)}$ are described by the category
\be
\Rep(S_3)
\ee
as both sides of the interface preserve the full $S_3$ and hence we can dress the interface with a representation of $S_3$.
\item The 1-endomorphisms of $\TQFT^{(\Z_2)}$ are described by the category
\be
\Rep(\Z_3)
\ee
as both sides of any underlying interface preserve $\Z_3$ and hence we can dress the interface with a representation of $\Z_3$.
\item The 1-endomorphisms of $\TQFT^{(\Z_3)}$ are described by the category
\be
\Rep(\Z_2)\oplus\Vec \,,
\ee
where the $\Rep(\Z_2)$ part comes from the vacuum pairs $(v_i,v_i)$ for which both sides of the interface preserve the same $\Z_2$ subgroup allowing us to dress it by a representation of $\Z_2$, and the $\Vec$ part comes from the vacuum pairs $(v_i,v_j)$ for $i\neq j$ for which the subgroups preserved on the two sides of an interface have no non-trivial intersection.
\item The 1-endomorphisms of $\TQFT^{(S_3)}$ are described by the category
\be
6\times\Vec\,,
\ee
because there are no preserved symmetries and the vacuum pairs form 6 orbits.
\een
Other 1-morphisms involving the identity object are $S_3$ symmetric boundary conditions which can be described for various cases as follows:
\bit
\item 1-morphisms from $\TQFT^{(\Z_2)}$ to $\TQFT^{(\id)}$ and from $\TQFT^{(\id)}$ to $\TQFT^{(\Z_2)}$ are described by
\be
\Rep(\Z_3)
\ee
as that is the subgroup preserved by $\TQFT^{(\Z_2)}$.
\item 1-morphisms from $\TQFT^{(\Z_3)}$ to $\TQFT^{(\id)}$ and from $\TQFT^{(\id)}$ to $\TQFT^{(\Z_3)}$ are described by
\be
\Rep(\Z_2)
\ee
as that is the subgroup preserved by $\TQFT^{(\Z_3)}$.
\item 1-morphisms from $\TQFT^{(S_3)}$ to $\TQFT^{(\id)}$ and from $\TQFT^{(\id)}$ to $\TQFT^{(S_3)}$ are described by
\be
\Vec
\ee
as no symmetry is preserved by $\TQFT^{(S_3)}$.
\eit
The remaining 1-morphisms are as follows:
\bit
\item 1-morphisms from $\TQFT^{(\Z_2)}$ to $\TQFT^{(\Z_3)}$ and from $\TQFT^{(\Z_3)}$ to $\TQFT^{(\Z_2)}$ are described by
\be
\Vec
\ee
as the subgroups preserved on the two sides of an interface have no non-trivial intersection, and all the interfaces lie in a single orbit.
\item 1-morphisms from $\TQFT^{(\Z_2)}$ to $\TQFT^{(S_3)}$ and from $\TQFT^{(S_3)}$ to $\TQFT^{(\Z_2)}$ are described by
\be
2\times \Vec
\ee
as the subgroups preserved on the two sides of an interface have no non-trivial intersection, and the interfaces form 2 orbits.
\item 1-morphisms from $\TQFT^{(\Z_3)}$ to $\TQFT^{(S_3)}$ and from $\TQFT^{(S_3)}$ to $\TQFT^{(\Z_3)}$ are described by
\be
3\times \Vec
\ee
as the subgroups preserved on the two sides of an interface have no non-trivial intersection, and the interfaces form 3 orbits.
\eit

The examples so far $\Gamma^{(0)}= \mathbb{Z}_2$ or $S_3$  did not have any non-trivial cocyles. 
In appendix \ref{App:Z2Z2} we discuss the structure of $\TwoRep(\Gamma^{(0)})$ the simplest example with non-trivial {cocycle}, where $\Gamma^{(0)}= \mathbb{Z}_2 \times \mathbb{Z}_2$. This 2-category is summarized in figure \ref{fig:category2RepZ2Z2}.

\subsection{2d TQFTs with 1-Form Symmetry}
\label{sec:2catGamma1}

\paragraph{Making a 2d TQFT 1-form symmetric.}
Let $\G1$ be a finite abelian group. A 2d TQFT $\TQFT_n$ characterized by an integer $n$ acquires a $\G1$ 1-form symmetry if one specifies a monoidal functor
\be
\cC_{\id}(\G1)\to\mathsf{Mat}_n(\Vec)\,,
\ee
where $\cC_\id(\G1)$ is the monoidal category defined in appendix \ref{App:Cats}, see also \cite{etingof2016tensor} sections 2.3 and 2.11.
This means that we specify, in $\TQFT_n$, a local operator $O_\gamma$ for every $\gamma\in\G1$ such that
\be
O_\gamma O_{\gamma'}=O_{\gamma\gamma'}
\ee
for all $\gamma,\gamma'\in\G1$. 

\paragraph{Classification of 2d TQFTs with 1-form symmetry and SPT phases.}
We can project these local operators to local operators $O_{\gamma,i}$ living in each vacuum $i$. These projected local operators also satisfy group multiplication for $\G1$. Thus, we can describe any $\G1$ symmetric 2d TQFT as a direct sum of \textit{2d SPT phases protected by $\G1$ 1-form symmetry}, which are $\G1$ symmetric 2d TQFTs having a single vacuum. For such a phase, we have $O_\gamma\in\C$ and hence 2d SPT phases protected by $\G1$ 1-form symmetry are classified by the group $\wh{\G1}$ of characters of $\G1$.



A general 2d TQFT with $\G1$ 1-form symmetry can be expressed as
\be
\bigoplus_{i=1}^n \SPT_i\,,
\ee
where $\SPT_i$ describes the SPT phase for 1-form symmetry appearing in the $i$-th vacuum of the TQFT, which is labeled by an element $\alpha_i\in\wh{\G1}$.

\paragraph{2-category of 2d TQFTs with 1-form symmetry.}
2d TQFTs with $\G1$ 1-form symmetry form the fusion 2-category $\mathsf{2Vec}_{\wh{\G1}}$, which is the 2-category formed by $\wh{\G1}$-graded finite semi-simple categories. The correspondence is again made via boundary conditions. We write a 2d TQFT with $\G1$ 1-form symmetry as
\be
\bigoplus_\alpha n_\alpha \SPT_\alpha\,,
\ee
where $\SPT_\alpha$ is the SPT phase associated to element $\alpha\in\wh{\G1}$. Its boundary conditions form a $\wh{\G1}$ graded finite semi-simple category, such that the finite semi-simple category in the grading $\alpha$ has $n_\alpha$ simple objects. The 1-morphisms of $\mathsf{2Vec}_{\wh{\G1}}$ describe topological interfaces invariant under $\G1$ 1-form symmetry. Such topological interfaces connect vacua with the same characters. Thus, there is a single simple 1-endomorphism of each $\SPT_\alpha$, while there is no 1-morphism from $\SPT_\alpha$ to $\SPT_\beta$ for $\alpha\neq\beta$.

\paragraph{Example: $\G1=\Z_2\times\Z_2$.}
Let us describe minimal 2d TQFTs with $\G1=\Z_2\times\Z_2$ 1-form symmetry. The group of characters is
\be
\wh{\G1}=\Z_2\times\Z_2\,.
\ee
We label these four characters by $(\pm1,\pm1)$ where a $+1$ entry means that the character for the corresponding $\Z_2$ in $\G1$ is trivial, while a $-1$ entry means that the character for the corresponding $\Z_2$ in $\G1$ is non-trivial.
Thus, there are four minimal 2d TQFTs with $\Z_2\times\Z_2$ 1-form symmetry
\be\label{SCV1fs}
\TQFT^{(\id)},~\TQFT^{(S)},~\TQFT^{(C)},~\TQFT^{(V)}
\ee
corresponding respectively to characters $(1,1)$, $(-1,1)$, $(1,-1)$ and $(-1,-1)$. The fusion rules of these TQFTs follows the multiplication of these characters.

\subsection{2d TQFTs with Split 2-Group Symmetry}\label{T2GS}

\subsubsection{2-Categorical Structure}

A split 2-group symmetry comprises of a finite 0-form symmetry group $\G0$ and a finite abelian 1-form symmetry group $\G1$, along with an action
\be
\rho:~\G0\to\mathsf{Aut}(\G1)
\ee
of 0-form symmetry $\G0$ on 1-form symmetry $\G1$. Such a 2-group symmetry corresponds to the monoidal category $\cC_{\G0}(\G1,\rho)$ defined in appendix \ref{App:Cats}, see also  \cite{etingof2016tensor} sections 2.3 and 2.11. 

\begin{figure}
\centering
\begin{tikzpicture}
\draw [ blue, thick] (0,0) --  (0,3) ;
\draw [teal, fill=teal] (1,1.5) ellipse (0.1 and 0.1);
\node[teal] at (1,2) {$O_{\gamma_1}$};
\node[blue] at (0,-0.5) {$L_{\gamma_0}$};
\node at (2.5,1.2) {=};
\draw [ blue, thick] (5,0) --  (5,3) ;
\draw [teal, fill=teal] (4,1.5) ellipse (0.1 and 0.1);
\node[teal] at (4,2) {$O_{\rho_{\gamma_0}(\gamma_1)}$};
\node[blue] at (5,-0.5) {$L_{\gamma_0}$};
\end{tikzpicture}
\caption{Permutation action of the lines $L_{\gamma_0}$ on the local operators $O_{\gamma_1}$, where $\gamma_i \in \Gamma^{(i)}$. Here $\rho_{\gamma_0}$ is the permutation action of $\Gamma^{(0)}$ on $\Gamma^{(1)}$. \label{fig:ZeroOnOne}}
\end{figure}
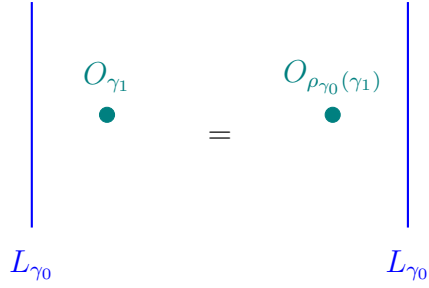

\paragraph{Equipping a 2d TQFT with split 2-group symmetry.}
A 2d TQFT $\TQFT_n$ characterized by an integer $n$ acquires a $\cC_{\G0}(\G1,\rho)$ split 2-group symmetry if one specifies a monoidal functor
\be
\cC_{\G0}(\G1,\rho)\to\mathsf{Mat}_n(\Vec)\,.
\ee
This means that we specify, in $\TQFT_n$, line operators $L_{\gamma_0}$ for all $\gamma_0\in\G0$, local operators $O_{\gamma_0,\gamma'_0}$ living at the junction of $L_{\gamma_0}$, $L_{\gamma'_0}$ and $L_{\gamma_0\gamma'_0}$, and local operators $O_{\gamma_1}$ for all $\gamma_1\in\G1$ such that:
\bit
\item The lines fuse according to the group law of the 0-form symmetry group $\G0$
\be
L_{\gamma_0}\circ L_{\gamma'_0}=L_{\gamma_0\gamma'_0}
\ee
\item The junction local operators $O_{\gamma_0,\gamma'_0}$ satisfy the associativity condition shown in figure \ref{fig:Assoc}. 
\item The local operators $O_{\gamma_1}$ fuse according to the group law of the 1-form symmetry group $\G1$
\be
O_{\gamma_1} O_{\gamma'_1}=O_{\gamma_1\gamma'_1}
\ee
\item $L_{\gamma_0}$ lines permute $O_{\gamma_1}$ local operators as shown in figure \ref{fig:ZeroOnOne}.
\eit


\paragraph{2d SPT phases protected by split 2-group symmetry.}
Let us discuss \textit{2d SPT phases protected by $\cC_{\G0}(\G1,\rho)$ 2-group symmetry}, which are 2d TQFTs with $\cC_{\G0}(\G1,\rho)$ 2-group symmetry having only a single vacuum. Such a 2-group SPT is specified by the following data:
\ben
\item An SPT phase $\alpha\in H^2\big(\G0,U(1)\big)$ for the $\G0$ 0-form symmetry.
\item An SPT phase for the 1-form symmetry which is left invariant by the $\G0$ action. Such an SPT phase is specified by an element 
\be
\beta\in\mathsf{Inv}_{\wh\rho}(\wh{\G1}) \,,
\ee
where $\mathsf{Inv}_{\wh\rho}(\wh{\G1})$ is the subgroup of $\wh{\G1}$ left invariant by the action $\wh\rho$ of $\G0$ on $\wh{\G1}$ dual to the action $\rho$ of $\G0$ on $\G1$.
\een

\paragraph{Minimal 2d TQFTs with split 2-group symmetry.}
Generalizing the above discussion, a minimal 2d TQFT with $\cC_{\G0}(\G1,\rho)$ 2-group symmetry is specified by the following data:
\ben
\item A minimal 2d TQFT $\TQFT_{{\G0}',\alpha}$ having $\G0$ 0-form symmetry.
\item SPT phases $\SPT_i$ for $\G1$ 1-form symmetry, where $1\le i\le \left|\G0/{\G0}'\right|$, arising in the various vacua of $\TQFT_{{\G0}',\alpha}$. The SPT phase $\SPT_i$ must be described by an element
\be
\beta_i\in\mathsf{Inv}_{\wh{\rho'_i}}(\wh{\G1}) \,,
\ee
where $\mathsf{Inv}_{\wh{\rho'_i}}(\wh{\G1})$ is the subgroup of $\wh{\G1}$ left invariant by the action $\wh{\rho'_i}$ of the spontaneously preserved subgroup ${\G0_i}'\cong{\G0}'$ of $\G0$ on $\wh{\G1}$ dual to the action $\rho'_i$ of ${\G0_i}'$ on $\G1$ obtained by restricting the action $\rho$ of $\G0$ on $\G1$ to ${\G0_i}'\subseteq\G0$. Moreover, we must have
\be
\beta_i=\gamma_i\cdot\beta_1\,,
\ee
where $\gamma_i\cdot\beta_1$ is obtained from $\beta_1$ by acting by the element $\gamma_i\in\G0$, which is a representative of the class $[\gamma_i]\in\G0/{\G0}'$ relating the vacuum $i$ to vacuum $1$.\\
In other words, in a particular vacuum $v$ of $\TQFT_{{\G0}',\alpha}$, we choose a 1-form SPT phase left invariant by the spontaneously preserved subgroup ${\G0}'$ of $\G0$. The 1-form SPT phase in any other vacuum $v'$ is obtained by simply acting on the SPT phase realized in $v$ by an element of $\G0$ taking $v$ to $v'$.
\een
Let us denote such a minimal $\cC_{\G0}(\G1,\rho)$ symmetric 2d TQFT as $\TQFT_{{\G0}',\alpha,\{{\G0_i}',\beta_i\}}$, where $\{{\G0_i}',\beta_i\}$ is the list of preserved 0-form symmetries and 1-form SPT phases in each vacuum. Then, an arbitrary 2d TQFT with $\cC_{\G0}(\G1,\rho)$ 2-group symmetry can be expressed as
\be
\bigoplus_{I}\TQFT_{{\G0_I}',\alpha_I,\{{\G0_{I,i}}',\beta_{I,i}\}}
\ee
by taking direct sums of the above minimal TQFTs.

\paragraph{2-category of 2d TQFTs with split 2-group symmetry.}
Such 2d TQFTs are expected to form the fusion 2-category of 2-representations of the 2-group which we denote by 
\be
\TwoRep\big(\cC_{\G0}(\G1,\rho)\big)\,.
\ee
In this paper, we do not describe the structure of this 2-category in detail, but it can be determined as in other cases using the knowledge of underlying TQFTs without symmetry. 

\paragraph{Fusion of objects.}
Let us illustrate this by describing the fusion of objects in $\TwoRep\big(\cC_{\G0}(\G1,\rho)\big)$ for abelian $\G0$. Consider stacking two minimal TQFTs $\TQFT_{{\G0_1}',\alpha_1,\{\beta_{1,i}\}}$ and $\TQFT_{{\G0_2}',\alpha_2,\{\beta_{2,a}\}}$, where we have labeled the vacua of first TQFT by $i$ and the vacua of second TQFT by $a$, and we have dropped the labels ${\G0_{1,i}}'$ and ${\G0_{2,a}}'$ as ${\G0_{1,i}}'={\G0_1}'$ for all $i$ and ${\G0_{2,a}}'={\G0_2}'$ for all $a$. Following our previous analysis, we learn that the resulting TQFT is a direct sum of $n_{12}$ number of minimal TQFTs, where $n_{12}$ is given in equation (\ref{n12}). We can thus write
\be
\TQFT_{{\G0_1}',\alpha_1,\{\beta_{1,i}\}}\otimes\TQFT_{{\G0_2}',\alpha_2,\{\beta_{2,a}\}}=\bigoplus_{I=1}^{n_{12}}\TQFT_{{\G0_{12}}',\alpha_{12},\{\beta_{I,x}\}}\,,
\ee
where ${\G0_{12}}'$ is given in (\ref{G012}), $\alpha_{12}$ is given in (\ref{a12}), $1\le x\le \left|\G0/{\G0_{12}}'\right|$ are the vacua of the $I$-theory, and $\beta_{I,x}$ are 1-form SPT phases in these vacua which can be easily determined as follows. As discussed earlier, the vacua of the $I$-th theory lie in a particular orbit of vacua $\{(i,a)\}$ under the $\G0$ action. Pick a vacuum $x=(i,a)$ in this orbit. Then we have
\be
\beta_{I,(i,a)}=\beta_{1,i}+\beta_{2,a}\,.
\ee

\paragraph{1-morphisms.}
Let us describe the 1-morphisms for abelian $\G0$ from $\TQFT_{{\G0_1}',\alpha_1,\{\beta_{1,i}\}}$ to $\TQFT_{{\G0_2}',\alpha_2,\{\beta_{2,a}\}}$. Such 1-morphisms form the 1-category
\be
n'_{12}\times\Rep_{\alpha'_{12}}\big({\G0_{12}}'\big) \,,
\ee
where $\alpha'_{12}$ is given in (\ref{ap12}), ${\G0_{12}}'$ is given in (\ref{G012}) and $n'_{12}$ are the number of orbits formed by vacuum pairs $(i,a)$ with $i$ being a vacuum of $\TQFT_{{\G0_1}',\alpha_1,\{\beta_{1,i}\}}$ and $a$ being a vacuum of $\TQFT_{{\G0_1}',\alpha_1,\{\beta_{1,i}\}}$, such that
\be
\beta_{1,i}=\beta_{2,a} \,.
\ee
This is because a $\cC_{\G0}(\G1,\rho)$ symmetric interface needs to also be $\G1$ symmetric, and hence can only arise between vacua carrying the same character for $\G1$.
The composition and monoidal product on 1-morphisms is determined in exactly the same way as for 2d TQFTs with $\G0$ 0-form symmetry.

\subsubsection{Example: $\G0=\Z_2$, $\G1=\Z_2\times\Z_2$, $\rho$ = exchange}\label{eg1}
Consider a 2-group comprising of 0-form symmetry group $\G0=\Z_2$, 1-form symmetry group $\G1=\Z_2\times\Z_2$ and the non-trivial element of $\G0$ acting on $\G1$ by exchanging the two $\Z_2$s, while leaving the diagonal $\Z_2$ invariant.

\paragraph{SPT phases.}
Let us first describe 2d SPT phases protected by this 2-group symmetry. There is no non-trivial 0-form SPT phase for $\Z_2$. On the other hand, the only characters in $\wh{\G1}=\Z_2\times\Z_2$ invariant under the $\G0$ action are $(1,1)$ and $(-1,-1)$. Thus we obtain two SPT phases that we denote as
\be
\TQFT^{(\id)},~\TQFT^{(V)} \,,
\ee
where $\TQFT^{(\id)}$ is trivial 2-group SPT, while $\TQFT^{(V)}$ is a non-trivial 2-group SPT.

\paragraph{Other minimal TQFTs.}
Other possibilities for minimal 2d TQFTs with this 2-group symmetry arise from the $\Z_2$ gauge theory discussed earlier, corresponding to choosing ${\G0}'$ to be the trivial group. This theory has two vacua with no 0-form symmetry preserved in either of the two vacua. Let us label these two vacua as $v_1$ and $v_2$. In vacuum $v_1$, we can choose any SPT phase for 1-form symmetry. The 1-form SPT phase appearing in $v_2$ is then obtained from the 1-form SPT phase in $v_1$ by acting by the $\Z_2$ action. Thus, we have four options:
\ben
\item $v_1$ carries characters $(1,1)$. $v_2$ carries characters $(1,1)$.
\item $v_1$ carries characters $(-1,-1)$. $v_2$ carries characters $(-1,-1)$.
\item $v_1$ carries characters $(1,-1)$. $v_2$ carries characters $(-1,1)$.
\item $v_1$ carries characters $(-1,1)$. $v_2$ carries characters $(1,-1)$.
\een
However, the third and fourth options describe the same TQFT as can be seen by relabelling the vacua. Thus, we only need to consider the first three options. We label the three resulting 2-group symmetric minimal 2d TQFTs as respectively
\be
\TQFT^{(\Z_2)},~\TQFT^{(V_{\Z_2})},~\TQFT^{(SC)}\,.
\ee

\paragraph{Fusion of objects.}
Let us compute the fusion rules. All the fusions are commutative, so we only display a single ordering of the objects being fused below. First of all,
\be
\TQFT^{(\id)}\otimes\TQFT^{(i)}=\TQFT^{(i)}
\ee
for any minimal TQFT $\TQFT^{(i)}$. Fusion with the non-trivial SPT phase changes the signs of all characters.
As a consequence, the SPT phases close under the fusion with
\be
\TQFT^{(V)}\otimes\TQFT^{(V)}=\TQFT^{(\id)}
\ee
and we have
\begin{align}
\TQFT^{(V)}\otimes\TQFT^{(\Z_2)}&=\TQFT^{(V_{\Z_2})}\\
\TQFT^{(V)}\otimes\TQFT^{(V_{\Z_2})}&=\TQFT^{(\Z_2)}\\
\TQFT^{(V)}\otimes\TQFT^{(\Z_2)}&=\TQFT^{(\Z_2)}\,.
\end{align}
Now let us consider the fusion of TQFTs having 2 vacua. As we know from before, we obtain two orbits. One of the orbits is $\{(v_1,v_1),(v_2,v_2)\}$, while the other orbit is $\{(v_1,v_2),(v_2,v_1)\}$. The character associated to vacuum $(v_i,v_a)$ is obtained by taking the character for $v_i$ in the first theory and multiplying it with the character for $v_a$ in the second theory. This leads to the following fusion rules
\begin{align}
\TQFT^{(\Z_2)}\otimes\TQFT^{(\Z_2)}&=2\TQFT^{(\Z_2)}\\
\TQFT^{(V_{\Z_2})}\otimes\TQFT^{(\Z_2)}&=2\TQFT^{(V_{\Z_2})}\\
\TQFT^{(V_{\Z_2})}\otimes\TQFT^{(V_{\Z_2})}&=2\TQFT^{(\Z_2)}\\
\TQFT^{(\Z_2)}\otimes\TQFT^{(SC)}&=2\TQFT^{(SC)}\\
\TQFT^{(V_{\Z_2})}\otimes\TQFT^{(SC)}&=2\TQFT^{(SC)}\\
\TQFT^{(SC)}\otimes\TQFT^{(SC)}&=\TQFT^{(\Z_2)}\oplus\TQFT^{(V_{\Z_2})}\label{mf}\,.
\end{align}

\paragraph{1-morphisms.}
The 1-morphisms are as follows. The simple 1-endomorphisms of $\TQFT^{(\id)}$ are
\be
\cI^{(\id)},~\cI^{(-)}
\ee
with $\cI^{(\id)}$ corresponding to the trivial representation of ${\G0}'=\Z_2$ and $\cI^{(-)}$ corresponding to the non-trivial irreducible representation of ${\G0}'=\Z_2$. Similarly, the simple 1-endomorphisms of $\TQFT^{(V)}$ are
\be
\cI^{(V)},~\cI^{(V,-)}
\ee
with $\cI^{(V)}$ corresponding to the trivial representation of ${\G0}'=\Z_2$ and $\cI^{(V,-)}$ corresponding to the non-trivial irreducible representation of ${\G0}'=\Z_2$. There are no 1-morphisms between $\TQFT^{(\id)}$ and $\TQFT^{(V)}$ because they carry different 1-form SPT phases.

There is a single simple 1-morphism
\be
\cI^{(\id),(\Z_2)}
\ee
from $\TQFT^{(\id)}$ to $\TQFT^{(\Z_2)}$, and a single simple 1-morphism
\be
\cI^{(\Z_2),(\id)}
\ee
from $\TQFT^{(\Z_2)}$ to $\TQFT^{(\id)}$, just as for 2d TQFTs with $\Z_2$ 0-form symmetry. Correspondingly, there is a single simple 1-morphism
\be
\cI^{(V),(V_{\Z_2})}
\ee
from $\TQFT^{(V)}$ to $\TQFT^{(V_{\Z_2})}$, and a single simple 1-morphism
\be
\cI^{(V_{\Z_2}),(V)}
\ee
from $\TQFT^{(V_{\Z_2})}$ to $\TQFT^{(V)}$. There are no 1-morphisms between $\TQFT^{(\id)}$ and $\TQFT^{(V_{\Z_2})}$, and between $\TQFT^{(V)}$ and $\TQFT^{(\Z_2)}$. For similar reasons, there are no 1-morphisms between $\TQFT^{(\id)}$ and $\TQFT^{(SC)}$, and between $\TQFT^{(V)}$ and $\TQFT^{(SC)}$.

As for the case of 2d TQFTs with $\Z_2$ 0-form symmetry, we have two simple 1-endomorphisms
\be
\cI^{(\Z_2)},~\cI^{(\Z_2,-)}
\ee
of $\TQFT^{(\Z_2)}$, with $\cI^{(\Z_2)}$ being the identity under composition. Similarly, we have two simple 1-endomorphisms
\be
\cI^{(V_{\Z_2})},~\cI^{(V_{\Z_2},-)}
\ee
of $\TQFT^{(V_{\Z_2})}$, with $\cI^{(V_{\Z_2})}$ being the identity under composition. On the other hand, we have a single simple 1-endomorphism
\be
\cI^{(SC)}
\ee
of $\TQFT^{(SC)}$ which is the identity under composition. This is essentially because the two vacua of $\TQFT^{(SC)}$ carry different 1-form SPT phases, unlike the vacua of $\TQFT^{(\Z_2)}$ and $\TQFT^{(V_{\Z_2})}$. Finally, there are no cross 1-morphisms between $\TQFT^{(\Z_2)}$, $\TQFT^{(V_{\Z_2})}$ and $\TQFT^{(SC)}$.

The compositions and monoidal products of these 1-morphisms can be figured out similarly to the case of 2d TQFTs with $\Z_2$ 0-form symmetry.

\subsubsection{Example: $\G0=\Z_2$, $\G1=\Z_4$, $\rho$ = charge conjugation}\label{eg2}
Consider now a 2-group comprising of 0-form symmetry group $\G0=\Z_2$, 1-form symmetry group $\G1=\Z_4$ and the non-trivial element of $\G0$ acting on $\G1$ by exchanging the two generators of $\Z_4$, while leaving the $\Z_2$ subgroup of $\Z_4$ invariant.

\paragraph{SPT phases.}
Let us label the characters in $\wh{\G1}=\Z_4$ as $\{1,i,-1,-i\}$. The only characters invariant under the $\G0$ action are $\pm1$. Thus we obtain two SPT phases that we denote by
\be
\TQFT^{(\id)},~\TQFT^{(V)}\,,
\ee
where $\TQFT^{(\id)}$ is trivial 2-group SPT corresponding to the character $+1$, while $\TQFT^{(V)}$ is a non-trivial 2-group SPT corresponding to the character $-1$.

\paragraph{Other minimal TQFTs.}
Other possibilities for minimal 2d TQFTs with this 2-group symmetry again arise from the $\Z_2$ gauge theory. Again, in vacuum $v_1$, we can choose any SPT phase for 1-form symmetry. The 1-form SPT phase appearing in $v_2$ is then obtained from the 1-form SPT phase in $v_1$ by acting by the $\Z_2$ action. Thus, we have four options:
\ben
\item $v_1$ carries character $1$. $v_2$ carries character $1$.
\item $v_1$ carries character $-1$. $v_2$ carries character $-1$.
\item $v_1$ carries character $i$. $v_2$ carries character $-i$.
\item $v_1$ carries character $-i$. $v_2$ carries character $i$.
\een
However, the third and fourth options describe the same TQFT as can be seen by relabelling the vacua. Thus, we only need to consider the first three options. We label the three resulting 2-group symmetric minimal 2d TQFTs as respectively
\be
\TQFT^{(\Z_2)},~\TQFT^{(V_{\Z_2})},~\TQFT^{(SC)}\,.
\ee
The reader can check that the fusion rules are the same as in the previous example. The spectrum of 1-morphisms, their compositions and monoidal products are also the same as in the previous example.

\subsubsection{Example: $\G0=S_3$, $\G1=\Z_2\times\Z_2$, $\rho$ = permutation}\label{eg3}
Next consider a 1-form symmetry $\Gamma^{(1)}=\mathbb{Z}_2^{(S)} \times \mathbb{Z}_2^{(C)}$, with the diagonal $\mathbb{Z}_2^{(V)}$, and a 0-form symmetry $\Gamma^{(0)}= S_3$ acting by $\rho$ which is permuting the labels $\{S, C, V\}$. An example of such a theory having this 2-group symmetry is given by a pure $\Spin(8)$ Yang-Mills theory in any spacetime dimension.

\paragraph{SPT phases.}
A 2d SPT phase protected by this 2-group symmetry must arise from the trivial SPT phase for the $S_3$ 0-form symmetry, which preserves all of $S_3$. The $\Z_2\times\Z_2$ 1-form SPT phase participating in such a 2-group SPT phase must be described by the character $(1,1)$ as it is the only character that is invariant under the full $S_3$ action.

Thus, we have only a single 2d SPT phase 
\be
\TQFT^{(\id)}
\ee
protected by this 2-group symmetry, which is the trivial SPT phase.

\paragraph{Other minimal TQFTs.}
Let us consider possible 2-group symmetric minimal 2d TQFTs arising from the 2d TQFT $\TQFT^{(\Z_2)}$ with $S_3$ 0-form symmetry. This theory has two vacua and in each vacuum, the $\Z_3$ subgroup of $S_3$ is preserved. Thus, the $\Z_2\times\Z_2$ character carried by each vacuum must be $(1,1)$ as it is the only $\Z_3$ invariant character. Thus, we obtain only a single 2-group symmetric minimal 2d TQFT which we denote by
\be
\TQFT^{(\Z_2)}\,.
\ee
Now consider 2-group symmetric minimal 2d TQFTs arising from the 2d TQFT $\TQFT^{(\Z_3)}$ with $S_3$ 0-form symmetry. This theory has three vacua with each vacuum preserving a different $\Z_2$ subgroup of $S_3$. One possible choice of characters is the trivial character $(1,1)$ in each vacuum, leading to the 2-group symmetric minimal 2d TQFT that we denote by
\be
\TQFT^{(\Z_3)}\,.
\ee
Another choice is to have character $(-1,-1)$ in vacuum $v_1$, which is invariant under $\Z_2^b$ as the element $b$ exchanges the two characters. This induces corresponding invariant characters $(1,-1)$ and $(-1,1)$ for the other two vacua. We are thus lead to another 2-group symmetric minimal 2d TQFT that we denote by
\be
\TQFT^{(SCV)}\,.
\ee
Finally, let us consider 2-group symmetric minimal 2d TQFTs arising from the 2d TQFT $\TQFT^{(S_3)}$ with $S_3$ 0-form symmetry. We again have the trivial choice of characters for each of the six vacua, leading to the 2-group symmetric minimal 2d TQFT that we denote by
\be
\TQFT^{(S_3)} \,.
\ee
Since no vacua has any non-trivial preserved symmetry, we can also choose a non-trivial character in a vacuum. Then all the other possible non-trivial characters arise in the other five vacua. So, we have a single 2-group symmetric minimal 2d TQFT coming from non-trivial choices of characters, that we denote by
\be
\TQFT^{(SCV_{\Z_2})} \,.
\ee

\paragraph{Fusion of objects.}
Let us compute the fusion rules. All the fusions are commutative, so we only display a single ordering of the objects being fused below. First of all,
\be
\TQFT^{(\id)}\otimes\TQFT^{(i)}=\TQFT^{(i)}
\ee
for any minimal TQFT $\TQFT^{(i)}$. Moreover, the fusion rules of $\TQFT^{(\Z_2)}$, $\TQFT^{(\Z_3)}$ and $\TQFT^{(S_3)}$ among themselves are the same as in section \ref{eg3p}. We only discuss other fusion rules below. These fusion rules are obtained by dressing the right hand sides of the fusion rules of section \ref{eg3p} by extra characters.

First of all, we have
\be
\TQFT^{(SCV)} \otimes \TQFT^{(\mathbb{Z}_2)} = \TQFT^{(SCV_{\mathbb{Z}_2})} \,.
\ee
To see this, note that every vacuum carries a non-trivial character and there is a single orbit of vacuum pairs.

Then, we have
\be
\TQFT^{(SCV)} \otimes \TQFT^{(\mathbb{Z}_3)} = 
\TQFT^{(SCV_{\mathbb{Z}_2})} \oplus \TQFT^{(SCV)} \,,
\ee
which can be understood by noting that any vacuum pair carries non-trivial characters. Similarly, we find
\be
\TQFT^{(SCV)} \otimes \TQFT^{(SCV)} = \TQFT^{(\mathbb{Z}_3)}\label{mf2} \oplus \TQFT^{(SCV_{\mathbb{Z}_2})} \,,
\ee
which can be understood by noting that a vacuum pair of the form $(v_i,v_i)$ carries trivial characters giving rise to $\TQFT^{(\mathbb{Z}_3)}$ on the right hand side, and a vacuum pair of the form $(v_i,v_j)$ for $i\neq j$ carries non-trivial characters giving rise to $\TQFT^{(SCV_{\mathbb{Z}_2})}$ on the right hand side.

Then, we have
\be
\TQFT^{(SCV)} \otimes \TQFT^{(S_3)} =  3 \TQFT^{(SCV_{\mathbb{Z}_2})} \,,
\ee
because we have a non-trivial character for each vacuum pair. Similarly, we have
\be
\TQFT^{(SCV)}\otimes \TQFT^{(SCV_{\mathbb{Z}_2})} = 
\TQFT^{(S_3)} \oplus 2 \TQFT^{(SCV_{\mathbb{Z}_2})} \,.
\ee
This follows from the fact that two of the orbits of vacuum pairs carry non-trivial characters, while one orbit does not.

Finally we have
\be
\TQFT^{(SCV_{\mathbb{Z}_2})}\otimes \TQFT^{(SCV_{\mathbb{Z}_2})} = 2 \TQFT^{(S_3)} \oplus 4 \TQFT^{(SCV_{\mathbb{Z}_2})}
\ee
as four out of six orbits have non-trivial characters.

\paragraph{1-morphisms.}
The 1-morphisms not discussed in section \ref{eg3p} are as follows:
\bit
\item There are no 1-morphisms between objects $\TQFT^{(SCV)},\TQFT^{(SCV_{\Z_2})}$ involving non-trivial characters and other objects involving trivial character.
\item Let us discuss the 1-endomorphisms of $\TQFT^{(SCV)}$. As discussed in section \ref{eg3p}, the underlying $S_3$ symmetric 2d TQFT $\TQFT^{(\Z_3)}$ has 1-morphisms given by $\Rep(\Z_2)$ coming from connecting vacua $(v_i,v_i)$, and 1-morphisms given by $\Vec$ coming from connecting vacua $(v_i,v_j)$ for $i\neq j$. Once the characters are included, we can make the connections $(v_i,v_i)$, but not the connections $(v_i,v_j)$. Thus, the 1-endomorphisms of $\TQFT^{(SCV)}$ are described by
\be
\Rep(\Z_2)\,.
\ee
Let us denote the two simple 1-endomorphisms of $\TQFT^{(SCV)}$ as
\be
\cI^{(SCV)},~\cI^{(SCV,-)}\,.
\ee
These simple 1-endomorphisms form a $\Z_2$ group under compositions, with $\cI^{(SCV,-)}$ being the non-trivial element.
\item Similarly, we can discuss the 1-morphisms between $\TQFT^{(SCV)}$ and $\TQFT^{(SCV_{\Z_2})}$. The underlying morphisms between $\TQFT^{(\Z_3)}$ and $\TQFT^{(S_3)}$ are $3\times\Vec$, out of which only a single $\Vec$ survives once the characters are taken into account. Thus, the 1-morphisms between $\TQFT^{(SCV)}$ and $\TQFT^{(SCV_{\Z_2})}$ are described by
\be
\Vec\,.
\ee
\item Finally, in a similar way we find that the 1-endomorphisms of $\TQFT^{(SCV_{\Z_2})}$ are described by
\be
2\times\Vec\,.
\ee
\eit

\paragraph{Fusion of 1-morphisms.}
In the remaining part of this subsection, let us describe how to compute monoidal product on above 1-morphisms. We will illustrate the procedure by computing the products of the 1-morphisms $\cI^{(SCV)},\cI^{(SCV,-)}$.

The result of any such product is a 1-endomorphism of $\TQFT^{(\Z_3)}$ plus a 1-endomorphism of $\TQFT^{(SCV_{\Z_2})}$, since there are no 1-morphisms between $\TQFT^{(\Z_3)}$ and $\TQFT^{(SCV_{\Z_2})}$. The resulting 1-endomorphism of $\TQFT^{(\Z_3)}$ must lie in the $\Rep(\Z_2)$ part of the 1-endomorphism category $\Rep(\Z_2)\oplus\Vec$ since both $\cI^{(SCV)}$ and $\cI^{(SCV,-)}$ connect $v_i$ with $v_i$. For the same reason, the resulting 1-endomorphism of $\TQFT^{(SCV_{\Z_2})}$ must lie in one of the $\Vec$ parts of the 1-endomorphism category $2\times\Vec$.

In more detail, we obtain
\begin{align}
\cI^{(SCV)}\otimes \cI^{(SCV)}&=\cI^{(\Z_3)}\oplus\cI^{(SCV_{\Z_2})}\\
\cI^{(SCV)}\otimes \cI^{(SCV,-)}&=\cI^{(\Z_3,-)}\oplus\cI^{(SCV_{\Z_2})}\\
\cI^{(SCV,-)}\otimes \cI^{(SCV,-)}&=\cI^{(\Z_3)}\oplus\cI^{(SCV_{\Z_2})}\,,
\end{align}
where $\cI^{(\Z_3)},\cI^{(\Z_3,-)}$ are the simple 1-endomorphisms of $\TQFT^{(\Z_3)}$ living in the $\Rep(\Z_2)$ part, and $\cI^{(SCV_{\Z_2})}$ is a simple 1-endomorphism of $\TQFT^{(SCV_{\Z_2})}$ living in one of the $\Vec$ parts. In particular, $\cI^{(\Z_3)}$ and $\cI^{(SCV_{\Z_2})}$ are the identity 1-endomorphisms under the composition of 1-endomorphisms. On the other hand, $\cI^{(\Z_3,-)}$ composes with itself to give rise to $\cI^{(\Z_3)}$.

As discussed at the end of the section \ref{appcomp}, the 2-category being discussed in this subsection arises as symmetry category of pure 4d $\Spin(8)\rtimes S_3$ gauge theory. This symmetry category was discussed by other methods in \cite{Bhardwaj:2022yxj}, but it is quite difficult to completely determine the fusions of $\cI^{(SCV)},\cI^{(SCV,-)}$ using the method described there. On the other hand, we are easily able to compute these fusions as described above. Moreover, we are able to include the condensation surface defects and topological lines living on them in these computations, which were not discussed at all in \cite{Bhardwaj:2022yxj}.




\subsection{2d TQFTs with Non-split 2-Group Symmetry}
\label{sec:2gpNonSplit}

A more general 2-group which is not split carries also a \textit{Postnikov class}
\be
[\omega]\in H^3_\rho\big(\G0,\G1\big) \,,
\ee
which is a degree three class in the group-cohomology  of $\G0$ with coefficients in $\G1$, twisted by the action of $\rho$. Such a 2-group symmetry corresponds to a monoidal category $\cC_{\G0}^{\omega}(\G1,\rho)$ defined in appendix \ref{App:Cats}, see also \cite{etingof2016tensor} sections 2.3 and 2.11., where $\omega\in Z^3_\rho\big(\G0,\G1\big)$ is a representative cocycle for the Postnikov class $[\omega]$. The categories $\cC_{\G0}^{\omega}(\G1,\rho)$ for different choices of representatives $\omega$ are equivalent. We make a specific choice of $\omega$ below, but the results obtained for different choices of $\omega$ are equivalent.

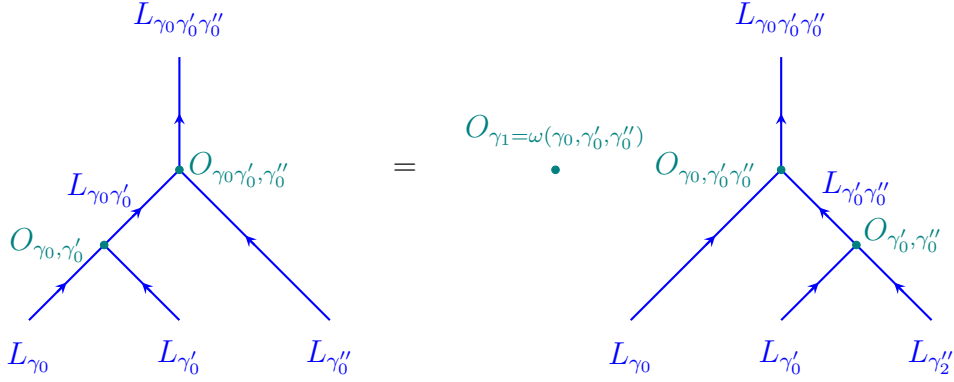
\begin{figure}
\centering
\begin{tikzpicture}
\draw [blue,thick](-2.5,-1.5) -- (-0.5,0.5) -- (-0.5,2) (-1.5,-0.5) -- (-0.5,-1.5) (-0.5,0.5) -- (1.5,-1.5);
\draw [teal,fill=teal] (-0.5,0.5) ellipse (0.05 and 0.05);
\draw [teal,fill=teal] (4.5,0.5) ellipse (0.05 and 0.05);
\draw [teal,fill=teal] (-1.5,-0.5) ellipse (0.05 and 0.05);
\draw [blue,thick,-stealth](-2.125,-1.125) -- (-2,-1);
\draw [blue,thick,-stealth](-0.875,-1.125) -- (-1,-1);
\draw [blue,thick,-stealth](-1.125,-0.125) -- (-1,0);
\draw [blue,thick,-stealth](0.5,-0.5) -- (0.375,-0.375);
\draw [blue,thick,-stealth](-0.5,1.125) -- (-0.5,1.25);
\node[blue] at (-2.5,-2) {$L_{\gamma_0}$};
\node[blue] at (-0.5,-2) {$L_{\gamma'_0}$};
\node[blue] at (1.5,-2) {$L_{\gamma''_0}$};
\node[blue] at (-0.5,2.5) {$L_{\gamma_0\gamma'_0\gamma''_0}$};
\node[blue] at (-1.55,0.2247) {$L_{\gamma_0\gamma'_0}$};
\node[teal] at (-2.25,-0.5) {{$O_{\gamma_0,\gamma'_0}$}};
\node[teal] at (0.3,0.5) {{$O_{\gamma_0\gamma'_0, \gamma''_0}$}};
\node at (2.5,0.5) {=};
\begin{scope}[shift={(8,0)}]
\draw [blue,thick](-2.5,-1.5) -- (-0.5,0.5) -- (-0.5,2) (0.5,-0.5) -- (-0.5,-1.5) (-0.5,0.5) -- (1.5,-1.5);
\draw [teal, fill=teal] (-0.5,0.5) ellipse (0.05 and 0.05);
\draw [teal,fill=teal] (0.5,-0.5) ellipse (0.05 and 0.05);
\draw [blue,thick,-stealth](-1.5,-0.5) -- (-1.375,-0.375);
\draw [blue,thick,-stealth](-0.125,-1.125) -- (0,-1);
\draw [blue,thick,-stealth](0.125,-0.125) -- (0,0);
\draw [blue,thick,-stealth](1.125,-1.125) -- (1,-1);
\draw [blue,thick,-stealth](-0.5,1.125) -- (-0.5,1.25);
\node[blue] at (-2.5,-2) {$L_{\gamma_0}$};
\node[blue] at (-0.5,-2) {$L_{\gamma'_0}$};
\node[blue] at (1.5,-2) {$L_{\gamma''_2}$};
\node[blue] at (-0.5,2.5) {$L_{\gamma_0\gamma'_0\gamma''_0}$};
\node[blue] at (0.5,0.25) {$L_{\gamma'_0\gamma''_0}$};
\node[teal] at (1.125,-0.375) {$O_{\gamma'_0,\gamma''_0}$};
\node[teal] at (-1.5,0.5) {$O_{\gamma_0, \gamma'_0\gamma''_0}$};
 \end{scope}
\node[teal] at (4.5,1) {$O_{\gamma_1=\omega(\gamma_0,\gamma'_0,\gamma''_0)}$};
\end{tikzpicture}
\caption{Associativity condition satisfied by line operators $L_{\gamma_0}$ and local operators $O_{\gamma_0,\gamma'_0}$ for a non-split 2-group symmetry. On the right hand side, we obtain an additional operator $O_{\gamma_1}$ where $\gamma_1=\omega(\gamma_0,\gamma'_0,\gamma''_0)\in\G1$.
\label{fig:Assoc2}}

\end{figure}

\paragraph{Equipping a 2d TQFT with non-split 2-group symmetry.}
A 2d TQFT $\TQFT_n$ characterized by a positive integer $n$ acquires a $\cC_{\G0}^\omega(\G1,\rho)$ 2-group symmetry if one specifies a monoidal functor
\be
\cC_{\G0}^\omega(\G1,\rho)\to\mathsf{Mat}_n(\Vec)\,.
\ee
This means that we specify, in $\TQFT_n$, line operators $L_{\gamma_0}$ for all $\gamma_0\in\G0$, local operators $O_{\gamma_0,\gamma'_0}$ living at the junction of $L_{\gamma_0}$, $L_{\gamma'_0}$ and $L_{\gamma_0\gamma'_0}$, and local operators $O_{\gamma_1}$ for all $\gamma_1\in\G1$ such that
\bit
\item The lines fuse according to the group law of the 0-form symmetry group $\G0$
\be
L_{\gamma_0}\circ L_{\gamma'_0}=L_{\gamma_0\gamma'_0}
\ee
\item The local operators $O_{\gamma_1}$ fuse according to the group law of the 1-form symmetry group $\G1$
\be
O_{\gamma_1} O_{\gamma'_1}=O_{\gamma_1\gamma'_1}
\ee
\item $L_{\gamma_0}$ lines permute $O_{\gamma_1}$ local operators as shown in figure \ref{fig:ZeroOnOne}.
\item The junction local operators $O_{\gamma_0,\gamma'_0}$ satisfy the associativity condition shown in figure \ref{fig:Assoc2}, which involves appearance of $O_{\gamma_1}$ local operators.
\eit

\paragraph{2d TQFTs with non-split 2-group symmetry.}
The above data can also be summarized as follows:
\ben
\item First of all, each $L_{\gamma_0}$ permutes the $n$ vacua in some way, giving us a homomorphism
\be
\sigma:~\G0\to S_n
\ee
from 0-form group $\G0$ to the permutation group $S_n$ of $n$ elements. This also provides an action of $\G0$ on the group $(\C^*)^n$ of all local operators of $\TQFT_n$, as each $\C^*$ subfactor of $(\C^*)^n$ corresponds to a particular vacuum.
\item The operators $O_{\gamma_1}$ provide a homomorphism
\be
\beta:~\G1\to(\C^*)^n\,,
\ee
which needs to commute with the action of $\G0$ on $\G1$ and $(\C^*)^n$.
\item Finally, the junction operators $O_{\gamma_0,\gamma'_0}$ provide a $\G0$ 2-cochain $\alpha$ valued in $(\C^*)^n$. The associativity condition requires that
\be
\delta_\rho\alpha=\beta(\omega)\,,
\ee
where $\delta_\rho\alpha$ is the $(\C^*)^n$ valued $\G0$ 3-cochain obtained by acting on $\alpha$ with the differential twisted by the action $\rho$ and $\beta(\omega)$ is a $(\C^*)^n$ valued $\G0$ 3-cochain obtained by applying the homomorphism $\beta$ on the $\G1$ valued $\G0$ 3-cochain $\omega$.
\een
Thus, 2d TQFTs with $\cC_{\G0}^\omega(\G1,\rho)$ 2-group symmetry are specified by quadruples
\be
(n,\sigma,\beta,\alpha)
\ee
satisfying the above conditions. Two TQFTs specified by quadruples $(n,\sigma,\beta,\alpha)$ and $(n',\sigma',\beta',\alpha')$ are isomorphic if the following conditions hold:
\ben
\item $n=n'$.
\item $\sigma'=\pi\sigma\pi^{-1}$ for some element $\pi\in S_n$.
\item $\beta'=\rho_{\pi}(\beta)$ where $\rho_\pi$ is the automorphism of $(\C^*)^n$ obtained from $\pi$ acting on the $n$ vacua.
\item $\alpha'+\delta_{\rho_{\sigma'}}c'=\rho_{\pi}(\alpha)+\delta_{\rho_{\sigma}}c$, where $\delta_{\rho_i}$ is the differential twisted by the action $\rho_i$ of $\G0$ on $(\C^*)^n$, and $c,c'$ are $(\C^*)^n$ valued $\G0$ 1-cochains.
\een
These 2d TQFTs with $\cC_{\G0}^\omega(\G1,\rho)$ 2-group symmetry form objects of the 2-category of 2-representations of the 2-group $\cC_{\G0}^\omega(\G1,\rho)$, which we denote as 
\be
\TwoRep\big(\cC_{\G0}^\omega(\G1,\rho)\big)\,.
\ee
The structure of this 2-category is discussed in \cite{Elgueta}, where the classification of the objects also appears.

\paragraph{2d SPT phases protected by non-split 2-group symmetry.}
In particular, the inequivalent 2d SPT phases protected by $\cC_{\G0}^\omega(\G1,\rho)$ 2-group symmetry, or in other words single-vacuum 2d TQFTs with $\cC_{\G0}^\omega(\G1,\rho)$ 2-group symmetry, are specified by:
\ben
\item A 1-form SPT phase left invariant by the 0-form action, which is specified by a character
\be
\beta\in\mathsf{Inv}_{\wh\rho}(\wh{\G1})
\ee
as in the previous subsection, satisfying the additional condition that
\be
\beta([\omega])=0\in H^3\big(\G0,U(1)\big) \,.
\ee
\item An analog of a 0-form SPT phase, specified by an equivalence class $[\alpha]$ of $U(1)$ valued $\G0$ 2-cocycles, such that any representative $\alpha$ of $[\alpha]$ satisfies
\be
\delta\alpha=\beta(\omega)
\ee
and two representatives $\alpha,\alpha'$ of $[\alpha]$ are related by
\be
\alpha'=\alpha+\delta c \,,
\ee
where $c$ is a $U(1)$ valued $\G0$ 1-cochain.
\een
This slightly refines the classification of 2-group SPT phases appearing in \cite{Kapustin:2013uxa}.




\section{Relationship to Condensation Defects}
\label{RCD}
We have studied above a 2-category 
\be
\cC_2^{\G0}=\TwoRep(\G0)
\ee
of dual symmetries arising in a $d$-dimensional theory $\fT/\G0$ obtained after gauging $\G0$ 0-form symmetry of a $d$-dimensional theory $\fT$. This 2-category contains a sub-1-category
\be
\cC_1^{\G0}=\Rep(\G0)
\ee
describing the Wilson line defects in the gauged theory $\fT/\G0$.

In this section, we argue that all the surface topological defects associated to objects of $\TwoRep(\G0)$ are actually condensation defects. That is, they can be obtained by performing a generalized gauging of the $\Rep(\G0)$ categorical symmetry associated to Wilson lines on a two-dimensional surface in spacetime. This procedure has also been dubbed as \textit{higher-gauging} in the recent literature \cite{Roumpedakis:2022aik, Choi:2022zal}.

This will be demonstrated explicitly for abelian $\G0$. We first show that the 2d TQFT with $\G0$ 0-form symmetry associated to a simple object of $\TwoRep(\G0)$ can be described as a 2d gauge theory with gauge group
\be
\wh\cG\subseteq\wh{\G0}
\ee
and discrete torsion given by an element
\be
[\alpha_{\wh\cG}]\in H^2\big(\wh\cG,U(1)\big) \,.
\ee
The corresponding topological surface defect can then be obtained by gauging the $\wh\cG$ subgroup of the $\wh{\G0}$ $(d-2)$-form symmetry generated by the Wilson line defects along a two-dimensional surface, with the discrete torsion for the gauging specified by $[\alpha_{\wh\cG}]$.

For non-abelian $\G0$, we will not be so explicit, but rather use some more mathematical ideas coming from category theory to argue that every simple object of $\TwoRep(\G0)$ is a condensation defect.

Finally, we will use these mathematical ideas to describe two-dimensional condensation defects in any arbitrary quantum field theory $\fT$. We argue that the two-dimensional condensation defects and topological interfaces between them in $\fT$ form the fusion 2-category $\mathsf{Mod}(\cB_\fT)$, namely the 2-category formed by module categories of $\cB_\fT$, where $\cB_\fT$ is the braided fusion category (with trivial braiding for $d>3$) formed by topological line defects of $\fT$.

\subsection{2d TQFTs with Abelian $\Gamma^{(0)}$ as Gauge Theories}
In this subsection, we show that any minimal 2d TQFT $\TQFT_{{\G0}',\alpha}$ with abelian $\G0$ 0-form symmetry can be described as a 2d gauge theory with discrete torsion.

\subsubsection{Revisiting SPT Phases Protected by 0-Form Symmetry}
We begin by describing a 2d SPT phase protected by abelian $\G0$ 0-form symmetry in a slightly non-traditional language following \cite{Gaiotto:2014kfa}. Consider such an SPT phase associated to an element
\be
[\alpha]\in H^2\big(\G0,U(1)\big)\,.
\ee
Let $\alpha\in Z^2\big(\G0,U(1)\big)$ be a representative of $[\alpha]$, and define
\be\label{ea}
\epsilon(g_1,g_2):=\alpha(g_1,g_2)\alpha(g_1^{-1},g_2^{-1})\,.
\ee
From the fact that $\alpha$ is a cocycle, we can deduce that $\epsilon$ is a bi-homomorphism
\be
\epsilon:~\G0\times \G0\to U(1)
\ee
with the additional property
\be\label{ap}
\epsilon(g_1,g_2)=\epsilon(g_2^{-1},g_1)\,.
\ee
It can be checked that changing the representative $\alpha$ does not change $\epsilon$. 

Conversely, given a bi-homomorphism $\epsilon$ satisfying (\ref{ap}), we can pick a $\G0$ 2-cochain $\alpha$ satisfying (\ref{ea}). The bi-homomorphism condition on $\epsilon$ implies that $\alpha$ is a cocycle, and different choices of $\alpha$ lie in the same cohomology class. Thus, in conclusion, a bosonic SPT phase protected by a finite abelian 0-form symmetry group $\G0$ is specified by a bi-character $\epsilon$ satisfying (\ref{ap}).

The data of $\epsilon$ is encoded beautifully in the spectrum of defects of the SPT phase. We have topological line defects $L_g$ labeled by group elements $g\in \G0$. Moreover, we have a topological local operator $O_g$ living at the end of $L_g$ satisfying the following conditions
\be
\begin{tikzpicture}
\draw [thick,blue](-4,-1) -- (-2,-1);
\draw [thick,blue](1.5,-1) -- (-0.5,-1);
\draw [red,fill=red] (-2,-1) node (v1) {} ellipse (0.05 and 0.05);
\draw [red,fill=red] (-0.5,-1) node (v1) {} ellipse (0.05 and 0.05);
\node[blue] at (-4.5,-1) {$L_g$};
\node[blue] at (2,-1) {$L_g$};
\node[red] at (-2,-1.5) {$O_g$};
\node[red] at (-0.5,-1.5) {$O_{g^{-1}}$};
\draw [thick,blue,-stealth](-3,-1) -- (-2.9,-1);
\draw [thick,blue,-stealth](0.6,-1) -- (0.7,-1);
\node at (3,-1) {=};
\begin{scope}[shift={(8.5,0)}]
\node[blue] at (-4.5,-1) {$L_g$};
\draw [thick,blue](-4,-1) -- (-1,-1);
\draw [thick,blue,-stealth](-2.5,-1) -- (-2.4,-1);
\end{scope}
\begin{scope}[shift={(3.5,-2)}]
\draw [thick,blue](-4,-1) -- (-2,-1);
\draw [red,fill=red] (-2,-1) node (v1) {} ellipse (0.05 and 0.05);
\node[blue] at (-3,-0.5) {$L_g$};
\node[red] at (-1.5,-1) {$O_g$};
\draw [thick,blue,-stealth](-3,-1) -- (-2.9,-1);
\draw [red,fill=red] (-4,-1) node (v1) {} ellipse (0.05 and 0.05);
\node[red] at (-4.65,-1) {$O_{g^{-1}}$};
\node at (-0.5,-1) {=};
\end{scope}
\draw [thick,blue,dashed](4.5,-3) -- (7.5,-3);
\node[blue] at (4,-3) {$L_1$};
\end{tikzpicture}
\ee
where the dashed line is the identity line. The bi-character $\epsilon$ captures the charge of the operators $O_g$ under the 0-form symmetry $\G0$. That is, we have
\be
\begin{tikzpicture}
\draw [thick,blue](-4.5,-1) -- (-2,-1);
\draw [red,fill=red] (-2,-1) node (v1) {} ellipse (0.05 and 0.05);
\node[blue] at (-5,-1) {$L_{g_2}$};
\node[red] at (-2,-1.5) {$O_{g_2}$};
\draw [thick,blue,-stealth](-3.5,-1) -- (-3.4,-1);
\begin{scope}[shift={(-2,-3)},rotate=-90]
\node[blue] at (-4,-1) {$L_{g_1}$};
\draw [thick,blue](-3.5,-1) -- (-1,-1);
\draw [thick,blue,-stealth](-2.5,-1) -- (-2.4,-1);
\end{scope}
\node at (-1,-1) {=};
\begin{scope}[shift={(7,0)}]
\draw [thick,blue](-4.5,-1) -- (-2,-1);
\draw [red,fill=red] (-2,-1) node (v1) {} ellipse (0.05 and 0.05);
\node[blue] at (-5,-1) {$L_{g_2}$};
\node[red] at (-2,-1.5) {$O_{g_2}$};
\draw [thick,blue,-stealth](-3.5,-1) -- (-3.4,-1);
\begin{scope}[shift={(0,-3)},rotate=-90]
\node[blue] at (-4,-1) {$L_{g_1}$};
\draw [thick,blue](-3.5,-1) -- (-1,-1);
\draw [thick,blue,-stealth](-2.5,-1) -- (-2.4,-1);
\end{scope}
\end{scope}
\node at (0.5,-1) {$\epsilon(g_1,g_2)~\times$};
\end{tikzpicture}
\ee
Now, the condition (\ref{ap}) follows from
\be
\begin{tikzpicture}
\draw [thick,blue](-4.5,-1) -- (-2,-1);
\draw [red,fill=red] (-2,-1) node (v1) {} ellipse (0.05 and 0.05);
\node[blue] at (-5,-1) {$L_{g_2}$};
\node[red] at (-2,-1.5) {$O_{g_2}$};
\draw [thick,blue,-stealth](-3.5,-1) -- (-3.4,-1);
\begin{scope}[shift={(-2,-3)},rotate=-90]
\node[blue] at (-4,-1) {$L_{g_1}$};
\draw [thick,blue](-3.5,-1) -- (-1,-1);
\draw [thick,blue,-stealth](-2.5,-1) -- (-2.4,-1);
\draw [red,fill=red] (-1,-1) node (v1) {} ellipse (0.05 and 0.05);
\node[red] at (-0.5,-1) {$O_{g_1}$};
\end{scope}
\node at (-1,-1) {=};
\begin{scope}[shift={(7,0)}]
\draw [thick,blue](-4.5,-1) -- (-2,-1);
\draw [red,fill=red] (-2,-1) node (v1) {} ellipse (0.05 and 0.05);
\node[blue] at (-5,-1) {$L_{g_2}$};
\node[red] at (-2,-1.5) {$O_{g_2}$};
\draw [thick,blue,-stealth](-3.5,-1) -- (-3.4,-1);
\begin{scope}[shift={(0,-3)},rotate=-90]
\node[blue] at (-4,-1) {$L_{g_1}$};
\draw [thick,blue](-3.5,-1) -- (-1,-1);
\draw [thick,blue,-stealth](-2.5,-1) -- (-2.4,-1);
\draw [red,fill=red] (-1,-1) node (v1) {} ellipse (0.05 and 0.05);
\node[red] at (-0.5,-1) {$O_{g_1}$};
\end{scope}
\end{scope}
\node at (0.5,-1) {$\epsilon(g_1,g_2)~\times$};
\begin{scope}[shift={(0,-5)}]
\node at (-1,-1) {=};
\begin{scope}[shift={(7,0)}]
\draw [thick,blue](-4.5,-3) -- (-2,-3);
\draw [red,fill=red] (-2,-3) node (v1) {} ellipse (0.05 and 0.05);
\node[blue] at (-5,-3) {$L_{g_2}$};
\node[red] at (-2,-3.5) {$O_{g_2}$};
\draw [thick,blue,-stealth](-3.5,-3) -- (-3.4,-3);
\begin{scope}[shift={(-2,-3)},rotate=-90]
\node[blue] at (-4,-1) {$L_{g_1}$};
\draw [thick,blue](-3.5,-1) -- (-1,-1);
\draw [thick,blue,-stealth](-2.5,-1) -- (-2.4,-1);
\draw [red,fill=red] (-1,-1) node (v1) {} ellipse (0.05 and 0.05);
\node[red] at (-0.5,-1) {$O_{g_1}$};
\end{scope}
\end{scope}
\node at (0.5,-1) {$\epsilon(g_2^{-1},g_1)~\times$};
\end{scope}
\end{tikzpicture}
\ee
where the top configuration on the right hand side arises by moving $O_{g_2}$ across $L_{g_1}$, and the bottom configuration on the right hand side arises by moving $O_{g_1}$ across $L_{g_2}$. Thus, the condition (\ref{ap}) can be understood as requiring mutual locality of operators $O_g$.

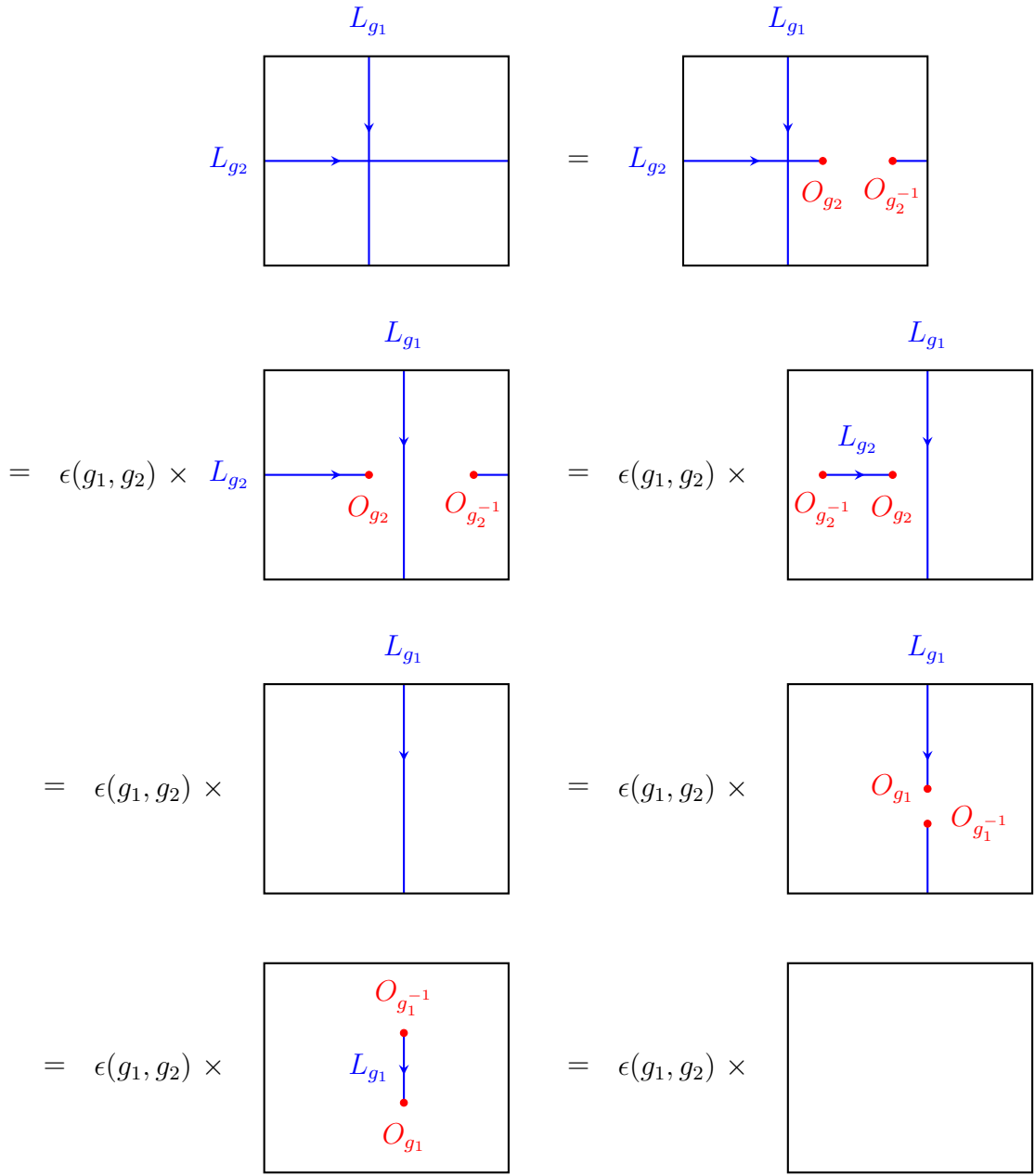
\begin{figure}
$$
\begin{tikzpicture}
\draw [thick,blue](-5,-1) -- (-1.5,-1);
\node[blue] at (-5.5,-1) {$L_{g_2}$};
\draw [thick,blue,-stealth](-4,-1) -- (-3.9,-1);
\begin{scope}[shift={(-2.5,-3)},rotate=-90]
\node[blue] at (-4,-1) {$L_{g_1}$};
\draw [thick,blue](-3.5,-1) -- (-0.5,-1);
\draw [thick,blue,-stealth](-2.5,-1) -- (-2.4,-1);
\end{scope}
\node at (-0.5,-1) {=};
\draw [thick](-5,0.5) -- (-5,-2.5) -- (-1.5,-2.5) -- (-1.5,0.5) -- cycle;
\begin{scope}[shift={(6,0)}]
\draw [thick,blue](-5,-1) -- (-3,-1);
\draw [thick,blue](-2,-1) -- (-1.5,-1);
\node[blue] at (-5.5,-1) {$L_{g_2}$};
\draw [thick,blue,-stealth](-4,-1) -- (-3.9,-1);
\begin{scope}[shift={(-2.5,-3)},rotate=-90]
\node[blue] at (-4,-1) {$L_{g_1}$};
\draw [thick,blue](-3.5,-1) -- (-0.5,-1);
\draw [thick,blue,-stealth](-2.5,-1) -- (-2.4,-1);
\end{scope}
\draw [thick](-5,0.5) -- (-5,-2.5) -- (-1.5,-2.5) -- (-1.5,0.5) -- cycle;
\draw [red,fill=red] (-2,-1) node (v1) {} ellipse (0.05 and 0.05);
\draw [red,fill=red] (-3,-1) node (v1) {} ellipse (0.05 and 0.05);
\node[red] at (-3,-1.5) {$O_{g_2}$};
\node[red] at (-2,-1.5) {$O_{g_2^{-1}}$};
\end{scope}
\begin{scope}[shift={(0,-4.5)}]
\draw [thick,blue](-5,-1) -- (-3.5,-1);
\draw [thick,blue](-2,-1) -- (-1.5,-1);
\node[blue] at (-5.5,-1) {$L_{g_2}$};
\draw [thick,blue,-stealth](-4,-1) -- (-3.9,-1);
\begin{scope}[shift={(-2,-3)},rotate=-90]
\node[blue] at (-4,-1) {$L_{g_1}$};
\draw [thick,blue](-3.5,-1) -- (-0.5,-1);
\draw [thick,blue,-stealth](-2.5,-1) -- (-2.4,-1);
\end{scope}
\draw [thick](-5,0.5) -- (-5,-2.5) -- (-1.5,-2.5) -- (-1.5,0.5) -- cycle;
\draw [red,fill=red] (-2,-1) node (v1) {} ellipse (0.05 and 0.05);
\draw [red,fill=red] (-3.5,-1) node (v1) {} ellipse (0.05 and 0.05);
\node[red] at (-3.5,-1.5) {$O_{g_2}$};
\node[red] at (-2,-1.5) {$O_{g_2^{-1}}$};
\node at (-8.5,-1) {=};
\node at (-7,-1) {$\epsilon(g_1,g_2)~\times$};
\end{scope}
\begin{scope}[shift={(7.5,-4.5)}]
\draw [thick,blue](-4.5,-1) -- (-3.5,-1);
\node[blue] at (-4,-0.5) {$L_{g_2}$};
\draw [thick,blue,-stealth](-4,-1) -- (-3.9,-1);
\begin{scope}[shift={(-2,-3)},rotate=-90]
\node[blue] at (-4,-1) {$L_{g_1}$};
\draw [thick,blue](-3.5,-1) -- (-0.5,-1);
\draw [thick,blue,-stealth](-2.5,-1) -- (-2.4,-1);
\end{scope}
\draw [thick](-5,0.5) -- (-5,-2.5) -- (-1.5,-2.5) -- (-1.5,0.5) -- cycle;
\draw [red,fill=red] (-4.5,-1) node (v1) {} ellipse (0.05 and 0.05);
\draw [red,fill=red] (-3.5,-1) node (v1) {} ellipse (0.05 and 0.05);
\node[red] at (-3.5,-1.5) {$O_{g_2}$};
\node[red] at (-4.5,-1.5) {$O_{g_2^{-1}}$};
\node at (-8,-1) {=};
\node at (-6.5,-1) {$\epsilon(g_1,g_2)~\times$};
\end{scope}
\begin{scope}[shift={(0,-9)}]
\begin{scope}[shift={(-2,-3)},rotate=-90]
\node[blue] at (-4,-1) {$L_{g_1}$};
\draw [thick,blue](-3.5,-1) -- (-0.5,-1);
\draw [thick,blue,-stealth](-2.5,-1) -- (-2.4,-1);
\end{scope}
\draw [thick](-5,0.5) -- (-5,-2.5) -- (-1.5,-2.5) -- (-1.5,0.5) -- cycle;
\node at (-8,-1) {=};
\node at (-6.5,-1) {$\epsilon(g_1,g_2)~\times$};
\end{scope}
\begin{scope}[shift={(7.5,-9)}]
\begin{scope}[shift={(-2,-3)},rotate=-90]
\node[blue] at (-4,-1) {$L_{g_1}$};
\draw [thick,blue](-3.5,-1) -- (-2,-1);
\draw [thick,blue,-stealth](-2.5,-1) -- (-2.4,-1);
\draw [thick,blue](-1.5,-1) -- (-0.5,-1);
\draw [red,fill=red] (-2,-1) node (v1) {} ellipse (0.05 and 0.05);
\draw [red,fill=red] (-1.5,-1) node (v1) {} ellipse (0.05 and 0.05);
\node[red] at (-2,-1.5) {$O_{g_1}$};
\node[red] at (-1.5,-0.25) {$O_{g_1^{-1}}$};
\end{scope}
\draw [thick](-5,0.5) -- (-5,-2.5) -- (-1.5,-2.5) -- (-1.5,0.5) -- cycle;
\node at (-8,-1) {=};
\node at (-6.5,-1) {$\epsilon(g_1,g_2)~\times$};
\end{scope}
\begin{scope}[shift={(0,-13)}]
\begin{scope}[shift={(-2,-3.5)},rotate=-90]
\node[blue] at (-2.5,-1.5) {$L_{g_1}$};
\draw [thick,blue](-3,-1) -- (-2,-1);
\draw [thick,blue,-stealth](-2.5,-1) -- (-2.4,-1);
\draw [red,fill=red] (-2,-1) node (v1) {} ellipse (0.05 and 0.05);
\draw [red,fill=red] (-3,-1) node (v1) {} ellipse (0.05 and 0.05);
\node[red] at (-1.5,-1) {$O_{g_1}$};
\node[red] at (-3.5,-1) {$O_{g_1^{-1}}$};
\end{scope}
\draw [thick](-5,0.5) -- (-5,-2.5) -- (-1.5,-2.5) -- (-1.5,0.5) -- cycle;
\node at (-8,-1) {=};
\node at (-6.5,-1) {$\epsilon(g_1,g_2)~\times$};
\end{scope}
\begin{scope}[shift={(7.5,-13)}]
\draw [thick](-5,0.5) -- (-5,-2.5) -- (-1.5,-2.5) -- (-1.5,0.5) -- cycle;
\node at (-8,-1) {=};
\node at (-6.5,-1) {$\epsilon(g_1,g_2)~\times$};
\end{scope}
\end{tikzpicture}
$$
\caption{$T^2$-partition function with insertion of two topological lines, with $\epsilon$ the bi-character. \label{fig:Moves}}
\end{figure}

$\epsilon(g_1,g_2)$ can also be understood as the partition function of the SPT phase on a torus $T^2$ with a $\G0$ background turned on such that we have $g_1$ holonomy along the $A$ cycle of $T^2$ and $g_2$ holonomy along the $B$ cycle of $T^2$. This can {be} seen from the  moves in figure \ref{fig:Moves},
where the last partition function on a bare $T^2$ is trivial for an SPT phase.

Similarly, one can compute the partition function of the SPT phase on a Riemann surface $\Sigma_g$ of genus $g$ and background holonomies $g^A_i, g^B_i$ along cycles $A_i$ and $B_i$ respectively, to be
\be
Z\left[\Sigma_g,\{g^A_i\},\{g^B_i\}\right]=\prod_{i=1}^g\epsilon\left(g_i^A,g_i^B\right)\,.
\ee

\subsubsection{2d $\wh{\G0}$ Gauge Theories}
Now, we shift gears a little bit and discuss 2d gauge theories with $\wh{\G0}$ gauge group in a similar language as above.
A 2d $\wh{\G0}$ gauge theory has two dynamical fields: $a_1$ which is a $\wh{\G0}$ valued 1-cochain, and $b_0$ which is a $\G0$ valued 0-cochain. The action takes the form
\be
S=2\pi\int a_1\cup\delta b_0\,,
\ee
where the cup product is taken using the natural pairing $\wh{\G0}\times\G0\to\R/\Z$. The theory has a $\G0$ 0-form symmetry and a $\wh{\G0}$ 1-form symmetry. The two symmetries have a mixed 't Hooft anomaly due to the fact that the topological local operators generating $\wh{\G0}$ 1-form symmetry are charged under the $\G0$ 0-form symmetry. That is, we have
\be
\begin{tikzpicture}
\draw [red,fill=red] (-2,-1) node (v1) {} ellipse (0.05 and 0.05);
\node[red] at (-2,-1.5) {$O_{\wh g}$};
\begin{scope}[shift={(-2,-3)},rotate=-90]
\node[blue] at (-4,-1) {$L_g$};
\draw [thick,blue](-3.5,-1) -- (-1,-1);
\draw [thick,blue,-stealth](-2.5,-1) -- (-2.4,-1);
\end{scope}
\node at (-1,-1) {=};
\begin{scope}[shift={(3.5,0)}]
\draw [red,fill=red] (-2,-1) node (v1) {} ellipse (0.05 and 0.05);
\node[red] at (-2,-1.5) {$O_{\wh g}$};
\begin{scope}[shift={(0,-3)},rotate=-90]
\node[blue] at (-4,-1) {$L_g$};
\draw [thick,blue](-3.5,-1) -- (-1,-1);
\draw [thick,blue,-stealth](-2.5,-1) -- (-2.4,-1);
\end{scope}
\end{scope}
\node at (0.5,-1) {$\wh g(g)~\times$};
\end{tikzpicture}
\ee
where we have picked arbitrary elements $g\in\G0$, $\wh g\in\wh{\G0}$, and $\wh g(g)\in U(1)$ is obtained by applying the homomorphism $\wh g:~\G0\to U(1)$ to $g\in \G0$.

The corresponding anomaly theory can be expressed in terms of the background fields $B_2$ and $A_1$ for the 1-form and 0-form symmetries as
\be
\cA=\text{exp}\left(2\pi i\int B_2\cup A_1\right) \,.
\ee
Note that there cannot exist a non-zero local operator living at the end of any topological line defect $L_g$ with $g\neq 1$, otherwise $g$ is forced to satisfy $\wh g(g)=1$ for all $\wh g$, which is a contradiction. Below we discuss a generalization where $L_g$ can end.

The generalization involves turning on a discrete theta angle, which modifies the action to
\be
S=2\pi\int\big(a_1\cup\delta b_0+a_1\cup\theta(a_1)\big) \,,
\ee
where $\theta$ is a homomorphism
\be
\theta:~\wh{\G0}\to \G0
\ee
such that
\be\label{apt}
\wh g_2\big(\theta\left(\wh g_1\right)\big)=\wh g_1^{\,\,-1}\big(\theta\left(\wh g_2\right)\big)
\ee
for arbitrary $\wh g_1,\wh g_2\in\wh{\G0}$.

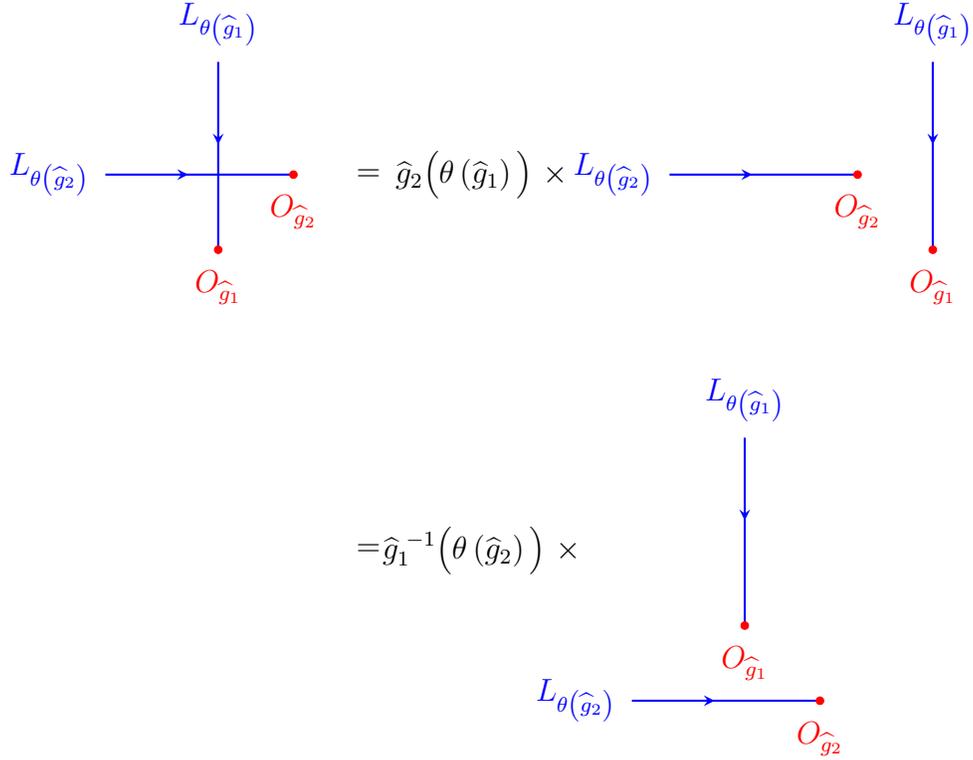
\begin{figure}
$$
\begin{tikzpicture}
\draw [thick,blue](-4.5,-1) -- (-2,-1);
\draw [red,fill=red] (-2,-1) node (v1) {} ellipse (0.05 and 0.05);
\node[blue] at (-5.25,-1) {$L_{\theta\left(\wh g_2\right)}$};
\node[red] at (-2,-1.5) {$O_{\wh g_2}$};
\draw [thick,blue,-stealth](-3.5,-1) -- (-3.4,-1);
\begin{scope}[shift={(-2,-3)},rotate=-90]
\node[blue] at (-4,-1) {$L_{\theta\left(\wh g_1\right)}$};
\draw [thick,blue](-3.5,-1) -- (-1,-1);
\draw [thick,blue,-stealth](-2.5,-1) -- (-2.4,-1);
\draw [red,fill=red] (-1,-1) node (v1) {} ellipse (0.05 and 0.05);
\node[red] at (-0.5,-1) {$O_{\wh g_1}$};
\end{scope}
\node at (-1,-1) {=};
\begin{scope}[shift={(7.5,0)}]
\draw [thick,blue](-4.5,-1) -- (-2,-1);
\draw [red,fill=red] (-2,-1) node (v1) {} ellipse (0.05 and 0.05);
\node[blue] at (-5.25,-1) {$L_{\theta\left(\wh g_2\right)}$};
\node[red] at (-2,-1.5) {$O_{\wh g_2}$};
\draw [thick,blue,-stealth](-3.5,-1) -- (-3.4,-1);
\begin{scope}[shift={(0,-3)},rotate=-90]
\node[blue] at (-4,-1) {$L_{\theta\left(\wh g_1\right)}$};
\draw [thick,blue](-3.5,-1) -- (-1,-1);
\draw [thick,blue,-stealth](-2.5,-1) -- (-2.4,-1);
\draw [red,fill=red] (-1,-1) node (v1) {} ellipse (0.05 and 0.05);
\node[red] at (-0.5,-1) {$O_{\wh g_1}$};
\end{scope}
\end{scope}
\node at (0.5,-1) {$\wh g_2\big(\theta\left(\wh g_1\right)\big)~\times$};
\begin{scope}[shift={(0,-5)}]
\node at (-1,-1) {=};
\begin{scope}[shift={(7,0)}]
\draw [thick,blue](-4.5,-3) -- (-2,-3);
\draw [red,fill=red] (-2,-3) node (v1) {} ellipse (0.05 and 0.05);
\node[blue] at (-5.25,-3) {$L_{\theta\left(\wh g_2\right)}$};
\node[red] at (-2,-3.5) {$O_{\wh g_2}$};
\draw [thick,blue,-stealth](-3.5,-3) -- (-3.4,-3);
\begin{scope}[shift={(-2,-3)},rotate=-90]
\node[blue] at (-4,-1) {$L_{\theta\left(\wh g_1\right)}$};
\draw [thick,blue](-3.5,-1) -- (-1,-1);
\draw [thick,blue,-stealth](-2.5,-1) -- (-2.4,-1);
\draw [red,fill=red] (-1,-1) node (v1) {} ellipse (0.05 and 0.05);
\node[red] at (-0.5,-1) {$O_{\wh g_1}$};
\end{scope}
\end{scope}
\node at (0.5,-1) {$\wh g_1^{\,\,-1}\big(\theta\left(\wh g_2\right)\big)~\times$};
\end{scope}
\end{tikzpicture}
$$
\caption{Mutual locality. \label{fig:mutual}}
\end{figure}

The theta angle term forces the topological local operator $O_{\wh g}$ to live at the end of the topological line operator $L_{\theta\left(\wh g\right)}$
\be
\begin{tikzpicture}
\draw [thick,blue](-4,-1) -- (-2,-1);
\draw [red,fill=red] (-2,-1) node (v1) {} ellipse (0.05 and 0.05);
\node[blue] at (-4.75,-1) {$L_{\theta\left(\wh g\right)}$};
\node[red] at (-2,-1.5) {$O_{\wh g}$};
\draw [thick,blue,-stealth](-3,-1) -- (-2.9,-1);
\end{tikzpicture}
\ee
while the charge of $O_{\wh g}$ under $g\in\G0$ is still $\wh g(g)\in U(1)$. From the mutual locality condition shown in figure \ref{fig:mutual}, 
we recover the condition (\ref{apt}) on $\theta$.

Let us define
\be
{\G0}':=\text{Im}(\theta)\subseteq {\G0} \,.
\ee
Here, ${\G0}'$ is the subgroup of topological line defects that can end. Moreover the topological local operators $O_{\wh g}$ for $\wh g\in \wh{{\G0}/{\G0}'}\subseteq\wh{\G0}$ are unattached to any $L_g$ line. In other words, $\wh{{\G0}/{\G0}'}$ is the 1-form symmetry group of the gauge theory.
The 't Hooft anomaly between ${\G0}$ 0-form symmetry and $\wh{{\G0}/{\G0}'}$ 1-form symmetry is
\be
\cA=\text{exp}\left(2\pi i\int \wh\pi(B_2)\cup A_1\right) \,,
\ee
where
\be
\wh\pi:~\wh{{\G0}/{\G0}'}\to\wh{\G0}
\ee
is the inclusion map.

Now consider the special case ${\G0}'={\G0}$. Since $\wh{\G0}\cong {\G0}$ for finite abelian ${\G0}$, in such a case $\theta$ becomes an isomorphism between ${\G0}$ and $\wh{\G0}$. All $O_{\wh g}$ operators are attached to the ends of $L_g$ lines, and all $L_g$ lines can end. Thus a $\wh{\G0}$ gauge theory with theta angle $\theta$ being an isomorphism must be a ${\G0}$-SPT phase. The bi-character $\epsilon$ of ${\G0}$ associated to the SPT phase is obtained by simply equating the charges under ${\G0}$ 0-form symmetry of the local operators living at the ends $L_g$ lines
\be
\epsilon(g_1,g_2)=\theta^{-1}(g_2)(g_1) \,,
\ee
where
\be
\theta^{-1}:~{\G0}\to\wh{\G0}
\ee
is the homomorphism inverse to $\theta$. The additional requirement (\ref{ap}) on $\epsilon$ is obtained from the additional requirement (\ref{apt}) on $\theta$.

It should be noted that we do not get all the ${\G0}$-SPT phases this way. For example, in the trivial SPT phase, none of the local operators living at the ends of $L_g$ lines are charged under ${\G0}$, while in a $\wh{\G0}$ gauge theory with $\theta$ being an isomorphism, every such operator carries a non-trivial charge under ${\G0}$.

The above $\wh{\G0}$ gauge theory can be understood as the theory obtained after gauging $\wh{\G0}$ 0-form symmetry of a $\wh{\G0}$ SPT phase associated to a bi-character
\be
\wh\epsilon:~\wh{\G0}\times\wh{\G0}\to U(1)
\ee
associated to $\theta$ via
\be
\wh\epsilon(\wh g_1,\wh g_2)=\wh g_2\big(\theta\left(\wh g_1\right)\big)\,.
\ee
The additional condition (\ref{ap}) that $\wh\epsilon$ needs to satisfy then follows from (\ref{apt}).

\subsubsection{General 2d TQFTs with 0-Form Symmetry}
 
Consider a minimal 2d TQFT with $\G0$ 0-form Symmetry described by a subgroup ${\G0}'$ of ${\G0}$ and an element
\be
[\alpha]\in H^2\big({\G0}',U(1)\big)\,.
\ee
The choice of $[\alpha]$ is equivalent to the choice of a bi-character
\be
\epsilon:~{\G0}'\times {\G0}'\to U(1)
\ee
satisfying the condition (\ref{ap}) with ${\G0}$ replaced by ${\G0}'$. This bi-character provides a homomorphism
\be
\phi:~{\G0}'\to\wh{{\G0}'}\,.
\ee
This theory has a $\wh{{\G0}/{\G0}'}$ 1-form symmetry group such that the topological local operators generating it are charged under the ${\G0}$ 0-form symmetry as
\be
\begin{tikzpicture}
\draw [red,fill=red] (-2,-1) node (v1) {} ellipse (0.05 and 0.05);
\node[red] at (-2,-1.5) {$O_{\wh g}$};
\begin{scope}[shift={(-2,-3)},rotate=-90]
\node[blue] at (-4,-1) {$L_g$};
\draw [thick,blue](-3.5,-1) -- (-1,-1);
\draw [thick,blue,-stealth](-2.5,-1) -- (-2.4,-1);
\end{scope}
\node at (-1,-1) {=};
\begin{scope}[shift={(4,0)}]
\draw [red,fill=red] (-2,-1) node (v1) {} ellipse (0.05 and 0.05);
\node[red] at (-2,-1.5) {$O_{\wh g}$};
\begin{scope}[shift={(0,-3)},rotate=-90]
\node[blue] at (-4,-1) {$L_g$};
\draw [thick,blue](-3.5,-1) -- (-1,-1);
\draw [thick,blue,-stealth](-2.5,-1) -- (-2.4,-1);
\end{scope}
\end{scope}
\node at (0.5,-1) {$\wh g\big(\pi(g)\big)~\times$};
\end{tikzpicture}
\ee
where $\wh g\in\wh{{\G0}/{\G0}'}$ and
\be
\pi:~{\G0}\to {\G0}/{\G0}'
\ee
is the natural surjective map.
The topological line defects $L_h$ for $h\in {\G0}'$ can end. Consider a topological local operator $O$ living at the end of $L_h$. Its charge under $\G0$ describes an element of the coset
\be
\wh\cG_h:=\wh i^{-1}\big(\phi(h)\big)\subseteq\wh G \,,
\ee
where
\be
\wh i:~\wh G\to\wh H
\ee
is the surjective map Pontryagin dual to the inclusion $i:~H\to G$. Fusing $O$ with operators $O_{\wh g}$ for $\wh g\in\wh{{\G0}/{\G0}'}$ leads to other local operators living at the end of $L_h$ whose charges under $\G0$ are associated to other elements of the coset $\wh\cG_h$.

We can describe the above theory as a $\wh\cG$ gauge theory, where
\be
\wh\cG=\wh i^{-1}\wh\cH\subseteq\wh{\G0}~;\qquad\wh\cH=\text{Im}(\phi)\subseteq\wh{\G0}'
\ee
The theta angle
\be\label{ta}
\theta:~\wh\cG\to\cG
\ee
is specified as
\be
\theta=i\circ\wt\phi^{-1}\circ\wh i \,,
\ee
where $\wt\phi^{-1}$ is the inverse of the isomorphism
\be
\wt\phi:~\cH\to\wh\cH
\ee
descending from the map $\phi$. Such a gauge theory naturally has a $\cG$ 0-form symmetry, but it can be treated as a theory with ${\G0}$ 0-form symmetry by using the surjective map ${\G0}\to\cG$ Pontryagin dual to the inclusion $\wh\cG\to\wh{\G0}$. The theta angle $\theta$ in (\ref{ta}) describes the discrete torsion $[\alpha_{\wh\cG}]\in H^2\big(\wh\cG,U(1)\big)$ involved in the $\wh\cG$ gauging.

\subsection{2d TQFTs with Non-Abelian $\Gamma^{(0)}$ as Generalized Gaugings}
Now we consider the case of non-abelian $\G0$. As we have seen in section \ref{T0F}, a minimal 2d TQFT with $\G0$ 0-form symmetry can be identified with an indecomposable module category of $\Vec_{\G0}$. Such an indecomposable module category can also be identified with an indecomposable module category of $\Rep(\G0)$. This is related to the fact that the categories $\Vec_{\G0}$ and $\Rep(\G0)$ are \textit{Morita equivalent} (see appendix \ref{App:Cats}).

On the other hand, an indecomposable module category of $\Rep(\G0)$ provides the information for performing a generalized gauging of a 2d topological system carrying $\Rep(\G0)$ categorical symmetry. Consequently, any minimal 2d TQFT with $\G0$ 0-form symmetry is obtained by performing a generalized gauging of the trivial 2d TQFT equipped with $\Rep(\G0)$ categorical symmetry.

We can also take the 2d topological system appearing in the previous paragraph to be the identity 2d defect in a $d$-dimensional QFT $\fT/\G0$ obtained by gauging $\G0$ 0-form symmetry of a $d$-dimensional QFT $\fT$. The Wilson line defects in $\fT/\G0$ coming from this $\G0$ gauging naturally equip the identity 2d defect of $\fT/\G0$ with $\Rep(\G0)$ categorical symmetry. 

Hence, any two-dimensional condensation defect arising from gauging this $\Rep(\G0)$ categorical symmetry is identified with an indecomposable module category for $\Vec_{\G0}$, which in turn is identified with a minimal 2d TQFT carrying $\G0$ 0-form symmetry. Hence, any 2d  defect in the 2-category $\cC_2^{\G0}=\TwoRep(\G0)$ of dual symmetries obtained by $\G0$ 0-form gauging is a condensation defect.

\subsection{2d Condensation Defects in Arbitrary QFTs}
We can generalize the mathematical arguments of the previous subsection to describe two-dimensional condensation defects in an arbitrary $d$-dimensional QFT $\fT$, which may or may not arise by gauging of some other $d$-dimensional QFT.

Such a two-dimensional condensation defect is obtained by performing generalized gauging of the categorical symmetry $\cB_\fT$ localized on the identity two-dimensional defect. The category $\cB_\fT$ can be identified with the category formed by topological line defects of $\fT$. For $d=3$, $\cB_\fT$ is a braided fusion category with a possibly non-trivial braiding. On the other hand, for $d>3$, since two lines cannot link, $\cB_\fT$ is a braided fusion category with a trivial braiding.

Now, as discussed in the previous subsection, such generalized gaugings are associated to module categories for $\cB_\fT$. Thus, we expect the two-dimensional condensation defects in $\fT$ to form the 2-category
\be
\mathsf{Mod}(\cB_\fT)\,,
\ee
namely the 2-category formed by module categories of the 1-category $\cB_\fT$. Moreover, since $\cB_\fT$ is a braided fusion category, the 2-category $\mathsf{Mod}(\cB_\fT)$ carries a monoidal structure and forms a fusion 2-category \cite{douglas2018fusion}. This is a consistency check because one should be able to fuse condensation defects and hence the 2-category formed by condensation defects must carry {a} monoidal structure.

In this paper, we have studied extensively the structure of $\mathsf{Mod}(\cB_\fT)$ for $\cB_\fT=\Rep(\G0)$, in which case we have
\be
\mathsf{Mod}(\cB_\fT)=\TwoRep(\G0) \,.
\ee
We have not studied the structure of $\mathsf{Mod}(\cB_\fT)$ for other $\cB_\fT$ in more detail, and doing so would be an interesting topic for future works.

\section{Gauging Sub-Symmetries and  Gauge Theory Applications}
\label{sec:GaugeTh}

In the previous sections, we discussed that the $(d-1)$-category $\cC_{d-1}^\cS$ describes symmetries in a $d$-dimensional theory $\fT/\cS$ obtained after gauging $\cS$ symmetry of a $d$-dimensional theory $\fT$, where $\cS$ is either a higher-form or a higher-group symmetry. We studied in detail the 2-categorical piece $\cC_2^\cS$ of $\cC_{d-1}^\cS$, which is non-trivial only if $\cS$ is either a 0-form, 1-form or a 2-group symmetry.

In this section, we discuss that these results for $\cC_2^\cS$ allow us to describe non-invertible 2-categorical symmetries of gauge theories in various spacetime dimensions. Some aspects of such non-invertible symmetries of gauge theories have been studied previously by other methods. Our description of $\cC_2^\cS$ is consistent with these previous results, and vastly extends the body of known results regarding these non-invertible symmetries.

We also discuss gauging sub-symmetries of higher-form and higher-group symmetries. In certain scenarios, the symmetry structure obtained after such a sub-symmetry gauging can again be described in terms of the higher-categories $\cC_{d-1}^\cS$. This allows us to describe in detail non-invertible 2-categorical symmetries of many other kinds of gauge theories. Our results are again consistent with previously known results, and vastly extend the body of known results regarding these non-invertible symmetries.



\subsection{2-Categorical Symmetries of Gauge Theories by $\cS$ Gauging}
It is easy to obtain finite 0-form, 1-form and split 2-group symmetries in gauge theories of arbitrary spacetime dimensions. A class of such symmetries is obtained as follows. Consider a $d$-dimensional gauge theory with a gauge group $G$ which is a product of simply connected Lie groups with all matter fields being scalars. The gauge theory has 1-form symmetry group $\G1$ which is a subgroup of the center of $G$ and a 0-form symmetry group $\G0$ descending from outer-automorphisms of the semi-simple gauge algebra. The 0-form symmetry group $\G0$ has an action $\rho$ on the 1-form symmetry group $\G1$ descending from the action of outer-automorphisms on the center of the gauge group $G$, giving rise to a split 2-group symmetry $\bG$.

Now we can discuss three types of gaugings:
\ben
\item Gauging $\G1$ leads to a $d$-dimensional gauge theory with gauge group $G/\G1$. We obtain dual symmetries given by
\be
\cC_{d-1}^{\G1}\,,
\ee
whose 2-categorical piece $\cC_{2}^{\G1}$ simply describes the dual $(d-2)$-form symmetry given by the group $\wh{\G1}$, which is the well-known magnetic $(d-2)$-form symmetry of the $G/\G1$ gauge theory.
\item Gauging $\G0$ leads to a $d$-dimensional gauge theory with disconnected gauge group $G\rtimes\G0$. We obtain dual symmetries given by
\be
\cC_{d-1}^{\G0}\,,
\ee
whose 2-categorical piece $\cC_{2}^{\G0}$ can be understood in great detail using the analysis presented in section \ref{T0F} of this paper.
\item Gauging $\bG$ leads to a $d$-dimensional gauge theory with disconnected gauge group $(G/\G1)\rtimes\G0$. We obtain dual symmetries given by
\be
\cC_{d-1}^{\bG}\,,
\ee
whose 2-categorical piece $\cC_{2}^{\bG}$ can be understood in detail using the analysis presented in section \ref{T2GS} of this paper.
\een
Out of these three cases, only the second and third contain non-invertible symmetries. Moreover, as discussed in section \ref{RCD}, the two-dimensional topological defects associated to the objects of $\cC_{2}^{\G0}$ are all condensation defects, i.e. all of them can be obtained from the identity two-dimensional defect by gauging the categorical symmetry described by Wilson lines. Thus all the information is contained in the Wilson lines for $\G0$, which form the category $\cC_{1}^{\G0}=\Rep(\G0)$. The most exciting case is the third one, as $\cC_{2}^{\bG}$ contains objects that correspond to non-invertible two-dimensional topological defects that are not condensation defects.

Let us consider the following examples realizing the third type of gauging discussed above.

\paragraph{Example 1: $PO(4N)$ Gauge Theory.}
Consider $d$-dimensional pure gauge theory with gauge group $PO(4N)$. This can be obtained by gauging the 2-group symmetry of the $\Spin(4N)$ gauge theory, which has $\G1=\Z_2\times\Z_2$ and $\G0=\Z_2$ acting on $\G1$ by exchanging the two $\Z_2$s. The 2-category $\cC_2^\bG$ of symmetries of the $PO(4N)$ theory is described in section \ref{eg1}. The object $\TQFT^{(V)}$ gives rise to an invertible topological two-dimensional defect in the $PO(4N)$ theory and $\TQFT^{(SC)}$ gives rise to a \textit{non-invertible} topological two-dimensional defect in the $PO(4N)$ theory. Both of these are non-condensation defects. On the other hand, $\TQFT^{(\Z_2)}$ gives rise to a non-invertible topological two-dimensional defect which is a condensation defect. Finally, the object $\TQFT^{(V_{\Z_2})}$ is a non-invertible topological two-dimensional defect which is not a condensation defect, but it can be obtained by performing a gauging localized along $\TQFT^{(V)}$.

\paragraph{Example 2: $PO(4N+2)$ Gauge Theory.}
Consider $d$-dimensional pure gauge theory with gauge group $PO(4N+2)$. This can be obtained by gauging the 2-group symmetry of the $\Spin(4N+2)$ gauge theory, which has $\G1=\Z_4$ and $\G0=\Z_2$ acting on $\G1$ by exchanging the two generators of $\Z_4$. The 2-category $\cC_2^\bG$ of symmetries of the $PO(4N+2)$ theory is described in section \ref{eg2}. The invertibility vs. non-invertibility and condensation vs. non-condensation properties of the resulting topological two-dimensional defects are exactly the same as in the previous example.

\paragraph{Example 3: $PSO(8)\rtimes S_3$ Gauge Theory.}
Consider $d$-dimensional pure gauge theory with gauge group $PSO(8)\rtimes S_3$. This can be obtained by gauging the 2-group symmetry of the $\Spin(8)$ gauge theory, which has $\G1=\Z_2\times\Z_2$ and $\G0=S_3$ acting on $\G1$ by permuting the three order 2 elements in $\Z_2\times\Z_2$. The 2-category $\cC_2^\bG$ of symmetries of the $PSO(8)\rtimes S_3$ theory is described in section \ref{eg3}. The objects $\TQFT^{(\Z_2)}$, $\TQFT^{(\Z_3)}$ and $\TQFT^{(S_3)}$ give rise to non-invertible topological two-dimensional defects in the $PSO(8)\rtimes S_3$ theory which are all condensation defects. On the other hand, $\TQFT^{(SCV)}$ gives rise to a non-invertible topological two-dimensional defect which is not a condensation defect. Finally, the object $\TQFT^{(SCV_{\Z_2})}$ gives rise to a non-invertible topological two-dimensional defect which can be obtained by performing a $\Z_2$ gauging localized along $\TQFT^{(SCV)}$.

\subsection{Gauging 0-Form Symmetry inside a $(d-1)$-Group}
\subsubsection{General Analysis}
Consider a $d$-dimensional theory $\fT$ with a finite $(d-1)$-group symmetry comprising of a finite 0-form symmetry group $\G0$ and a finite abelian $(d-2)$-form symmetry group $\G{d-2}$, along with an action $\rho$ of $\G0$ on $\G{d-2}$.

Let us gauge $\G0$ resulting in the theory $\fT/\G0$. We claim that this gauging procedure converts the $(d-1)$-group symmetry into a $(d-1)$-categorical symmetry described by the $(d-1)$-category 
\be
\cC_{d-1}^{\wt{\G0}}\,,
\ee
where
\be
\wt{\G0}:=\wh{\G{d-2}}\rtimes_{\wh\rho}\G0
\ee
is an auxiliary 0-form symmetry group. The semi-direct product $\rtimes_{\wh\rho}$ is taken according to the action $\wh\rho$ of $\G0$ on $\wh{\G{d-2}}$ dual to the action $\rho$ of $\G0$ on $\G{d-2}$.

This can be seen by performing the $\G0$ gauging in two steps:
\ben
\item We first gauge the $(d-2)$-form symmetry $\G{d-2}$, resulting in the theory $\fT/\G{d-2}$.
The theory $\fT/\G{d-2}$ has a dual 0-form symmetry given by the group $\wh{\G{d-2}}$, which combines with the already present 0-form symmetry given by the group $\G0$, to form a combined 0-form symmetry group given by group $\wt{\G0}$. No other dual symmetries are created as $(d-1)$ is the minimum dimension at which a theory (in particular a TQFT) can have a $(d-2)$-form symmetry.
\item Then we gauge the $\wt{\G0}$ 0-form symmetry of the theory $\fT/\G{d-2}$. This leads to the desired theory $\fT/\G0$.
\een
The construction of $\fT/\G0$ as $\wt{\G0}$ 0-form gauging of $\fT/\G{d-2}$ makes it manifest that the $(d-1)$-group symmetry of $\fT$ becomes a $(d-1)$-categorical symmetry given by $\cC_{d-1}^{\wt{\G0}}$ in the theory $\fT/\G0$.

In particular, the 2-categorical piece 
\be
\cC_2^{\wt{\G0}}=\TwoRep(\wt{\G0})
\ee
of the $\cC_{d-1}^{\wt{\G0}}$ symmetry of $\fT/\G0$ can be analyzed following the results in section \ref{T0F} of this paper.

We know from section \ref{RCD} that all the objects of $\cC_2^{\wt{\G0}}$ describe condensation defects. So, all the essential information regarding $\cC_2^{\wt{\G0}}$ is captured in the 1-categorical piece $\cC_{1}^{\wt{\G0}}$ of $\cC_{d-1}^{\wt{\G0}}$, which is given by
\be\label{cf}
\cC_{1}^{\wt{\G0}}=\Rep(\wt{\G0}) \,.
\ee

\subsubsection{Application to 3d Gauge Theories}

The above general analysis can be applied to 3d gauge theories by recognizing $\G{d-2}=\G1$ as the 1-form symmetry of the gauge theory. 
From the analysis of the previous subsection, we know that gauging $\G0$ symmetry in the gauge theory leads to a 2-category $\cC_2^{\G0}$ of dual symmetries in the resulting $G\rtimes\G0$ gauge theory. The analysis of this subsection describes what happens to the 1-form symmetry $\G1$ under the $\G0$ gauging process. We learn that 1-form symmetry $\G1$ provides extra contributions that {enhance} the 2-categorical symmetry $\cC_2^{\G0}$ to a larger 2-categorical symmetry described by the 2-category $\cC_2^{\wt{\G0}}$.

As we discussed above, all the essential information regarding the 2-categorical symmetry $\cC_2^{\wt{\G0}}$ of the 3d $G\rtimes\G0$ gauge theory is captured in the 1-category of symmetries obtained by restricting the 2-category $\cC_2^{\wt{\G0}}$ to its identity object. A different method for computing this 1-categorical part was described in \cite{Bhardwaj:2022yxj}. The results obtained there are consistent with our result (\ref{cf}).


Let us illustrate the above general procedure below via a few examples. We also discuss the match with the results of \cite{Bhardwaj:2022yxj}.

\paragraph{Example 1: 3d $\Pin^{+}(4N)$ Gauge Theory.}
Consider the pure 3d gauge theory with gauge group Pin$^+(4N)$. This theory can be constructed by gauging $\Z_2$ outer-automorphism 0-form symmetry of the pure 3d $\Spin(4N)$ gauge theory. 
In this case we have
\be
\wt{\G0}=(\Z_2\times\Z_2)\rtimes\Z_2=D_8
\ee
namely the dihedral group of order 8. Thus the symmetries of the Pin$^+$(4N) gauge theory are described by the 2-category
\be
\cC_{2}^{D_8}=\TwoRep(D_8)
\ee
with 1-categorical part being
\be
\cC_{1}^{D_8}=\Rep(D_8) \,.
\ee
Indeed it was found in \cite{Bhardwaj:2022yxj} that the categorical symmetry of pure 3d Pin$^+$(4N) gauge theory is described by $\Rep(D_8)$.

\paragraph{Example 2: 3d Pin$^+(4N+2)$ Gauge Theory.}
Similarly, consider the pure 3d gauge theory with gauge group Pin$^+(4N+2)$. This theory can be constructed by gauging $\Z_2$ outer-automorphism 0-form symmetry of the pure 3d $\Spin(4N+2)$ gauge theory. In this case we have
\be
\wt{\G0}=\Z_4\rtimes\Z_2=D_8\,.
\ee
Thus the symmetries of the Pin$^+(4N+2)$ gauge theory are described by the 2-category
\be
\cC_{2}^{D_8}=\TwoRep(D_8)\,.
\ee
with 1-categorical part being
\be
\cC_{1}^{D_8}=\Rep(D_8)\,.
\ee
Indeed it was found in \cite{Bhardwaj:2022yxj} that the categorical symmetry of pure 3d Pin$^+$(4N+2) gauge theory is described by $\Rep(D_8)$.

\paragraph{Example 3: 3d $\Spin(8)\rtimes S_3$ Gauge Theory.}
Consider the pure 3d $\Spin(8)\rtimes S_3$ gauge theory, which can be constructed by beginning with pure $\Spin(8)$ 3d gauge theory and gauging the $\G0=S_3$ outer-automorphism 0-form symmetry. In this case we have
\be
\wt{\G0}=(\Z_2\times\Z_2)\rtimes S_3=S_4\,.
\ee
Thus the symmetries of the $\Spin(8)\rtimes S_3$ gauge theory are described by the 2-category
\be
\cC_{2}^{D_8}=\TwoRep(S_4)
\ee
with 1-categorical part being
\be
\cC_{1}^{D_8}=\Rep(S_4)\,.
\ee
It was found in \cite{Bhardwaj:2022yxj} that the fusion ring of the categorical symmetry of pure 3d $\Spin(8)\rtimes S_3$ gauge theory coincides with the fusion ring of $\Rep(S_4)$.

\subsection{Gauging the 0-Form Symmetry inside a $(d-2)$-Group}
\subsubsection{General Analysis}
Consider a $d$-dimensional theory $\fT$ with a finite $(d-2)$-group symmetry comprising of a finite 0-form symmetry group $\G0$ and a finite abelian $(d-3)$-form symmetry group $\G{d-3}$ with an action $\sigma$ of $\G0$ on $\G{d-3}$.

Let us gauge $\G0$ resulting in the theory $\fT/\G0$. We claim that this gauging procedure converts the $(d-1)$-group symmetry into a $(d-1)$-categorical symmetry $\cC_{d-1}$ containing the $(d-1)$-category 
\be
\cC_{d-1}^{\bG}
\ee
as a sub-$(d-1)$-category, where $\bG$ is an auxiliary split 2-group comprising of an auxiliary 0-form symmetry $\G0$
and an auxiliary 1-form symmetry
\be
\wt{\G1}=\wh{\G{d-3}}
\ee
with the action $\wh\sigma$ of ${\G0}$ on $\wt{\G1}$.

Moreover, the $(d-2)$-category obtained by restricting the symmetry $(d-1)$-category $\cC_{d-1}$ of $\fT/\G0$ to its identity object is described precisely by
\be
\cC_{d-2}^{\bG}\,.
\ee

This can be seen by performing the $\G0$ gauging in two steps:
\ben
\item We first gauge the $\G{d-3}$ $(d-3)$-form symmetry of the theory $\fT$, resulting in the theory $\fT/\G{d-3}$. We obtain a dual 1-form symmetry described by the group $\wh{\G{d-3}}$, which combines with ${\G0}$ 0-form symmetry to form the split 2-group $\bG$. 
We also obtain new codimension-1 topological defects, that we ignore in the subsequent analysis. This is the reason why $\cC_{d-1}^{\bG}$ turns out to be only a sub-$(d-1)$-category of the symmetry $(d-1)$-category $\cC_{d-1}$ of $\fT/\G0$ coming from the $(d-1)$-group symmetry of $\fT$. However, $\cC_{d-2}^{\bG}$ describes the full symmetry $(d-2)$-category obtained by forgetting non-identity codimension-1 topological defects. 
\item Then we gauge the 2-group symmetry $\bG$ leading to the desired theory $\fT/\G0$.
\een
This construction of $\fT/\G0$ as $\bG$ 2-group gauging of $\fT/\G{d-3}$ justifies our above claimed results regarding the symmetry higher-category of $\fT/\G0$.

In particular, the 2-categorical piece 
\be\label{2cf}
\cC_2^{\bG}=\TwoRep(\bG)
\ee
of the $\cC_{d-1}^{\bG}$ symmetry of $\fT/\G0$ can be described in detail following the analysis presented in section \ref{T2GS} of this paper.

The 1-categorical piece of $\cC_{d-1}^{\bG}$ is
\be
\cC_{1}^{\bG}=\Rep(\G0) \,.
\ee
Some aspects of the 2-categorical piece were studied in detail in \cite{Bhardwaj:2022yxj} using other methods, in the context of 3d gauge theories, but the calculation presented there is valid for any $d$. The results obtained there are consistent with (\ref{cf}).

\subsubsection{Application to 4d Gauge Theories}\label{appcomp}
The above general analysis can be applied to 4d gauge theories by recognizing $\G{d-3}=\G1$ as the 1-form symmetry of the gauge theory. From the analysis of the previous subsections, we know that gauging $\G0$ symmetry in the gauge theory leads to a 2-category $\cC_2^{\G0}$ of dual symmetries in the resulting $G\rtimes\G0$ gauge theory. The analysis of this subsection describes what happens to the 1-form symmetry $\G1$ under the $\G0$ gauging process. We learn that 1-form symmetry $\G1$ provides extra contributions that enhances the 2-categorical symmetry $\cC_2^{\G0}$ to a larger 2-categorical symmetry described by the 2-category $\cC_2^{\bG}$.

A different method for computing some aspects of this 2-categorical part was described in \cite{Bhardwaj:2022yxj}. The results obtained there are consistent with our result (\ref{2cf}).

Let us illustrate the above general procedure below via a few examples. We also discuss the match with the results of \cite{Bhardwaj:2022yxj}.

\paragraph{Example 1: 4d $\Pin^+(4N)$ Gauge Theory.}
Consider the pure 4d gauge theory with gauge group Pin$^+(4N)$. This theory can be constructed by gauging $\Z_2$ outer-automorphism 0-form symmetry of the pure 4d $\Spin(4N)$ gauge theory. 
In this case $\bG$ is the 2-group discussed in section \ref{eg1}. Thus the 2-categorical part of the symmetries of the Pin$^+$(4N) gauge theory is described by the 2-category $\cC_2^\bG$ discussed in detail in section \ref{eg1}.

The object $\TQFT^{(SC)}$ gives rise to a topological surface defect in the Pin$^+(4N)$ gauge theory which is non-invertible and is not a condensation defect. This defect and its fusion properties were discussed in \cite{Bhardwaj:2022yxj}, where this defect was labeled as $D_2^{(SC)}$. Its fusion with itself appearing in (\ref{mf}) matches with the result obtained in \cite{Bhardwaj:2022yxj}.

The object $\TQFT^{(V)}$ was also discussed in \cite{Bhardwaj:2022yxj}, where it was denoted as $D_2^{(V)}$. Its fusion rules with itself and with $\TQFT^{(SC)}$ described in section \ref{eg1} also match the fusion rules computed in \cite{Bhardwaj:2022yxj}.

Other objects $\TQFT^{(\Z_2)}$ and $\TQFT^{(V_{\Z_2})}$ can be obtained by performing condensation on $\TQFT^{(\id)}$ and $\TQFT^{(V)}$ respectively, and are not discussed in \cite{Bhardwaj:2022yxj}. Our analysis provides fusion rules for these objects as well.

\paragraph{Example 2: 4d $\Pin^+(4N+2)$ Gauge Theory.}
Consider the pure 4d gauge theory with gauge group 
Pin$^+(4N+2)$. This theory can be constructed by gauging $\Z_2$ outer-automorphism 0-form symmetry of the pure 4d $\Spin(4N+2)$ gauge theory. 
In this case $\bG$ is the 2-group discussed in section \ref{eg2}. Thus the 2-categorical part of the symmetries of the Pin$^+(4N+2)$ gauge theory is described by the 2-category $\cC_2^\bG$ discussed in section \ref{eg2}.

\paragraph{Example 3: 4d $\Spin(8)\rtimes S_3$ Gauge Theory.}
Consider the pure 4d gauge theory with gauge group $\Spin(8)\rtimes S_3$. This theory can be constructed by gauging $S_3$ outer-automorphism 0-form symmetry of the pure 4d $\Spin(8)$ gauge theory. 
In this case $\bG$ is the 2-group discussed in section \ref{eg3}. Thus the 2-categorical part of the symmetries of the $\Spin(8)\rtimes S_3$ gauge theory is described by the 2-category $\cC_2^\bG$ discussed in section \ref{eg3}.

The object $\TQFT^{(SCV)}$ gives rise to a topological surface defect in the $\Spin(8)\rtimes S_3$ gauge theory which is non-invertible and is not a condensation defect. This defect and its fusion properties were discussed in \cite{Bhardwaj:2022yxj}, where this defect was labeled as $D_2^{(SCV)}$. Its fusion with itself appearing in (\ref{mf2}) matches with the result obtained in \cite{Bhardwaj:2022yxj}.

Other objects can be obtained by performing condensation on $\TQFT^{(\id)}$ and $\TQFT^{(SCV)}$ respectively, and are not discussed in \cite{Bhardwaj:2022yxj}. Our analysis provides fusion rules for these objects as well.

Moreover, we are able to determine the fusion (monoidal product, not composition) of lines living on the $\TQFT^{(SCV)}$ surface, which is not easy to determine using the method of \cite{Bhardwaj:2022yxj} and in fact is not fully determined there\footnote{The fusion of lines described in \cite{Bhardwaj:2022yxj} is incomplete in the following sense. There the fusion is described in terms of a bimodule of an algebra. The object underlying the bimodule is easy to compute and was computed in \cite{Bhardwaj:2022yxj}. However, the morphisms underlying the bimodule are more difficult to compute and were not computed explicitly in \cite{Bhardwaj:2022yxj}.}.

\section{Conclusions and Future Directions}

It is an irrefutable fact that higher-category theory is {intimately} intertwined with symmetries of Quantum Field Theories. The initial step towards this generalization was made in \cite{Gaiotto:2014kfa} by postulating that topological defects in a physical (not necessarily topological) QFT should be interpreted as symmetry generators. 
The extension to higher-form and higher-group symmetries retains the group-like composition, but recent insights have shown that this is only a special subcase of a vastly larger symmetry structure which naturally has a characterization in terms of a higher fusion category: i.e. topological defects of different dimensionality, with a fusion structure, and an intricate interconnection (e.g. in terms of interfaces and boundary conditions). 

This field of higher fusion-categories has seen progress not only in the physics literature, where the approach -- like in this paper -- has been more constructive, guided by the properties of topological defects in QFTs. In tandem, the mathematics community has also been making strikes in developing a rigorous mathematical foundation (see e.g. \cite{douglas2018fusion,Johnson-Freyd:2020usu,decoppet2021weak, decoppet2022multifusion, decoppetRelativeDeligneTensor2022} for very recent works). 

In this paper we derived a universal 2-categorical subcategory of the symmetry category, i.e. a universal sector of topological surfaces, lines and point operators. 
These arise whenever one gauges a 0-form, 1-form or 2-group symmetry. We determine the complete 2-category structures, including the fusion, and the interrelation between 2d defects, 1d interfaces and 0d junctions between interfaces. 

We also proposed that more generally a gauging of a higher-form or higher-group symmetry $\mathcal{S}$ (not restricted to 0-/1-form and 2-groups) in a $d$-dimensional QFT should result in a full $(d-1)$-category, see section \ref{sec:DualCat}. Obviously it would be very desirable to construct this category. 
A systematic way to proceed is to include a 3-category layer on top of the 2-category that we determined here and subsequently increase the dimensionality. In particular for 4d gauge theories, this 3-categorical layer would complete the determination of the universal sector that arises from the dual symmetry after gauging. However, in this case, the 3d TQFTs, which would play the role of objects have a much richer structure -- in terms of modular tensor categories. It will be exciting to determine the symmetries that arise when attaching such MTCs with global symmetries $\mathcal{S}$ to gauge theories. 

Another application of our work is to condensation defects, which are another type of universal symmetries. There is a non-trivial intersection between the two types of universal symmetries: namely the dual symmetries arising from gauging, and the symmetries generated by condensation defects. In fact, we showed that the objects of the 2-category of dual symmetries arising from gauging of a finite 0-form symmetry are all condensation defects. However, this is not true for objects of the 2-category of dual symmetries arising from gauging of a finite 2-group symmetry. We also described the 2-categorical part of condensation defects in any arbitrary quantum field theory in any spacetime dimension.

\subsection*{Acknowledgements}
We thank Thibault D\'ecoppet for numerous explanations, and Mathew Bullimore, Andrea Ferrari, Sahand Seifnashri, Apoorv Tiwari and Matthew Yu for discussions. This work is supported by the  European Union's Horizon 2020 Framework through the ERC grants 682608 (LB, SSN, JW) and 787185 (LB) and in part by the  Simons Foundation Collaboration on ``Special Holonomy in Geometry, Analysis, and Physics", Award ID: 724073, Schafer-Nameki, (SSN, JW).

\paragraph{Note.} 
Related work \cite{Lin:2022xod} and \cite{Bartsch:2022mpm} will appear at the same time and we thank these authors for coordinating submission. 

\appendix

\section{Categorical Matters}
\label{App:Cats}
In this appendix we collect some background and definitions of categorical nature. For in depth discussions of higher categories see e.g. \cite{etingof2016tensor}.

A $1$-category $\mathcal{C}$ consists of a set $\mathrm{Obj}(\cC)$ of objects and a set $\hom(a,b)$ of morphisms for any two objects $a,b\in \mathrm{Obj}(\cC)$. Moreover, for any three objects $a,b,c\in \mathrm{Obj}(\cC)$, one can compose the morphisms
\begin{equation}
    \begin{aligned}
     \hom(a,b) \times \hom(b,c) &\rightarrow \hom(a,c)\\
     f\times g &\mapsto f\circ g
\end{aligned}
\end{equation}
such that the composition is associative, i.e. $(h\circ g) \circ f = h \circ (g \circ f)$. 

A \emph{monoidal} category $\mathcal{C}$ is a category equipped with a functor
\begin{equation}
    \otimes: \qquad \mathcal{C} \times \mathcal{C} \rightarrow \mathcal{C}
\end{equation}
together with a monoidal unit. The monoidal product is associative up to a natural isomorphism
\begin{equation}
    \alpha(a,b,c):\qquad  (a\otimes b)\otimes c \rightarrow a\otimes (b\times c)\,.
\end{equation}
A category $\mathcal{C}$ is said to be \emph{linear} if for any two objects $a,b\in \mathcal{C}$, the set $\hom_{\mathcal{C}}(a,b)$ forms a vector space over $\mathbb{C}$.

An object in $\mathcal{C}$ is said to be \emph{simple} if its endomorphism algebra is one dimensional. An object is said to be \emph{semisimple} if it can be written as a direct sum of finitely many simple objects. A category is said to be \emph{semisimple} if every object is semisimple. In particular, the endomorphism algebra of any object in a semisimple category is semisimple.

We refer the readers to the textbook \cite{etingof2016tensor} for a more careful exposition and \cite{Bhardwaj:2017xup,Chang:2018iay} for  physicists-friendly reviews. We now list the definitions of $1$-categories that we encounter in this article.

\paragraph{Definition: $\Vec$.}
$\Vec$ denotes the category of vector spaces, whose objects are vector spaces (over $\mathbb{C}$) and morphisms are given by linear maps between them. The isomorphism classes of the objects are labelled by the dimension of the vector spaces and there is only $1$ simple object given by the one dimensional vector space $\mathbb{C}$.

\paragraph{Definition: $\cC_{\G0}(\G1)$ and $\Vec_{\G0}$.} Let $\G0$ be a finite group and $\G1$ be a finite abelian group.  $\cC_{\G0}(\G1)$ is a monoidal category (sections 2.3 and 2.11 of \cite{etingof2016tensor}) whose objects $\delta_g$ are labeled by elements $g$ of $\G0$. There are no nontrivial morphisms between different isomorphism classes of objects, i.e. $\hom(\delta_{g_1},\delta_{g_2}) = \emptyset, g_1\neq g_2$,
and 
\be
{\hom(\delta_g,\delta_g) = \G1\,,\qquad g\in \Gamma^{(0)} \,.}
\ee
The associativity isomorphism is the identity. The monoidal structure is given by the functor $\otimes$ defined by $\delta_g\otimes \delta_h = \delta_{gh}$ for objects and $a\otimes b = ab$ for morphisms. 

If $\G1$ is trivial, then we call the corresponding category $\cC_{\G0}$. On the other hand, if $\G0$ is trivial, then we call the corresponding category $\cC_\id(\G1)$.

The special case 
\be
\cC_{\G0}(\mathbb{C}^\times) \,,
\ee 
can be given a $\mathbb{C}$-linear category structure, often denoted as $\Vec_{\G0}$, i.e. the category of finite dimensional $G$-graded vector spaces. It is a pointed fusion category with the tensor product $\otimes$ and the unit object $\mathbb{C}$. 

\paragraph{Definition: 2-Group $\mathbb{G}$} A (weak) 2-group $\mathbb{G}$ is a monoidal category whose objects are all invertible (with respect to the monoidal product $\otimes$) and all morphisms are isomorphisms \cite{etingof2016tensor}. 
It is known that the equivalence classes of 2-groups are specified by a quadruple 
\be
(\G0,\G1,\rho,[\omega])\,,
\ee
where $\G0$, referred to as 0-form symmetry group, is a finite group and $\G1$, referred to as 1-form symmetry group, is a finite abelian $\G0$-module, with the action $\rho$
\be
\rho:\quad \G0\ \to \ \mathsf{Aut}(\G1)\,.
\ee
Furthermore, $[\omega]\in H^3_\rho(\G0,\G1)$ is a twisted 3-cocycle. A 2-group $\bG$ is said to be of split type if $[\omega]$ is the trivial cohomology class. Otherwise it is said to be non-split.

The corresponding monoidal category, denoted as $\cC_{\G0}^{\omega}(\G1,\rho)$  (sections 2.3 and 2.11 of \cite{etingof2016tensor}), is a generalization of $\cC_{\G0}(\G1)$ defined previously, where 
\begin{itemize}
    \item the monoidal product of the morphisms is modified as follows: if $a: \delta_{g} \rightarrow \delta_{g}$ and $b: \delta_{h} \rightarrow \delta_{h}$ then $a \otimes b=a g(b)$, where $g(b):=\rho(g) b$.
    \item the associativity isomorphism is not the identity,  but given by a twisted 3-cocycle $\omega\in Z^3_\rho\big(\G0,\G1\big)$.
\end{itemize}
 
\paragraph{Definition: $\Rep(G)$.}
Again let $G$ be a finite group. 
$\Rep(G)$ denotes the category of representations of $G$. Namely, the objects are finite dimensional linear representations of $G$ and morphisms are $G$-intertwiners. This is a  monoidal category with the tensor product given by the usual tensor product of representations and the unit object is the trivial representation.

\paragraph{Definition: Module 1-Category.}

Let $\mathcal{C}$ be a semisimple monoidal 1-category. A left module category  over $\mathcal{C}$, denoted by $\mathcal{M}_{\mathcal{C}}$, is a semi-simple 1-category equipped with an action of $\mathcal{C}$, i.e. a functor $\mathcal{C}\times \mathcal{M}_{\mathcal{C}} \rightarrow \mathcal{M}_{\mathcal{C}}$ and a natural isomorphism
\begin{equation}
m_{X, Y, M}:(X \otimes Y) \otimes M \stackrel{\sim}{\longrightarrow} X \otimes(Y \otimes M), \quad X, Y \in \mathcal{C}, M \in \mathcal{M}_{\mathcal{C}}\,,
\end{equation}
satisfying the module associativity constraint (see e.g. \cite{etingof2016tensor}). In particular, when $\mathcal{C} = \Vec_G$ for a finite group $G$, indecomposable semisimple module categories are known  to be labelled by a pair $(H,\alpha)$, with $H$ is a (conjugacy class of) subgroup of $G$, and $\alpha \in H^2(H,U(1))$ is a 2-cocycle. We will denote such a  module category by $\mathcal{M}_{\Vec_G}(H,\alpha)$.

Two categories $\mathcal{C}$ and $\mathcal{D}$ are said to be \emph{Morita equivalent} if there exists a module category $\mathcal{M}_\mathcal{C}$ over $\mathcal{C}$ such that $\mathcal{D}$ is the category of $\mathcal{C}$-module functors from $\mathcal{M}_{\mathcal{C}}$ to itself. In particular, it is known that $\Vec(G)$ and $\Rep(G)$ are Morita equivalent for any finite group $G$.

\paragraph{Definition: (Weak) 2-Categories.} Another key definition are (weak) 2-categories. Roughly speaking, a $2$-category consists of a collection of objects, a collection of $1$-morphisms between objects and a collection of $2$-morphisms between $1$-morphisms. More precisely, between any two objects $A,B$ in a $2$-category $\mathcal{C}$, there is a 1-category $\hom_\mathcal{C}(A,B)$, whose objects are referred to as 1-morphisms from $A$ to $B$. Given any two 1-morphisms $f,g\in \hom_\mathcal{C}(A,B)$, we have a set  $\hom_{\hom_\mathcal{C}(A,B)}(f,g)$ of 2-morphisms. Two 2-morphisms $\alpha,\beta \in \hom_{\hom_\mathcal{C}(A,B)}(f,g)$ can compose vertically $\alpha\circ \beta \in \hom_{\hom_\mathcal{C}(A,B)}(f,g)$.

Furthermore, for any three object $A,B,C\in \mathrm{Obj}(\mathcal{C})$, there is a composition functor 
\begin{align}
    \circ_{A,B,C}: \hom(A,B) \times \hom(B,C) & \  \rightarrow \ \hom(A,C)\\
    f\times g & \ \mapsto \ f\circ g\,.
\end{align}
Unlike 1-categories, this composition functor is not required to be associative. Rather, for any three 1-morphisms $f\in \hom(A,B)$, $g\in \hom(B,C)$ and $h \in \hom(C,D)$ between four objects $A,B,C,D$, there is a 2-isomorphism
\begin{equation}
\alpha_{f, g, h} \in \hom_{\hom(A, D)}((f \circ g) \circ h, f \circ(g \circ h))\,.
\end{equation}
Using the composition functor, one can also compose the 2-morphisms `horizontally', i.e.
given $\alpha_1 \in \hom_{\hom_\mathcal{C}(A,B)}(f_1,g_1)$ and $\alpha_2 \in \hom_{\hom_\mathcal{C}(B,C)}(f_2,g_2)$, one get $\alpha_1\circ \alpha_2 \in \hom_{\hom_{\mathcal{C}}(A,C)}(f_1\circ f_2,g_1\circ g_2)$.

A 2-category is said to be \emph{linear} if sets of 2-morphisms are all vector spaces over $\mathbb{C}$. A linear 2-category is said to be \emph{locally semisimple} if all of its $\hom$ categories are semisimple \cite{douglas2018fusion}. An object $X$ in a locally semisimple 2-category is said to be \emph{simple} if its identity 1-morphism $1_X\in \hom_\mathcal{C}(X,X)$ is simple. Note that in a 1-category,  the endomorphism algebra of any simple object is $\mathbb{C}$ and there are no nontrivial morphisms between inequivalent \emph{simple} objects. This is not the case for 2-categories and there can be nontrivial morphisms between inequivalent simple objects. A prototypical example is given by $\mathrm{Mod}(\Vec(\mathbb{Z}_2))$ to be defined below and its 2-category structure is depicted in \eqref{tikz:celldiagram2RepZ2}.

We will omit the technical details and again we will refer the readers to \cite{baez1997introduction,simpson2011homotopy,douglas2018fusion} for a more careful introduction. Examples of 2-categories that we will encounter in this article are as follows:
\paragraph{Definition: 2-Category $\TwoVec$.}
$\TwoVec$ denotes the category of 2-vector spaces, whose objects are given by finite semisimple 1-categories, 1-morphisms are linear functors and 2-morphisms are natural transformations.

\paragraph{Definition: 2-Category $\ModC(\mathcal{C})$ of Module Categories.}
The 2-category $\ModC(\mathcal{C})$ of module categories over a fusion category $\mathcal{C}$ is a finite semisimple 2-category.
Its objects are module categories $\mathcal{M}_{\mathcal{C}}$ defined above. 
In particular, given two module categories $\mathcal{M}_{\mathcal{C}}$ and $\mathcal{N}_{\mathcal{C}}$ over $\mathcal{C}$, it is known that the category $\text{Hom}_{\ModC(\mathcal{C})}(\mathcal{M}_{\mathcal{C}},\mathcal{N}_{\mathcal{C}})$ is finite and semisimple \cite{douglas2018fusion}. If two 1-categories $\mathcal{C}$ and $\mathcal{D}$ are Morita equivalent, then $\ModC(\mathcal{C})$ and $\ModC(\mathcal{D})$ are the same (more precisely there is a 2-equivalence between them)\cite{naiduCategoricalMoritaEquivalence2006,etingof2016tensor}. The 2-category $\TwoRep(G)$ we are mostly interested in is then defined as 
\begin{equation}
    \TwoRep(G) := \ModC(\Vec(G))\cong \ModC(\Rep{(G)})\,.
\end{equation}

\paragraph{Definition: 2-Representations of 2-Group $\TwoRep_{\mathcal{C}}\big(\mathbb{G}\big)$.}
The objects in $\TwoRep_{\mathcal{C}}\big(\mathbb{G}\big)$ are representations of a 2-group $\mathbb{G}$ on a 2-category $\mathcal{C}$, which is known \cite{Elgueta} to be given by a quadruple $(X,\sigma,\beta,\alpha)$: $X$ is an object of $\mathcal{C}$; $\sigma: \G0\rightarrow \pi_0(\mathrm{Aut}(X))$ and $\beta:\G1\rightarrow \pi_1(\mathrm{Aut}(X))$ are group morphism to the homotopy groups of the autoequivalences of $X$. Furthermore, $\alpha\in C^2(\G0,\pi_1(\mathrm{Aut}(X) )$ is a 2-cochain such that
\begin{equation}
\delta \alpha = \beta(\omega)-\omega_X(\sigma,\sigma,\sigma)
\end{equation}
with $\omega_X$ a 3-cocycle of $\mathrm{Aut}(X)$. 
The 2-category of 2-representations of $\mathbb{G}$ will be denoted by $\TwoRep_{\mathcal{C}}\big(\mathbb{G}\big)$. In the main text, we study $\mathbb{G} = \cC_{\G0}^\omega(\G1,\rho)$ and $\mathcal{C} = 2\Vec$. An object $X\in 2\Vec$ is a semisimple 1-category specified by a nonnegative integer $n$. Furthermore, we have $\sigma: \G0\rightarrow S_n$, $\beta: \G1\rightarrow (\mathbb{C}^*)^n$ and 3-cochain $\alpha$ satisfies that
\begin{equation}
    \delta_\rho \alpha = \beta(\omega)\,.
\end{equation}
The corresponding 2-category will be denoted as $\TwoRep(\cC_{\G0}^\omega(\G1,\rho))$.

\section{Example: $\Gamma^{(0)} = \mathbb{Z}_2 \times \mathbb{Z}_2$ and $\TwoRep(\G0 = \Z_2\times \Z_2)$}
\label{App:Z2Z2}

The simplest example with nontrivial second cohomology group is $\Gamma^{(0)} = \mathbb{Z}_2\times \mathbb{Z}_2$, realized as
\begin{equation}
    \Gamma^{(0)} = \langle x,y\mid x^2 = y^2 = e, xy = yx \rangle\,.
\end{equation}
There are five subgroups up to conjugation $\id$,$\langle x \rangle$,$\langle y \rangle$,$\langle xy \rangle$ and $\G0$. The nontrivial cohomology group is 
\be
H^2(\mathbb{Z}_2\times \mathbb{Z}_2,U(1)) = \mathbb{Z}_2
\ee
with the only nontrivial 2-cochain
\begin{equation}
    \mu(y,x) = -1\,.
\end{equation}
There are 6 objects in the $2$-category
\begin{equation}
    \TQFT^{(\id)}, \ \TQFT^{(\id,\mu)}, \ \TQFT^{(\G0/\langle x \rangle)},\  \TQFT{^{(\G0/\langle y \rangle)}},\  \TQFT{^{(\G0/\langle xy \rangle)}}, \ \TQFT{^{(\G0)}}\,,
\end{equation}
where again we label the TQFT by superscript $\G0/{\G0}'$ and we have a TQFT $ \TQFT^{(\id,\mu)}$ corresponding to ${\G0}' = \G0$ with nontrivial 2 cocycle $\mu$. The order of the double cosets is
\begin{equation}
\begin{array}{|c||c|c|c|c|c|}\hline
|\Gamma^{(0)'}_1\backslash \G0 /\Gamma^{(0)'}_2| & \id & \langle x\rangle & \langle y\rangle & \langle x y\rangle & \G0 \\ \hline
\hline \id & 4 & 2 & 2 & 2 & 1  \\
\hline \langle x\rangle & 2 & 2 & 1 & 1 & 1 \\
\hline \langle y\rangle & 2 & 1 & 2 & 1 & 1  \\
\hline\langle x y\rangle & 2 & 1 & 1 & 2 & 1  \\
\hline \G0 & 1 & 1 & 1 & 1 & 1  \\
\hline
\end{array}
\end{equation}
The fusion of the objects are given by \eqref{eq:fusionofobjectsBurnside} (or equivalently \eqref{G0fus}) is: 
\begin{equation*}
\begin{array}{|c||c|c|c|c|c|c|}\hline
 & \TQFT{^{(\G0)}} & \TQFT^{(\G0/\langle x \rangle)} & \TQFT{^{(\G0/\langle y \rangle)}} & \TQFT{^{(\G0/\langle xy \rangle)}} & \TQFT^{(\id)} & \TQFT^{(\id,\mu)} \\\hline
\hline\TQFT{^{(\G0)}} & 4 \TQFT{^{(\G0)}} & 2\TQFT{^{(\G0)}}& 2 \TQFT{^{(\G0)}} & 2 \TQFT{^{(\G0)}} &\TQFT{^{(\G0)}} & \TQFT{^{(\G0)}} \\ 
\hline\TQFT^{(\G0/\langle x \rangle)} & 2\TQFT{^{(\G0)}} & 2\TQFT^{(\G0/\langle x \rangle)} & \TQFT{^{(\G0)}} & \TQFT{^{(\G0)}} & \TQFT^{(\G0/\langle x \rangle)} & \TQFT^{(\G0/\langle x \rangle)} \\
\hline \TQFT{^{(\G0/\langle y \rangle)}} & 2\TQFT{^{(\G0)}}  & \TQFT{^{(\G0)}} & 2\TQFT{^{(\G0/\langle y \rangle)}} & \TQFT{^{(\G0)}} & \TQFT{^{(\G0/\langle y \rangle)}} & \TQFT{^{(\G0/\langle y \rangle)}} \\
\hline\TQFT{^{(\G0/\langle xy \rangle)}} & 2\TQFT{^{(\G0)}} & \TQFT{^{(\G0)}} & \TQFT{^{(\G0)}}  & 2\TQFT{^{(\G0/\langle xy \rangle)}} & \TQFT{^{(\G0/\langle xy \rangle)}} & \TQFT{^{(\G0/\langle xy \rangle)}} \\
\hline\TQFT^{(\id)} & \TQFT{^{(\G0)}} & \TQFT{^{(\G0/\langle x\rangle)}}  & \TQFT{^{(\G0/\langle y\rangle)}}  & \TQFT{^{(\G0/\langle xy\rangle)}} & \TQFT^{(\id)} & \TQFT^{(\id,\mu)} \\
\hline\TQFT^{(\id,\mu)} & \TQFT{^{(\G0)}} & \TQFT{^{(\G0/\langle x\rangle)}} & \TQFT{^{(\G0/\langle y\rangle)}}   & \TQFT{^{(\G0/\langle xy\rangle)}}  & \TQFT^{(\id,\mu)} & \TQFT^{(\id)} \\
\hline
\end{array}
\end{equation*}
Note that the table is symmetric so the fusion is commutative. In particular, $\TQFT^{(\id)}$ is the monoidal unit.

\begin{figure}[t]
    \centering
\begin{tikzcd}
{\begin{array}{c}\bullet\\ \TQFT^{(\id)}\end{array}} \arrow["\mathrm{Rep}(\G0)"', loop, distance=2em, in=125, out=55] \arrow[rrrr, leftrightarrow, "\mathrm{Rep}_\mu(\G0)" description, bend left] \arrow[dddd, "\mathrm{Rep}(\mathbb{Z}_2)" description, bend right, leftrightarrow] \arrow[rdd, "\mathrm{Rep}(\mathbb{Z}_2)" description, leftrightarrow] & &  &   & {\begin{array}{c}\bullet\\ \TQFT^{(\id,\mu)}\end{array}} \arrow["\mathrm{Rep}(\G0)"', loop, distance=2em, in=125, out=55] \arrow[dddd, "\mathrm{Rep}(\mathbb{Z}_2)" description, bend left, leftrightarrow] \arrow[ldd, leftrightarrow,  "\Vec" description] \arrow[llldd,  leftrightarrow, "\mathrm{Rep}(\mathbb{Z}_2)" description] \\&                         &  &   &\\& {\begin{array}{c}\bullet\\ \TQFT^{(\G0/\langle xy\rangle)}\end{array}} \arrow["2\times \Rep(\Z_2)"', loop, distance=2em, in=215, out=145] \arrow[rr,leftrightarrow,"\Vec(\Z_2)" description] \arrow[rrrdd,  leftrightarrow, "\Vec" description] &  & {\begin{array}{c}\bullet\\ \TQFT^{(\G0)}\end{array}} \arrow["\Vec(\G0)"', loop, distance=2em, in=35, out=325] \arrow[llluu, "\Vec" description, leftrightarrow] \arrow[llldd, "\Vec(\Z_2)" description, leftrightarrow] &   \\
&  &  & &   \\
{\begin{array}{c}\bullet\\ \TQFT^{(\G0/\langle x\rangle)}\end{array}} \arrow["2\times\Rep(\mathbb{Z}_2)"', loop, distance=2em, in=305, out=235, leftrightarrow] \arrow[rrrr, "\Vec" description, bend right, leftrightarrow] \arrow[ruu, "\Vec" description, leftrightarrow]&  &  &                 & {\begin{array}{c}\bullet\\ \TQFT^{(\G0/\langle y\rangle)}\end{array}} \arrow["2\times\Rep(\Z_2)"', loop, distance=2em, in=305, out=235] \arrow[luu, "\Vec{\mathbb{Z}_2}" description, leftrightarrow]               \end{tikzcd}
\caption{2-category structure of $\TwoRep(\G0 = \Z_2\times \Z_2)$. There are $6$ simple objects. The categories of 1-morphisms are computed from \eqref{eq:1morphisms} and labelled on the arrow.} 
\label{fig:category2RepZ2Z2}
\end{figure}
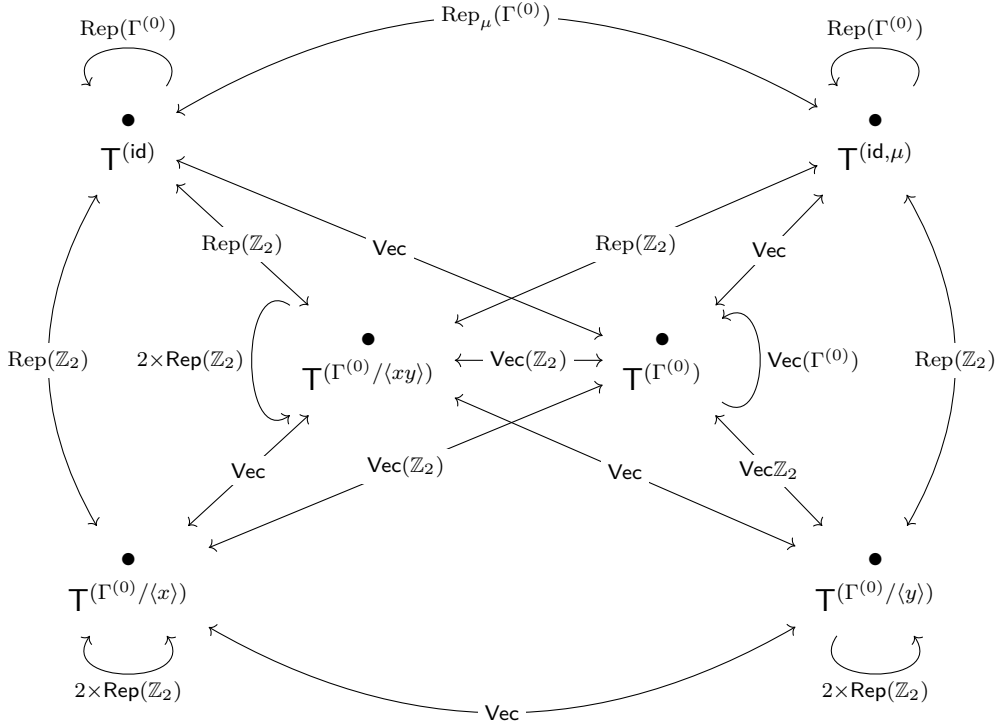

1-morphisms are given by \eqref{eq:1morphisms}. Recall that $\G0 = \mathbb{Z}_2\times \mathbb{Z}_2$ has four inequivalent linear irreducible representations and one projective representation. Denote the category of linear rep of $\G0$ as $\mathrm{Rep}(\G0)$ and category of projective rep with nontrivial Schur multiplier $\mu$ as $\mathrm{Rep}_\mu(\G0)$. Note that usually two projective representations are considered equivalent if they differ by a one-dimensional representation, but we don't make such identification here. So both $\mathrm{Rep}(\G0)$ and $\mathrm{Rep}_\mu(\G0)$ have four objects. The 2-category structure is then shown in figure \ref{fig:category2RepZ2Z2}.

We exemplify some compositions of the morphisms:
\begin{equation}
 \mathcal{I}^{(\id), (\G0)} \circ \mathcal{I}^{(\G0), (\id)} = \sum_{R_j\in \mathrm{Rep}(\G0)}\mathcal{I}^{(\id),(\id)}(R_j) \,.
\end{equation}
This is indeed, the direct analogue of \eqref{eq:fusionZ2xy} in $\TwoRep(\mathbb{Z}_2)$.

Next, the interface between $\TQFT^{(\id)}$ and $\TQFT^{(\id,\mu)}$ is labelled by a (projective) representation $R_j^{-\mu}$ of $\G0$, with $j \in \mathbb{Z}_2\times \mathbb{Z}_2$ and a nontrivial Schur multiplier $-\mu$, i.e. an element in $ \mathrm{Rep}_{-\mu}(\G0)$.
\begin{equation}
 \mathcal{I}^{(\id), (\id,\mu)}(R_j^{-\mu}) \circ \mathcal{I}^{(\id,\mu), (\id,0)}(R_k^{\mu}) = \mathcal{I}^{(\id,0), (\id,0)}(R_j^{-\mu}\otimes R_k^{\mu})\,, \label{eq:G0GmuG0}
\end{equation}
where $R_j^{-\mu}\otimes R_k^{\mu}$ is thought of as a (genuine) representation of $\G0$.

Next, look at the morphisms between $\TQFT^{(\G0/\langle x\rangle)}$ and $\TQFT^{(\G0/\langle y\rangle)}$. Both have two vacua and all four vacuum pairs are in a single orbit of $\G0$, so there is a single morphism $\mathcal{I}^{(\G0/\langle x\rangle), (\G0/\langle y\rangle)}$ from $\TQFT^{(\G0/\langle x\rangle)}$ to $\TQFT^{(\G0/\langle y\rangle)}$, and a single morphism $\mathcal{I}^{(\G0/\langle y\rangle), (\G0/\langle x\rangle)}$ in the opposite direction. The composition of these two gives us an endomorphism in $\TQFT^{(\G0/\langle x\rangle)}$, which from \eqref{eq:1morphisms} is an element in two copies of $\Rep(\mathbb{Z}_2)$ 
\begin{align}
 \mathcal{I}^{(\G0/\langle x\rangle), (\G0/\langle y\rangle)} \circ \mathcal{I}^{(\G0/\langle y\rangle), (\G0/\langle x\rangle)} = \sum_{X= 1,2}\sum_{R\in \mathrm{Rep}(\mathbb{Z}_2)}\mathcal{I}_X^{(\G0/\langle x\rangle),(\G0/ \langle x\rangle)}(R) \,,
\end{align}
where $X$ labels two different copies of $\Rep(\mathbb{Z}_2)$, corresponding to two different orbits of vacuum pairs.

It is also instructive to look at a case with ${\G0_{12}}'$,${\G0_{13}}'$,${\G0_{23}}'$ all being trivial. For example, the category of 1-morphisms between any two of $\TQFT^{(\G0/\langle x\rangle)}$,  $\TQFT^{(\G0/\langle y\rangle)}$ and $\TQFT^{(\G0/\langle xy\rangle)}$ is $\Vec$ since all the vacuum pairs are in a single orbit. From \eqref{eq:compositionof1morphismvs}, two vacua of $\TQFT^{(\G0/\langle y\rangle)}$ give us identical compositions, and hence we have
\begin{equation}
        \mathcal{I}^{(\G0/\langle x\rangle),(\G0/\langle y\rangle)} \circ \mathcal{I}^{(\G0/\langle y\rangle), (\G0/\langle xy\rangle)} =  2\mathcal{I}^{(\G0/\langle x\rangle),(\G0/\langle xy\rangle)} \,.
\end{equation}
However, let us emphasize it is not true that one always gets a sum of 1-morphism corresponding to the vacua of the TQFT in the middle, as shown already in \eqref{eq:compositionof1morphismZ2Z2id} and \eqref{eq:compositionof1morphismidZ2Z2}. 
Let us further illustrate this by looking at the composition of 1-morphism between $\TQFT^{(\id)}$ and $\TQFT^{(\G0/\langle xy\rangle)}$ and 1-morphism between $\TQFT^{(\G0/\langle xy\rangle)}$ and $\TQFT^{(\G0)}$. The former is labelled by a representation $R$ of $\mathbb{Z}_2 = \langle xy\rangle$ and the latter is labelled by the orbits $X$ of vacuum pairs, with $X=1,2$. The composition is a single copy of $\mathcal{I}^{(\id),(\G0)}$ regardless of the representation $R$ and the orbit $X$:
\begin{equation}
    \mathcal{I}^{(\id), (\G0/\langle xy\rangle)}(R) \circ \mathcal{I}^{(\G0/\langle xy\rangle), (\G0)}_X = \mathcal{I}^{(\id),(\G0)}, \quad \forall R\in \Rep{(\Z_2)},X = 1,2 \,.
\end{equation}


\bibliographystyle{ytphys}
\let\bbb\bibitem\def\bibitem{\itemsep4pt\bbb}
\bibliography{2dTFT}
\end{document}